\documentclass[12pt]{article}
\usepackage{array}
\usepackage{tabularx}
\usepackage{amsmath,amssymb,amsfonts,epsfig,cite,setspace,bigstrut,framed,eufrak}
\usepackage[all]{xy}
\usepackage{color}
\usepackage{pifont}
\usepackage{xcolor}

\usepackage{mathtools}

\makeatletter \@addtoreset{equation}{section} \makeatother
\renewcommand{\theequation}{\thesection.\arabic{equation}}

\begin{document}

\vskip 0.25in

\newcommand{\todo}[1]{{\bf\color{blue} !! #1 !!}\marginpar{\color{blue}$\Longleftarrow$}}
\newcommand{\nn}{\nonumber}
\newcommand{\comment}[1]{}
\newcommand\T{\rule{0pt}{2.6ex}}
\newcommand\B{\rule[-1.2ex]{0pt}{0pt}}

\newcommand{\CO}{{\cal O}}
\newcommand{\cI}{{\cal I}}
\newcommand{\cM}{{\cal M}}
\newcommand{\cW}{{\cal W}}
\newcommand{\cN}{{\cal N}}
\newcommand{\cR}{{\cal R}}
\newcommand{\cH}{{\cal H}}
\newcommand{\cK}{{\cal K}}
\newcommand{\cT}{{\cal T}}
\newcommand{\cZ}{{\cal Z}}
\newcommand{\cO}{{\cal O}}
\newcommand{\cQ}{{\cal Q}}
\newcommand{\cB}{{\cal B}}
\newcommand{\cC}{{\cal C}}
\newcommand{\cD}{{\cal D}}
\newcommand{\cE}{{\cal E}}
\newcommand{\cF}{{\cal F}}
\newcommand{\cA}{{\cal A}}
\newcommand{\cX}{{\cal X}}
\newcommand{\IA}{\mathbb{A}}
\newcommand{\IP}{\mathbb{P}}
\newcommand{\IQ}{\mathbb{Q}}
\newcommand{\IH}{\mathbb{H}}
\newcommand{\IR}{\mathbb{R}}
\newcommand{\IC}{\mathbb{C}}
\newcommand{\IF}{\mathbb{F}}
\newcommand{\IS}{\mathbb{S}}
\newcommand{\IV}{\mathbb{V}}
\newcommand{\II}{\mathbb{I}}
\newcommand{\IZ}{\mathbb{Z}}
\newcommand{\re}{{\rm Re}}
\newcommand{\im}{{\rm Im}}
\newcommand{\tr}{\mathop{\rm Tr}}
\newcommand{\ch}{{\rm ch}}
\newcommand{\rk}{{\rm rk}}
\newcommand{\ext}{{\rm Ext}}
\newcommand{\bi}{\begin{itemize}}
\newcommand{\ei}{\end{itemize}}
\newcommand{\beq}{\begin{equation}}
\newcommand{\eeq}{\end{equation}}
\newcommand{\bea}{\begin{eqnarray}}
\newcommand{\eea}{\end{eqnarray}}
\newcommand{\ba}{\begin{array}}
\newcommand{\ea}{\end{array}}

\newcommand{\CN}{{\cal N}}
\newcommand{\y}{{\mathbf y}}
\newcommand{\z}{{\mathbf z}}
\newcommand{\C}{\mathbb C}\newcommand{\R}{\mathbb R}
\newcommand{\CA}{\mathbb A}
\newcommand{\CP}{\mathbb P}
\newcommand{\cP}{\mathcal P}
\newcommand{\tmat}[1]{{\tiny \left(\begin{matrix} #1 \end{matrix}\right)}}
\newcommand{\mat}[1]{\left(\begin{matrix} #1 \end{matrix}\right)}
\newcommand{\diff}[2]{\frac{\partial #1}{\partial #2}}
\newcommand{\gen}[1]{\langle #1 \rangle}

\newtheorem{theorem}{\bf THEOREM}
\newtheorem{proposition}{\bf PROPOSITION}
\newtheorem{observation}{\bf OBSERVATION}

\def\theequation{\thesection.\arabic{equation}}
\newcommand{\setall}{
	\setcounter{equation}{0}
}
\renewcommand{\thefootnote}{\fnsymbol{footnote}}

\begin{titlepage}
\vfill
\begin{flushright}
{\tt\normalsize KIAS-P21044}\\

\end{flushright}
\vfill
\begin{center}
{\Large\bf Aspects of 5d Seiberg-Witten Theories on $\IS^1$}

\vskip 1.5cm

Qiang Jia and Piljin Yi
\vskip 5mm

{\it School of Physics,
Korea Institute for Advanced Study, Seoul 02455, Korea}

\end{center}
\vfill

\begin{abstract}
We study the infrared physics of 5d $\cN=1$ Yang-Mills theories compactified on $\IS^1$, with a view toward 4d and 5d limits.
Global structures of the simplest Coulombic moduli spaces are outlined, with an emphasis on how multiple planar 4d Seiberg-Witten geometries are embedded in the cigar geometry of a single 5d theory on $\IS^1$. The Coulomb phase boundaries in the decompactification limit are given particular attention and related to how the wall-crossings by 5d BPS particles turn off. On the other hand, the elliptic genera of magnetic BPS strings do wall-cross and retain the memory of 4d wall-crossings, which we review with the example of dP$_2$ theory. Along the way, we also offer a general field theory proof of the odd shift of electric charge on Sp$(k)_\pi$ instanton solitons, previously observed via geometric engineering for low-rank  supersymmetric theories.

\end{abstract}

\vfill
\end{titlepage}

\tableofcontents

\section{An Overview}

When one realizes 5d $\cN=1$ theories by geometric engineering as M-theory on a local Calabi-Yau \cite{Seiberg:1996bd,Morrison:1996xf,Douglas:1996xp,Intriligator:1997pq,Katz:1996fh}, BPS objects are realized by M2 and M5 branes wrapping 2-cycles and 4-cycles respectively. The former gives electrically charged particles, including dyonic instantons, while the latter gives magnetic strings. Compared to their 4d counterpart\cite{Seiberg:1994rs, Seiberg:1994aj}, these 5d theories look very simple; the 5d prepotential is at most a piecewise cubic function of the real Coulombic vev's, while on $\IS^1$, one must deal with the special K$\ddot{\textrm{a}}$hler geometry of complex vacuum expectation values. In particular, the ubiquitous wall-crossing phenomena of 4d \cite{Seiberg:1994rs, Ferrari:1996sv} turns off in 5d, as far as particle-like BPS states are concerned.

The simplicity of 5d $\cN=1$ theories is gratifying but at times appears too simple in that the 5d theory and the same theory compactified on a circle $\IS^1$ seem superficially very disparate. The latter acquires a much richer character. This is partly because the compactification produces particle-like monopoles from magnetic strings wrapping $\IS^1$ and allows wall-crossing in the 4d sense. When compactifying a theory on a circle, many properties of the theory change discontinuously in the zero radius limit. One of the more well-known such is the Witten index for supersymmetric theories \cite{Witten:1982df}. The Witten index of a theory is often computed by putting the field theory on a torus, modulo some topological subtleties \cite{Witten:2000nv, Tachikawa:2014mna}, and, as such, it remains the same for any finite radius. However, when one collapses a compactified spatial direction to a zero radius, the Witten index is often not preserved.

A canonical example is 4d $\cN=1$ pure Yang-Mills theory with the index
equal to the dual Coxeter number \cite{Witten:1982df} versus their 3d
$\cN=2$ counterpart with no supersymmetric vacua at zero Chern-Simons level \cite{Witten:1999ds}. More recently, such differences between the entire class of 2d $\cN=(2,2)$ elliptic genus \cite{Benini:2013xpa} and their 1d counterpart \cite{Hori:2014tda} have been systematically understood; Only in the latter, wall-crossings under D-term deformation can occur so the two cannot, in general, be continuously connected. For gauge theories, such discontinuities in the zero radius limit are now understood quantitatively via the notion of the holonomy saddles \cite{Hwang:2018riu}. For the compactified 5d theories, however, one finds the wall-crossing turning off in exactly the opposite limit of the infinite radius, instead. One purpose of this note is to dissect this situation.

The IIA/M-theory picture of the compactified theory, in which the KK modes on the circle are interpreted as D0 branes, gives us a universal handle on how one might study the resulting particle-like BPS spectra and the wall-crossing thereof. The D0 probe theory may be interpreted as the BPS quiver \cite{Closset:2018bjz} whose individual nodes represent certain D4-D2-D0 bound states with fractional D0 charges. One may naively think that these quiver quantum mechanics would lift to $d=1+1$ linear sigma models in the 5d, given how the nodes of such quivers carry magnetic charges and correspond to string-like objects. However, one can see immediately how this is incorrect.

The BPS quiver quantum mechanics arise from a nonrelativistic approximation in which the binding energy, if any, must be small compared to the rest mass. Note that the rest masses of elementary objects scale with $\sim R_5$ in the large $R_5$ limit. On the other hand, the D0 brane with its mass $\sim 1/R_5$ is represented by a particular rank assignment to the quiver nodes, say, $(1,1,\dots,1)$ in case of toric Calabi-Yau's \cite{Closset:2018bjz, Duan:2020qjy}. Combined, this means that in the large radius limit of $\IS^1$, the nonrelativistic requirement breaks down spectacularly.

This is, of course, related to how the $\IS^1$ momenta are reflected in the rank information of the quiver. In 5d, where the sum over all KK modes are obligatory, the correct description of these magnetic objects is, in fact, well known to be governed by  $(0,4)$ nonlinear sigma models \cite{Maldacena:1997de, Minasian:1999qn}, instead. The BPS quiver quantum mechanics does not naturally elevate to $d=1+1$ dynamics when we have to consider magnetic monopoles as extended objects and, as such, would not tell us much about how the wall-crossing turns off.

A better handle for understanding the large radius limit is the Seiberg-Witten description of the Coulomb phase. With finite $R_5$, the prepotential receives an infinite number of additive contributions membrane instantons \cite{Lawrence:1997jr}, whose exponents are all proportional to $R_5$. The latter drops out entirely in the $R_5\rightarrow \infty$ limit, leaving behind the aforementioned cubic prepotential. As such, how 5d theory is connected to 5d theory compactified on a circle can be more easily kept track of with the prepotential, and this should allow us to probe exactly what happens to various marginal stability walls that persists for any finite $\IS^1$ size.

In this note, we wish to study 5d rank one theories compactified on a circle to study how the decompactification limit emerges in the large radius limit and how the complicated wall-crossing turns off and characterizes the co-dimension-one boundary of the Coulomb phase. We will see shortly the two are closely related to each other, and one recovers the known fact that 5d Coulomb phases end where a BPS magnetic string becomes tensionless. For some theories, this tensionless limit coincides with the symmetry restoration of the non-Abelian gauge symmetry, but not always.

On the other hand, the absence of wall-crossing really addresses the stability of 5d BPS particle states, whose central charges are all ``real."  4d monopoles uplift to monopole strings, and the central charge density thereof is by and large imaginary. In fact, the Kaluza-Klein particles also come with an imaginary central charge. This explains why, upon $\IS^1$ compactification, the wall-crossing turns on immediately. For a finite size of this circle $\IS^1$, the magnetic strings remain extended, so the relevant counting should be given by the elliptic genus\cite{deBoer:2006vg, Gaiotto:2006wm,Gaiotto:2007cd, Denef:2007vg,Kraus:2006nb, Dabholkar:2005dt, Alim:2010cf}. In the last part of this note, we will take a close look at the wall-crossing of this elliptic genus across a point where a charged matter becomes massless, or equivalently, across a flop transition.

One well-known fact about 5d Sp gauge theories is the possibility of turning on a discrete theta angle, associated with $\pi_4({\rm Sp}(k))=\IZ_2$ \cite{Witten:1982fp}. For the example of pure Sp(1), this angle distinguishes F1 theory from F0 theory. A lesser-known fact is how the instanton soliton can carry different electric charges in these two theories by a half shift that is the same as the charge of a quark. This can be seen from $(p,q)$ five-brane constructions most easily \cite{Aharony:1997bh}, and has also been noted from the localization computation of the instanton partition functions more recently \cite{Hwang:2014uwa}. Nevertheless, a direct field theory understanding of this shift seems unavailable in literature, as far as the authors can see. Note how this phenomenon sounds similar to the 4d Witten effect, where magnetic solitons pick up electric charges proportional to the continuous 4d $\theta$ angle \cite{Witten:1979ey}. We extend the mechanism to this 5d version and give a clean and universal field theory proof of such a shift for arbitrary ${\rm Sp}(k)$. This fact plays a crucial role in understanding F1 theory.

Another curiosity one encounters for 5d theories is the possibility of a negative bare coupling-squared. Innocently represented in the $(p,q)$ fivebrane picture as a vertically elongated internal face, as opposed to a horizontally elongated one, how such a parameter regime connects to 4d Seiberg-Witten theory could be a little confusing initially. In particular, the naive expectation that the 5d Coulomb phase would end where the massive charged vector becomes massless does not hold in such theories.

What happens at such an endpoint of the Coulomb phase is that the magnetic string becomes tensionless. The same happens with positive bare coupling-squared, in fact, and both are per how wall-crossing of BPS particles must disappear in the decompactification limit. At the same time, for the F0 theory with negative bare coupling-squared, one sees a dyonic instanton with a "unit" electric charge becoming massless. The latter means that the non-Abelian gauge symmetry is restored, albeit via a different massless vector multiplet becoming massless and constituting the Yang-Mills field. However, one must not confuse this with the usual strong-weak duality, as we will delineate in a later section.

How 4d Seiberg-Witten geometries are embedded in the small radius limit of the former deserves more careful attention also. While one might think that taking the 4d limit is merely a matter of decoupling KK modes on the circle, the reality is more complex;  the gauge holonomy along $\IS^1$ itself lives in a circle, and a 4d theory emerges only after one expands the gauge field $A_5$ around some fixed holonomy and treat the fluctuation as an adjoint scalar field. The resulting dimensionally reduced gauge theory with supersymmetry intact is not unique, and the diverse possibilities thereof have been dubbed ``holonomy saddles."

With eight supercharges, this reduction produces a continuous family of infinitely many different 4d theories, a generic example of which is decoupled free Abelian gauge theory. More interesting holonomy saddles emerge by fine-tuning the gauge holonomy to allow non-Abelian gauge group and/or matter content visible at a scale much lower than $1/R_5$. For example, 5d gauge theories with quarks may reduce to 4d interacting Seiberg-Witten theories with no quarks or to theories with smaller gauge group or some combinations thereof. As such, the 4d limit of the 5d Seiberg-Witten geometry is rather complex and full of surprises.

This note is organized as follows. Section 2 will, in part, review basic structures, all well-known, necessary to understanding 5d Seiberg-Witten theory and the circle compactification thereof. Section 3 will pose to consider the implication of the discrete theta angle, also studied elsewhere. However, we will treat the issue entirely from the field theory viewpoint and give a simple and universal explanation of the 5d Witten effect. In Section 4, we investigate some global characters of the Coulombic moduli space, emphasizing how one recovers the 4d Seiberg-Witten theory. More specifically, we take the holonomy saddles into account and illustrate how the 5d Seiberg-Witten geometry on $\IS^1$ is generically capped by more than one 4d Seiberg-Witten geometries.

In Section 5, we will revisit the question of the wall-crossing starting from the circle-compactified theory and addressing what to expect from the decompactification limit. Exactly what we mean by the absence of 5d wall-crossing is mulled over, from which we draw the anticipation that the boundary of the 5d Coulomb phase should be universally characterized by tensionless BPS strings. Section 6 will take the three simplest types of rank one non-Abelian theories on a circle and investigate the Coulomb phase in quantitative detail. Here we will see how, in the decompactification limit, the 5d Coulomb phase boundary is realized to make sure that certain marginal walls collapse, as suggested by the general discussion on wall-crossings. Also addressed are how the flop transitions of the local Calabi-Yau's that geometrically engineer such theories manifest in the Seiberg-Witten moduli space. The wall-crossing of magnetic BPS string across such a flop transition is the topic for Section 7, where we illustrate using the example of dP$_2$ near a massless quark point.

\section{Preliminaries}

In this section, we start by gathering and reviewing well-known basic facts about 5d Yang-Mills theories, their Coulomb phases, and the Seiberg-Witten descriptions upon a circle compactification.

\subsection{Review of $(p,q)$ Fivebrane Constructions}

Let us start by recalling the Coulomb phase of 5d $\mathcal{N}=1$ supersymmetric gauge theory has a real scalar field $\phi$ in the vector multiplet. On a generic point in the Coulomb branch, the gauge group $G$ is broken to its Cartan part ${\rm U}(1)^{r_G}$, where $r_G$ is the rank of the group $G$. The prepotential in the Coulomb branch is one-loop exact and can be written as \cite{Intriligator:1997pq}:
\begin{equation}\label{5d IMS prepotential}
	\mathcal{F}_{\textrm{IMS}}(\phi) = \frac{1}{2} {\mu}_0 h_{ij} \phi^i \phi^j + \frac{\kappa}{6} d_{ijk} \phi^i \phi^j \phi^k + \frac{1}{12} \left( \sum_{e \in \textrm{Roots}} |e\cdot \phi|^3 - \sum_f \sum_{\omega \in \mathbf{R}_f} |\omega \cdot \phi - \mu_f|^3 \right),
\end{equation}
which is usually called the Intriligator-Morrison-Seiberg (IMS) prepotential.  Here ${\mu}_0 \equiv 8\pi^2 / g_5^2$ is the 5d inverse coupling-squared and also the instanton mass and $h_{ij} = \textrm{Tr}(T_j T_j)$ and $d_{ijk} = \frac{1}{2} \textrm{Tr}(T_i \left\{T_j,T_k \right\})$.
The cubic terms in the parenthesis are the one-loop contribution from the vector multiplet and from the matters, respectively.

We follow the convention of Ref.~\cite{Intriligator:1997pq}, and in particular, for Sp(1) theories which are the primary example of ours, $d_{ijk}$ would be absent while $h_{ij}\rightarrow 2$. The way how the prepotential enters the
5d action is different from that of 4d by a factor of $i/2\pi$; this is related to how the normalization of $\cF_{\rm IMS}$ is tied to
the integer-quantized bare Chern-Simons level $\kappa$. For example,  $\partial^2 {\mathcal{F}}_{\textrm{IMS}} / \partial \phi^i \partial \phi^j$ gives $-i2\pi\tau_{ij}$ where $\tau_{ij}$ is the pure imaginary 5d coupling.

For a pure Sp(1) theory, e.g., we have
\begin{equation}\label{5d Sp(1) prepotential}
	\mathcal{F}_{\textrm{IMS}}(\phi) = \mu_0 \phi^2 + \frac{4}{3} \phi^3 \ .
\end{equation}
The first derivative gives the monopole string tension $T_{\rm mono}$ as,
\begin{equation}\label{5d Sp(1) monopole}
	i T_{\textrm{mono}} = \frac{i}{2\pi}\frac{\partial \mathcal{F}_{\textrm{IMS}}}{\partial \phi} = \frac{i 2\phi ( \mu_0 + 2 \phi )}{2\pi},
\end{equation}
where we used $i T_{\textrm{mono}}$ on the left hand side as a reminder that
the 4d monopole central charge upon a compactification is imaginary in the large radius limit.
The second derivative produces the pure imaginary 5d coupling as
\begin{equation}\label{5d Sp(1) coupling}
	\tau_{\textrm{5d}} = \frac{i}{2\pi}\frac{\partial^2 \mathcal{F}_{\textrm{IMS}}}{\partial \phi^2}=\frac{i(\mu_0 + 4 \phi)}{\pi},
\end{equation}
with $\tau_{\textrm{5d}} = i 8\pi / g_{5,\textrm{eff}}^2$.

In five dimensions, the gauge theory is often realized via embedding into M/string-theory. Such embeddings are not unique and can generate different UV theories, even though they might share the same prepotential. F0 and F1 theories, both of which generated the same rank-one prepotential above, are the canonical pair and we will dwell on these two examples much. For this purpose, we will rely on the $(p,q)$ 5-brane web realizations \cite{Aharony:1997bh} and later also its M/IIA geometric counterpart.

\begin{figure}[htpb]
\centering
\includegraphics[scale=0.4]{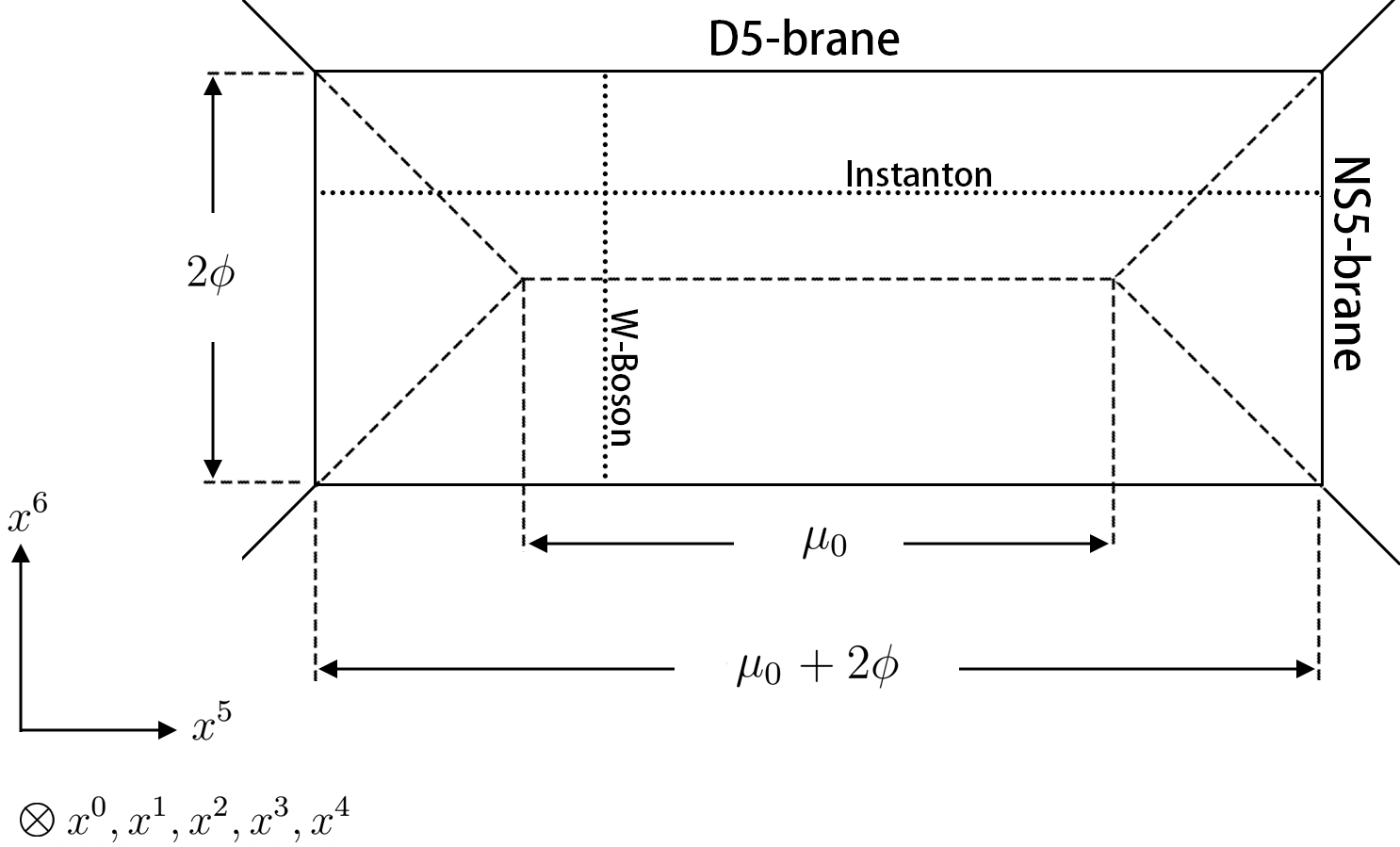}
\caption{F0 case : Brane web for 5d pure Sp(1) gauge theory with zero $\theta$-angle.}
\label{Fig-big-F0-brane}
\end{figure}

The brane diagram for F0-theory is depicted in figure \ref{Fig-big-F0-brane}. As is the usual convention, the vertical lines represent NS5 branes, while the horizontal ones are D5 branes. Then, the W-bosons are the vertical F-string stretching between D5-branes, while the monopole strings are D3-branes spanning a common internal face. There are also dyonic instanton solitons carrying two units of {\rm U}(1) charges, equivalent to that of a W-boson, corresponding to the D-strings stretching between NS5-branes.
From the diagram, it is straightforward to see that the mass of the W-boson/dyonic instanton is proportional to the distance between D5/NS5-branes, and also the tension of the monopole string proportional to the area of the common face.

The $1/2\pi$ factor in \eqref{5d Sp(1) monopole} may be also understood as follows. Note that $2\phi$ and $2\phi + \mu_0$ equal the corresponding lengths of the rectangle in figure \ref{Fig-big-F0-brane} times the D1/F1 tensions $T_{\textrm{D1}}$ and $T_{\textrm{F1}}$, respectively. On the other hand the monopole string tension $T_{\textrm{mono}}$ is the area of the rectangle times the D3-brane tension $T_{\textrm{D3}}$. Using the relation $T_{\textrm{D3}} = T_{\textrm{D1}} T_{\textrm{F1}}/2\pi $, it is easy to see that the monopole string tension $T_{\textrm{mono}}$ is $2\phi ( \mu_0 + 2 \phi )/2\pi$.

\begin{figure}[htpb]
\centering
\includegraphics[scale=1]{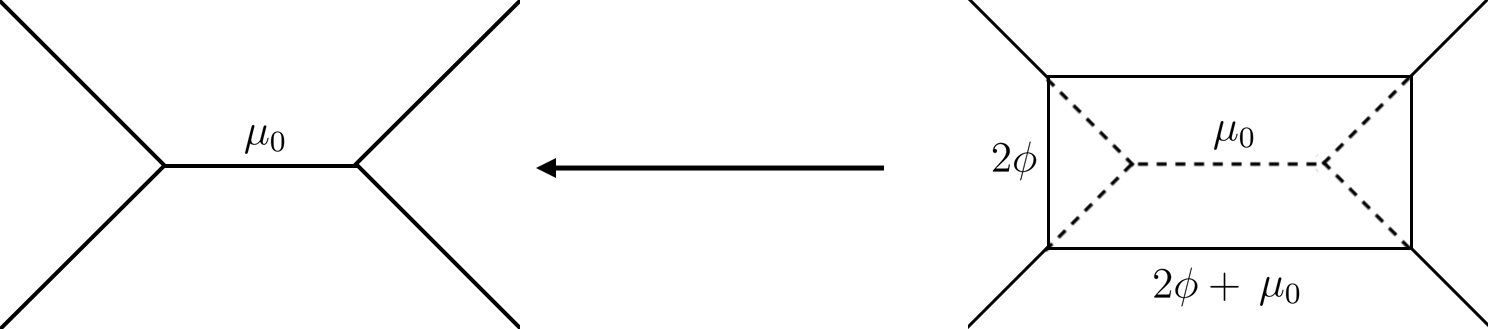}
\caption{F0 theory with positive bare coupling-squared  ($\mu_0>0$).}
\label{Fig-F0-moduli-positive}
\end{figure}

If the bare coupling-squared $g_5^2 \sim \mu_0^{-1}$ is positive, the distance between two D5-branes will become smaller as $\phi$ decreases and merge at the endpoint of the moduli space. 
On the other hand, if $\mu_0$ is negative, as $\phi$ decreases, the distance between two NS5-branes will become smaller and merge at the endpoint of the moduli space\ref{Fig-F0-moduli-positive}.  In both cases, the Sp(1) gauge symmetry is restored at the endpoint of the moduli space.

\begin{figure}[htpb]
\centering
\includegraphics[scale=0.4]{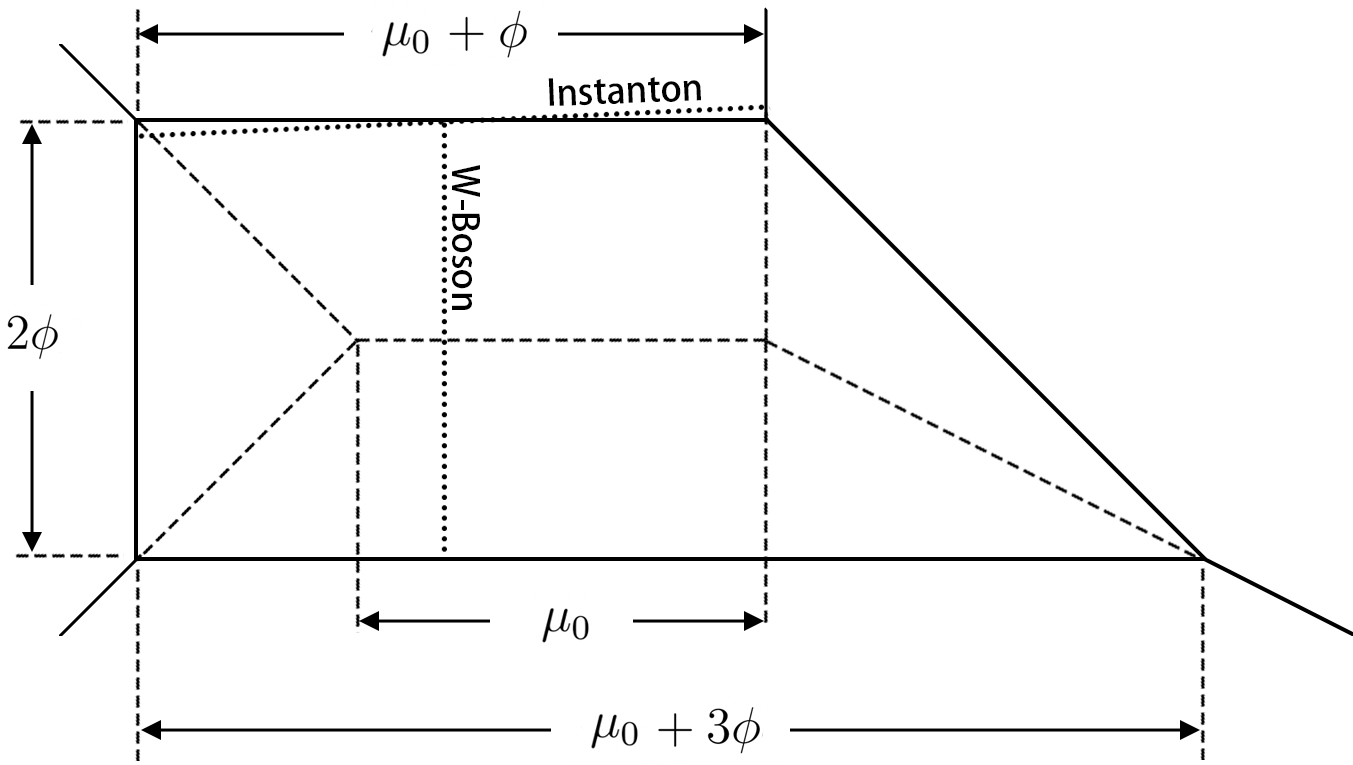}
\caption{F1 case, Brane web for 5d pure Sp(1) gauge theory with $\theta = \pi$.}
\label{Fig-Big-F1-brane}
\end{figure}

The brane diagram for F1-theory is depicted in figure \ref{Fig-Big-F1-brane} for a positive $\mu_0$, and it also corresponds to the 5d pure Sp(1) gauge theory with a discrete $\theta$ angle associated with $\pi_4({\rm Sp}(1))$ turned on. The W-bosons are still F-strings stretching between two D5-branes, and the monopole strings are D3-branes wrapping the common internal face. The dyonic instantons differ from those in the F0 case; they are D-strings stretching between one internal NS5-brane and another external NS5-brane and carrying only one {\rm U}(1) charge. We will discuss the field theory origin of this charge difference in the next section. Nevertheless, the prepotential is the same as the F0 case \eqref{5d Sp(1) prepotential} and is not sensible to the $\theta$-angle.

\begin{figure}[htpb]
\centering
\includegraphics[scale=0.4]{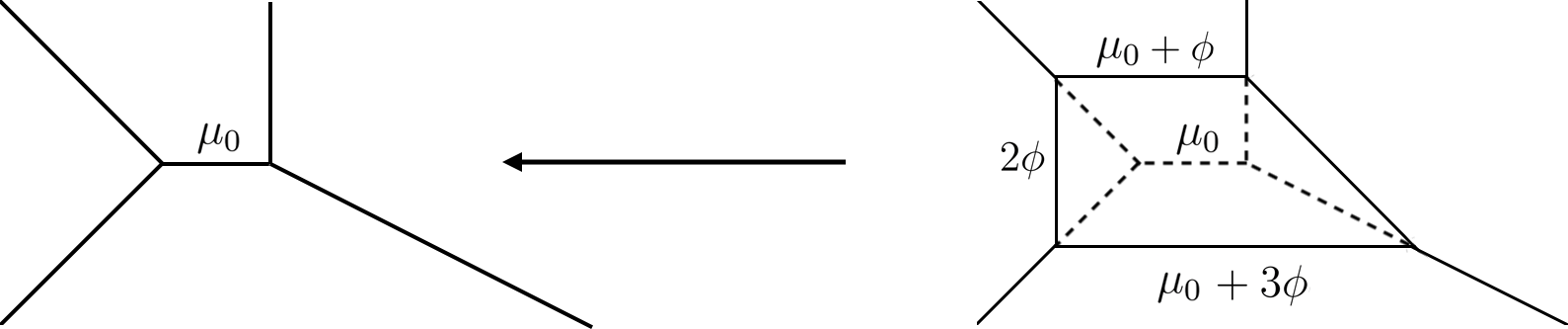}
\caption{F1 theory with positive bare coupling-squared ($\mu_0>0$).}
\label{Fig-F1-moduli-positive}
\end{figure}

\begin{figure}[htpb]
\centering
\includegraphics[scale=1]{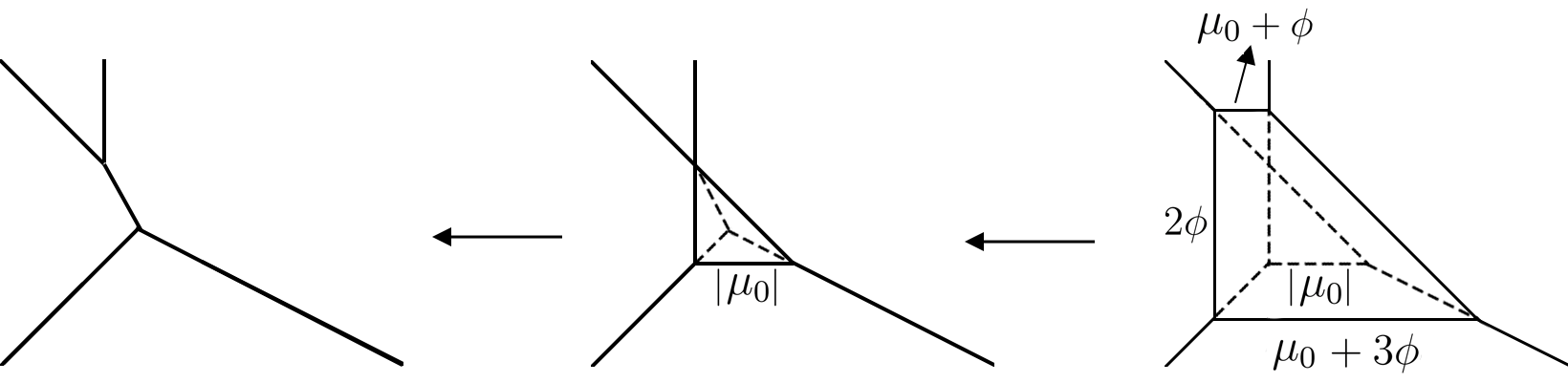}
\caption{F1 theory with negative bare coupling-squared ($\mu_0<0$).}
\label{Fig-F1-moduli-negative}
\end{figure}

For a positive $\mu_0$, the distance between two D5-branes becomes smaller as $\phi$ decreases and merges at the endpoint of the moduli space where the Sp(1) gauge symmetry is also restored, just as in the F0 case. The brane web is given in figure \ref{Fig-F1-moduli-positive}. However, the story is quite different for negative bare coupling-squared ($\mu_0 < 0$), as shown in figure \ref{Fig-F1-moduli-negative}. As $\phi$ becomes smaller, the upper D5-brane will firstly shrink to zero, and the common internal face is a triangle at the point. However, this is not the endpoint of the moduli space; the triangle can further shrink and eventually becomes a tree diagram, which is the actual endpoint of the moduli space. The endpoint theory is called the $E_0$ theory, where no gauge symmetry restoration occurs at the boundary point \cite{Morrison:1996xf}.

\subsection{Exact Prepotentials with $\IS^1$}

If we compactify the 5d theory on a circle $\IS^1$ of radius $R_5$, the prepotential will receive extra contributions from the worldline instantons along the circle. 
Here, M/IIA theory version via a local Calabi-Yau 3-fold $\hat{X}$ is more versatile and gives us the exact nonperturbative prepotential straightforwardly. 

\begin{figure}[pbth]
\centering
\includegraphics[scale=1]{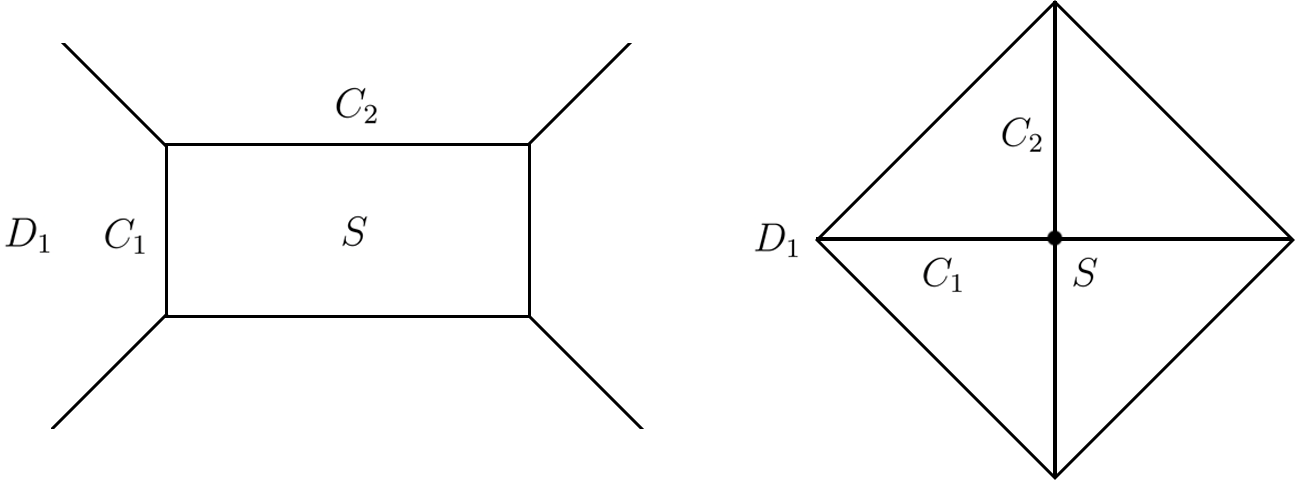}
\caption{The toric diagram dual to the F0 brane web.}
\label{Fig-F0-brane-toric}
\end{figure}

\begin{figure}[pbth]
\centering
\includegraphics[scale=1]{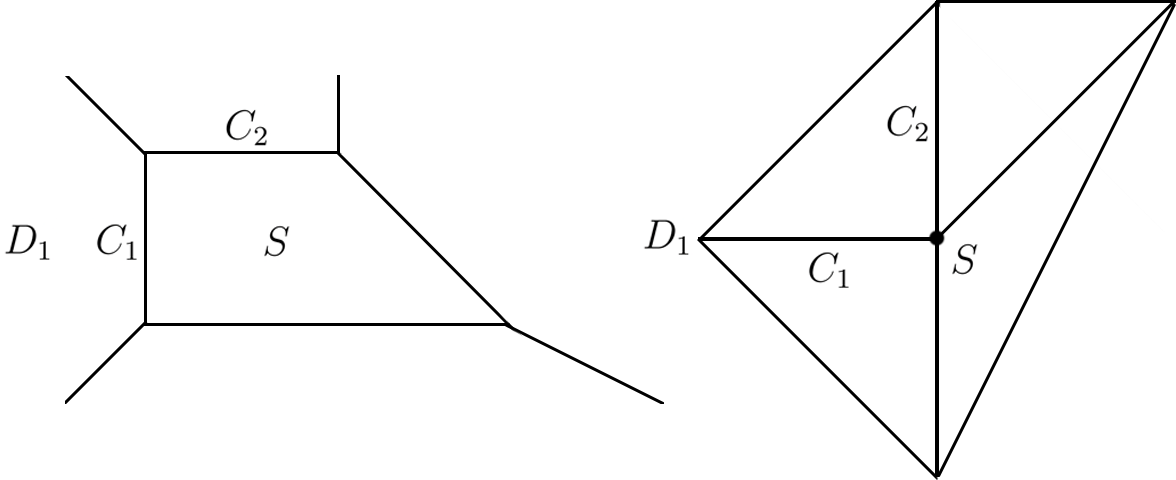}
\caption{The toric diagram dual to the F1 brane web.}
\label{Fig-F1-brane-toric}
\end{figure}

Let us first review the 5d gauge theory modeled by M-theory compactified on a toric local Calabi-Yau 3-fold $\hat{X}$, a crepant resolution of the singular Calabi-Yau 3-fold $X$, which is directly related to the $(p,q)$ web constructions earlier. The geometry contains various finite 2-cycles and 4-cycles (divisors), which can be read from the toric diagram. The rules of brane webs/toric diagram duality can be found in \cite{Aharony:1997bh}, and here we again give the example of F0,F1 in figure \ref{Fig-F0-brane-toric} and \ref{Fig-F1-brane-toric} for later use \cite{Closset:2018bjz}.
In the toric diagram, each line corresponds to a compact/non-compact holomorphic 2-cycle, and each vertex corresponds to a compact/non-compact holomorphic 4-cycle (divisor) in the local Calabi-Yau $\hat{X}$.
The Coulomb moduli $\{\phi^i \}$ are related to the K$\ddot{\textrm{a}}$hler moduli dual to the compact divisors $\{ S_i \}$, and the mass deformations are related to the K$\ddot{\textrm{a}}$hler parameters $\{\mu^a\}$ dual to the non-compact divisors $\{ D_a \}$. Then the K$\ddot{\textrm{a}}$hler form $[J]$ can be expanded as:
\begin{equation}
	[ J ] = \sum_{a} \mu^a [D_a] - \sum_i \phi^i [S_i],
\end{equation}
where $[J], [D_a]$ and $[S_i]$ denote the Poincar$\acute{\textrm{e}}$ dual 2-forms of the divisor classes $J$, $D_a$ and $S_i$. 

The 5d prepotential $\mathcal{F}_{\textrm{IMS}}$ on the Coulomb branch can be read from the geometry as \cite{Intriligator:1997pq}:
\begin{equation}\label{5d-prepotential-toric}
	\mathcal{F}_{\textrm{IMS}} (\phi^i, \mu^a) = - \frac{1}{6} \int_{\hat{X}} [J] \wedge [J] \wedge [J] = - \frac{1}{6} J \cdot J \cdot J,
\end{equation}
and the third derivatives of the prepotential is the triple intersections among the divisors, for example:
\begin{equation}
	\frac{\partial^3 \mathcal{F}_{\textrm{IMS}}(\phi^i, \mu^a)  }{\partial \phi^i \partial \phi^j \partial \phi^k} =  S_i \cdot S_j \cdot S_k.
\end{equation}
The BPS spectrum consists of M2 or M5-branes wrapping on different compact 2-cycles or divisors. The M5-branes wrapped on the compact divisors give the 5d monopole string charged under the {\rm U}(1) associated with the corresponding divisors. The BPS particles are given by M2-branes wrapped on some compact 2-cycle $C$ and the electric charge under $\textrm{{\rm U}(1)}_i$ is given by the intersection number $C\cdot S_i$. For F0 and F1 theory the (independent) compact 2-cycles/divisor are $C_1,C_2/S$ as depicted in figure \ref{Fig-F0-brane-toric} and \ref{Fig-F1-brane-toric}. M2-branes wrapped on $C_1/C_2$ give the 5d W-bosons/dyonic instantons, and M5-branes wrapped on $S$ give the 5d monopole string.

After compactification on $\mathbb{S}^1$, i.e., in IIA theory on the same local Calabi-Yau, the Coulomb moduli $\{ \phi^i \}$ are complexified, and the above prepotential should be modified by the membrane instantons given by M2-branes wrapped over compact 3-cycles $C_I \times \mathbb{S}^1$, or worldsheet instantons in the type IIA language. The total contribution to the prepotential via the third derivatives is given by \cite{Katz:1996fh, Lawrence:1997jr,Candelas:1990rm,Chiang:1999tz,Katz:1996ht,Klemm:1997gg},
\begin{equation}\label{Instanton prepotential}
	\frac{\partial^3 F_{\rm exact}(t^i,v^a)}{\partial t^i \partial t^j \partial t^k} = S_i \cdot S_j \cdot S_k +  \sum_{\eta \in H_2(X)} \frac{q^{\eta}}{1-q^{\eta}} N_{\eta} (S_i \cdot \eta)(S_j \cdot \eta)(S_k \cdot \eta),
\end{equation}
The sum  is over all integral classes of compact 2-cycles $\{C_I \}$
with non-negative coefficients in the local Calabi-Yau 3-fold $\hat{X}$ and $N_{\eta}$ is an invariant which counts the number of such holomorphic 2-cycles of class $\eta$, where the exact numbers can be found in Ref.~\cite{Chiang:1999tz} for F0 and F1.
$q^{\eta}$ is computed by the complexified and $R_5$-rescaled K$\ddot{\textrm{a}}$hler class as
\begin{equation}
	q^{\eta} = \exp \left( 2\pi i \eta \cdot \left(\sum_a v^a D_a + \sum_i t^i S_i\right) \right)\ ,
\end{equation}
where the new parameters $v^a$ and $t^i$ both carry a factor of $R_5$ and are related to $\mu^a$ and $\phi^i$ as follows.

The real scalar $\phi$'s of the IMS prepotential connect with complex $t$'s as
\bea
\phi=\lim_{R_5\rightarrow 0} it/R_5\ ,
\eea
with the accompanying relation,
\begin{equation} \label{4d-5d-relation}
\mathcal{F}_{\textrm{IMS}}= \lim_{R_5\rightarrow \infty}  \left( \frac{i}{R_5}\right)^3 F_\textrm{exact}\ .
\end{equation}
For this reason, we will later adopt the complexification of $\phi$ as
\bea
a\equiv it/R_5\ ,
\eea
which, for Sp(1), represents half of the W-boson central charge. Similarly,
the external parameters $v$'s are related to $\mu$'s as
\bea
\mu=-iv/R_5\ ,
\eea
where we should be mindful that, unlike the Coulombic scalars, these external parameters $v$'s would remain pure imaginary real, or equivalently $\mu$'s real, regardless of $\IS^1$ compactification.

The compactification introduces an energy scale $1/2\pi R_5$, across which the character of the theory is from 5d to 4d. Also, with the Coulomb vev in the strongly coupled region, which would be well below $1/2\pi R_5$, the membrane instanton sum \eqref{Instanton prepotential} is divergent and must be resumed. One way to probe such a strongly coupled region which is located well below this scale is to use mirror symmetry, mapping the type IIA string theory compactified on Calabi-Yau 3-fold $\hat{X}$ to a type IIB string theory compactified on the mirror Calabi-Yau 3-fold $\hat{X}'$.

Before closing this section, we point out that the third derivatives \eqref{Instanton prepotential} along with the relation \eqref{4d-5d-relation} are not enough to determine the full prepotential. There are additional terms in $F_{\textrm{exact}}$ given by the integral of the second and third Chern classes in the Calabi-Yau $\hat{X}$\cite{Chiang:1999tz}:
\begin{equation}
	\frac{1}{24} \int_{\hat{X}} c_2(\hat{X}) \wedge \left(\sum_a v^a [D_a] + \sum_i t^i [S_i]\right) - i \frac{\zeta(3)}{2 (2\pi)^3} \int_{\hat{X}} c_3(\hat{X}),
\end{equation}
which can not be detected by the third derivatives of $F_{\textrm{exact}}$. Also these terms will not survive in $\mathcal{F}_{\textrm{IMS}}$ when we take the $R_5 \rightarrow \infty$ limit according to \eqref{4d-5d-relation}.
Such terms do not affect the coupling but they will contribute a shift $\sim \frac{1}{R_5^2}\int_{\hat{X}} c_2(\hat{X})\wedge [S] $ to the tension of monopole string.

On a closer look, the latter effect represents a fractional shift of D0 charge and
the subsequent 4d central charge shift of $\sim 1/R_5$. With the low energy dynamics viewpoint to be
discussed in the last Section, this D0 charge shift can be traced to the usual regulated
sum of the half-integral vacuum charges due to the infinite number of KK modes associated
with the left-moving chiral fields on the monopole string. For large $R_5$, the effect is subleading relative
to the magnetic tension while, for small $R_5$ where the 5d KK gauge field decouples, the shift can
be absorbed into the renormalized 4d mass of the magnetic monopole.

Another potential issue related to this linear term is that it shifts discontinuously across
a flop transition, e.g., at those Coulomb vev where a quark becomes massless. This is in turn
related to how the number of the left-moving chiral fields on the monopole string jumps
across a flop, and seemingly implies a monopole tension that is discontinuous across
a flop. However, no such discontinuity of a central charge is expected in general;
A canceling discontinuity occurs in the infinite sum (\ref{Instanton prepotential}),
so that $\cF_{\rm exact}$ and its derivatives are themselves continuous. In this sense,
this linear piece is a crucial piece for the consistency of $\cF_{\rm exact}$, but with little
consequence for most part of this note where $\tau_{\rm 5d}$ is the main object of interest.
As such, we will ignore this contribution in the following.

\section{5d Witten Effect}

Although 4d theories tend to be richer in their infrared behavior, one inherently 5d phenomenon is the discrete theta angle associated with $\pi_4({\rm Sp}(k))=\IZ_2$. As such, Sp$(k)$ theories have two types of 5d reincarnations, labeled by a discrete $\theta$ angle that takes values 0 or $\pi$. For pure Sp(1), $\theta=0$ is known as the F0 theory, while the F1 theory has this angle turned on. This discrete theta angle is known to affect the electric charges on the instanton soliton, much like how 4d theta angle shifts the electric charge on magnetic monopoles, early on from the $(p,q)$ five-brane realization \cite{Aharony:1997bh}, and also more recently via instanton partition functions \cite{Hwang:2014uwa}.

Nevertheless, a direct field theory understanding has not been available. In particular, this observed quantization is so far tied to the minimal supersymmetry and low-rank examples. Here, before we venture into the study of F0 and F1 theories below, we wish to delineate a simple and very general derivation of this charge quantization, based entirely on field theory argument for all ${\rm Sp}(k)$ theories and regardless of supersymmetry.

A word of caution on the normalization of the electric charges is in order. Because much of the mechanisms are described in theory with adjoint fields only, the notation here is well-tailored to the charge convention where Sp(1) W boson has the unit electric charge under the unbroken U(1). This is also the convention taken in Ref.~\cite{Witten:1979ey}. As such, for this section only, the definition of quantized electric charge will be half of the rest of the note.

\subsection{4d Witten Effect Revisited}

In four dimensional gauge theory, one may add a $\theta$-term to the Lagrangian which will violate the parity as
\begin{equation}
	\mathcal{L}_{\theta} = \frac{\theta}{8\pi^2}\int F\wedge F \ .
\end{equation}
Witten pointed out \cite{Witten:1979ey} that this $\theta$ has a non-trivial effect on the magnetic monopole and will shift the allowed electric charge by an amount of $\theta / 2\pi$. Let us first review this well-known 4d Witten effect. It turns out that one of the two approaches to the effect, both illustrated by Witten in his seminal paper \cite{Witten:1979ey}, generalizes to the 5d analog for instanton solitons.

The simplest version of the 4d Witten effect appears already with ${\rm U}(1)$
theory
\bea
S_{{\rm U}(1)} = -\frac{1}{4g^2}\int F_{\mu\nu}F^{\mu\nu} +\frac{\theta}{8\pi^2}\int F\wedge F\ ,
\eea
for which we have the Dirac quantization
\bea
n_e n_m \in 2\pi\IZ \ ,
\eea
for the electric and the magnetic charges, $n_e$ and $n_m$. When we
say this, it is important to remember that the integer-quantized
electric charge is defined by the conjugate momenta of the gauge field, $\pi^i$,
\bea
n_e\equiv -\int_{\IS^2_\infty} \pi^i\, dS_i
\eea
at spatial infinity. On the other hand,
\bea
 \pi^i=\frac{\delta S}{\delta \dot A_i}=-\frac{1}{g^2}E^i+\frac{\theta}{4\pi^2}B^i\ ,
\eea
with $E^i=-F_{0i}$ and $B_1=F_{23}$ etc, so the charge as measured by the electric field $E^i$ is
\bea
q\equiv \frac{1}{g^2}\int_{\IS^2_\infty} E^i\cdot dS_i = n_e + \frac{\theta}{4\pi^2}\int_{\IS^2_\infty} B^i\,dS_i= n_e\  +\frac{\theta}{2\pi}n_m
\eea
with
\bea
\int_{\IS^2_\infty} B^i\,dS_i =2\pi n_m\ ,
\eea
from the topological quantization of the ${\rm U}(1)$ magnetic field. Although
there are more than one ways to understand this shift, the most pertinent
for us is how it is $\pi_i$  that underlies the gauge transformation;
the non-integral shift of $q$ as measured by $E^i$ is itself  a consequence
of the gauge-invariance.

This extends to solitonic magnetic monopoles, say, in
$Sp(1)$ theories with an adjoint scalar,
\bea
S=-\frac{1}{g^2}\int\left(\frac12{\rm tr}\; \cF_{\mu\nu}\cF^{\mu\nu}+{\rm tr}\; D_\mu\Phi D^\mu\Phi\right) +\frac{\theta}{8\pi^2}\int {\rm tr}\;\cF\wedge \cF\ ,
\eea
similarly, with the gauge generators normalized as
\bea
T_a=\sigma_a/2\ ,
\eea
with the Pauli matrices $\sigma$'s. Now the magnetic solitons obey
\bea
D_k\Phi=\cB_{k}\equiv \frac12\epsilon_{ijk}\cF_{ij} \ .
\eea
Taking the unitary gauge
\bea
\Phi=\phi\frac{\sigma_3}{2} \ , \qquad
\cA = A \frac{\sigma_3}{2} + W^\pm \sigma_\pm\ ,
\eea
with $W^\pm$ unit-charged under $A$ and $\mathbf{v}\equiv \phi(\infty)>0$.
The topological quantization gives
\bea
\int_{\IS_\infty^2} B^idS_i=4\pi n_m\ ,
\eea
where $n_m\in \IZ$ with the total energy $4\pi  n_m \mathbf{v}/{g^2}$ as usual.
The effective ${\rm U}(1)$ theory has
\bea
S_{{\rm U}(1)}=-\frac{1}{4g^2}\int\left(F_{\mu\nu} F^{\mu\nu}+ 2\partial_\mu\phi \partial^\mu\phi\right) +\frac{\theta}{16\pi^2}\int F\wedge F+\cdots\ ,
\eea
so looking back at the Abelian case, we find again that
\bea
q=n_e+ \frac{\theta}{2\pi}n_m
\eea
now with $n_e \in \IZ/2$ if we are to allow matters in the fundamental representations.\footnote{This charge content refers to physical states, elementary or coherent. It is sometimes said that non-local objects such as `t Hooft lines allow distinction between ``Sp(1) theory" and, say, ``$SO(3)_\pm$  theories" that admit half-integral $n_m$. However, such objects are external, costing infinite energy even if allowed by the Dirac quantization via restricted gauge representations and cannot be built from elementary fields. The choices of the Lie group, rather than the Lie algebra, are meaningful for the gauge theory proper only if the spacetime carries nontrivial second homology \cite{Witten:2000nv}.}

Now a slightly different perspective \cite{Witten:1979ey} on this charge shift will be
useful for the 5d generalization. With the conjugate momenta
\bea
\Pi^{ai} \equiv -\frac{1}{g^2}\cE^{ai} +\frac{\theta}{8\pi^2}\cB^{ai}
\eea
with $\cE_i=-\cF_{0i}$, the operator
\bea
\mathbf G\equiv -\frac{1}{\mathbf{v}} \int_{\IR^3} \left(D_i\Phi^a \cdot \Pi^{ai}\right)
\eea
generates the unbroken ${\rm U}(1)$ transformation,
\bea
[\mathbf G,\cA_i^a] = -iD_i(\Phi^a/\mathbf{v})\ ,
\eea
with the right hand side solving the zero-mode equation around
the magnetic monopoles.
However, we must recall that the unitary gauge is actually singular at
origin. A better choice would be the Hedgehog gauge, with which the
gauge function
\bea
U\equiv  e^{2\pi i \Phi/\mathbf{v}} \approx e^{\pi i\,\sigma_a \hat x^a\,(\phi/\mathbf{v})}
\eea
far away carries the $n_m$ winding number of $\pi_3(Sp(1))$, and
\bea
e^{2\pi i \mathbf G}[\cA]=iU\cA U^{-1}+ UdU^{-1}
\eea
shift the topological number of the vacuum by $-n_m$.

Recall that $\theta$ can be equivalently considered as the
parameter that  defines the vacuum made up of an infinite sum over the
distinct topological sectors, $\vert n\rangle$,
\bea
\vert \theta\rangle =\sum e^{in\theta}\vert n\rangle\ .
\eea
If the sum over the topological sectors are handled this way,
the conjugate momentum and the ${\rm U}(1)$ charge operator would
have the term proportional to $\theta$ missing. If we denote
this charge operator $\mathbf Q$, we find
\bea
e^{2\pi i\mathbf Q} =e^{2\pi i \mathbf G}\,\biggr\vert_{\theta\rightarrow 0}\ ,
\eea
which still shifts $\vert n\rangle \rightarrow \vert n-n_m\rangle $,
since the difference between $\mathbf Q$ and $\mathbf G$ is the purely
magnetic term. As such,
\bea
e^{2\pi i \mathbf Q} \vert \theta\rangle =e^{in_m \theta} \vert \theta\rangle\ ,
\eea
which shifts eigenvalues of $\mathbf Q$, otherwise integral or
half-integral, by $n_m\theta/2\pi$.

\subsection{5d Witten Effect for Instanton Solitons}

All that happened in the second approach above is that the would-be Gauss constraint built
from $\mathbf Q$ is not really the full gauge generator when $\theta\neq 0$.
This alternate viewpoint is more suitable in 5d where the analog of
$\theta$ angle is discrete, due to $\pi_4({\rm Sp}(k))=\IZ_2$, but
has no local term representing it in the action. With a nontrivial
discrete $\theta$ angle, again, the naive Gauss constraint built from
the  action does not correctly impose the gauge invariance and must
be augmented by how it acts as the two topological vacua.

First note how we are dealing with the theory in the Coulomb phase
almost exclusively, even though the instanton itself exists	 regardless of
the Coulombic vev. This means the electric charge we refer to must be
that of the unbroken ${\rm U}(1)$, which is in turn determined by the single adjoint scalar $\Phi(x)$. The classical 5d dyonic instanton solution
in the singular gauge is \cite{Lambert:1999ua}:
\begin{align}\label{instanton solution}
	\mathcal{A}_{\mu}(x) =\frac{\rho^2}{x^2 (x^2 + \rho^2)} \eta^a_{\mu \nu} x_{\nu}\sigma_a\ ,\qquad
	\Phi (x)= \mathbf{v} \frac{x^2}{x^2 + \rho^2} \frac{\sigma^3}{2},
\end{align}
where $x^{\mu} (\mu = 1,2,3,4)$ are the spatial coordinates, $a=1,2,3$ are the Sp(1) gauge indices, $\rho$ is the size of the instanton and $\eta^{a}_{\mu \nu}$ is the 't Hooft matrix.

The solution is analogous to the unitary gauge version of the
solitonic monopole, singular at origin,
so we first need to perform a gauge transformation to remove the artificial singularity of the solution \eqref{instanton solution} at the origin.
The necessary gauge transformation is:
\begin{equation}
	U(x) = \frac{i \bar{\sigma}_{\mu} x _{\mu}}{|x|},
\end{equation}
where the 4d $\sigma$-matrices are defined by: $\sigma_{\mu} = (\vec{\sigma},-i)$ and $\bar{\sigma}_{\mu} = (\vec{\sigma},i)$.
and the resulting regular soliton is
\begin{align}\label{instanton solution-nonsingular}
	\mathcal{A}_{\mu} = \bar{\eta}_{a \mu \nu} \frac{x_{\nu}\sigma^a}{x^2 + \rho^2}\ , \qquad
	\Phi = \frac{\mathbf{v}}{2} \frac{ \vec{\tilde{x}} \cdot \vec{\sigma}}{x^2 + \rho^2}\ ,
\end{align}
where $\vec{\tilde{x}}$ is a three-dimensional vector
\begin{equation}\label{Hopf map}
	(\tilde x_1,\tilde x_2,\tilde x_3) =
(2x_1 x_3 + 2 x_2 x_4 , -2 x_1 x_4 + 2 x_2 x_3, -x_1^2 - x_2^2 + x_3^2 + x_4^2)\ ,
\end{equation}
given by the Hopf map. The ${\rm U}(1)$ gauge operator is again given by
\begin{equation}
	\mathbf{Q} \equiv -\frac{1}{\mathbf{v}} \int_{\mathbb{R}^4} \left( D_{\mu} \Phi^a \cdot \Pi^{a \mu} \right),
\end{equation}
and defines the ${\rm U}(1)$ charge operator, the analog of $\mathbf{Q}$ of 4d case.
Unlike in 4d, there is no realization of the discrete theta angle via a local
term in the action, so $\mathbf(G)$ operator with the contribution from the
theta angle embedded is no longer available; we have $\mathbf{Q}$ only.
As such, the ${\rm U}(1)$ operator $\exp(2\pi i \mathbf{Q})$ generates the usual gauge transformation with
the gauge function,
\begin{equation}\label{gauge rotation-instanton}
	U(x) \equiv e^{2\pi i \Phi / \mathbf{v}} = \exp \left( \frac{i \pi \vec{\tilde{x}}\cdot \vec{\sigma}}{x^2 + \rho^2} \right),
\end{equation}
The main point that leads to the promised charge shift is that
this gauge function represents the non-trivial element of $\pi_4({\rm Sp}(1)) = \mathbf{Z}_2$.

This can be seen in the following way. Since $U(\infty) = -1$ we may compactify $\mathbb{R}_4$ to $S_4$, and consider any continuous curve $\gamma(s)$ ($s\in[0,1]$) in the Sp(1) manifold connecting the unit element $1$ and the $-1$, which means $\gamma(0) = 1$ and $\gamma(1) = -1$. On the other hand, from the gauge transformation function $U(x)$ one can see that $U(0) = 1$ and $U(\infty) = -1$, therefore the inverse image $U^{-1}(\gamma(0))$ and $U^{-1}(\gamma(1))$ are two points in $S_4$. However, for generic $s \neq 0,1$, the inverse image $U^{-1}(\gamma(s))$ must be a circle in $S_4$ since the $\tilde{x}$ defined in \eqref{Hopf map} is the image of a Hopf map. Combined them together, the inverse image of the curve $U^{-1} (\gamma(s))$ must be a 2-sphere inside $S_4$. However, this 2-sphere is not shrinkable, because the curve $\gamma(s)$ in Sp(1) manifold is not shrinkable, therefore the map $U(x)$ is a non-trivial map, which must be the non-trivial element of $\pi_4(Sp(1)) = \mathbf{Z}_2$.

Let us denote the two 5d topological vacua  due to $\pi_4 ({\rm Sp}(1)) = \mathbf{Z}_2$
as $|0\rangle$ and $|1\rangle$.
The gauge transformation $\exp(2\pi i \mathbf{Q})$ will swap the two vacua as:
\begin{equation}
	e^{2\pi i \mathbf{Q}} |0\rangle = |1\rangle,\quad e^{2\pi i \mathbf{Q}} |1\rangle = |0\rangle \ .
\end{equation}
There are two discrete analogs of the 4d $\theta$-vacua,
\begin{equation}
	|+ \rangle = |0\rangle + |1\rangle, \quad |-\rangle = | 0 \rangle - | 1 \rangle \ .
\end{equation}
on which $e^{2\pi i \mathbf{Q}}$ acts as

\begin{equation}
	e^{2\pi i \mathbf{Q}} | +\rangle = | +\rangle,\quad e^{2\pi i \mathbf{Q}} | -\rangle = - | - \rangle,
\end{equation}
and therefore leave the $\theta=0$ vacuum invariant and shift the $\theta = \pi$ vacuum by an additional phase $e^{\pi i}$.
The electric charge of the unit instanton soliton is thus ``integer" quantized for the $\theta = 0$
vacuum $| +\rangle $, but should be shifted by half a unit in the $\theta = \pi$ vacuum $| -\rangle $.
The half charge here is on par with those of the defining representation, as cautioned
at the head of this section, so the lightest dyonic instanton of F1 theory, for example, would form an Sp(1) doublet.

In summary, the key to such charge shifts, both in 4d and in 5d, is how the presence of a topological soliton induces a topological winding number of $e^{2\pi i\Phi/\mathbf{v}}$ as a map from $\IR^{d-1}$ to $G$, valued in $\pi_{d-1}(G)$. Note that the latter are not the topological numbers that characterize the soliton itself; the 4d monopole is defined by the {\it asymptotic} winding of $\Phi$ valued in $\pi_2({\rm Sp}(1)/{\rm U}(1))$ while the 5d instanton carries the second Chern class of the gauge fields. The latter 5d case may look a little odder since the instanton does not need $\Phi$ for its existence, but the salient point is how the electric charge here really refers to the unbroken ${\rm U}(1)$'s. So, a symmetry breaking $\Phi$ must enter and, once the vev of $\Phi$ is turned on, an odd instanton number induces winding of the exponentiated $\Phi$ valued in $\pi_4({\rm Sp}(k))=\IZ_2$.

Although we performed the above computation for Sp(1), the same argument follows for the instanton of Sp$(k)$ theory by embedding an Sp(1) instanton. The only difference for the latter is how the latter, in general, comes with more angular moduli for orientation in the gauge group, which does not interfere with these topological questions.

\section{Subtleties with the 4d End}

As we saw in the preliminary materials, the 5d Seiberg-Witten on a circle comes with cylindrical Coulombic moduli spaces. The circular directions are generated by the gauge field $A_5$ along the circle, constituting the holonomy torus with periods $\sim 1/R_5$, which collapses in the decompactification limit of $R_5\rightarrow \infty$. This leaves behind 5d Coulombic moduli space whose structure is well captured by the $(p,q)$ fivebrane web, and the 5d Seiberg-Witten geometry on a circle is, asymptotically at least, a Cartan torus fibred over this strictly 5d moduli space.

At the lower end, with scale much less than $1/R_5$, this cylindrical geometry would be capped by its 4d counterparts. However, the structure at such a 4d end of the Coulombic moduli space is not as simple as one might imagine. In a later section, we will spend some time on the large $R_5$ limit, but here we dwell on exactly how 4d Seiberg-Witten geometries cap the 5d Coulombic cylinder by taking $1/R_5$ to a large value, effectively acting as the UV cut-off in 4d sense, and investigating how 4d Seiberg-Witten geometries enter at the low energy scale much less $1/R_5$.

To be more quantitative, let us declare the
complexified couplings $\tau_{\rm 4d}$ in 4d and $\tau_{\rm 5d}$ in 5d,
respectively. One useful definition in 4d is \cite{Seiberg:1994aj},
\begin{equation}
\tau_{\textrm{4d}}=  \frac{\theta}{\pi} + \frac{8 \pi i}{g^2_{4,\textrm{eff}}}\ ,
\end{equation}
whose normalization is also the one we chose for $\tau_{\rm 5d}$. This leads to
we find
\bea
a_D\approx \tau_{\rm 4d}\cdot a
\eea
in the asymptotic region of the 4d Sp(1) Seiberg-Witten moduli space. This differs from more standard $\tau$ by a factor 2, but has the advantage of having asymptotic monodromy being integer shifts of $\tau$. It is also consistent with our convention that $2a$ is the central charge of the W-boson.
On the other hand, given $1/ g^2_{4,\textrm{eff}} =2\pi R_5/ g^2_{5,\textrm{eff}}$
one
\begin{equation}\label{4d-coupling-large-a}
	\tau_{\textrm{4d}} = 2\pi R_5 \tau_{\textrm{5d}} \,
\end{equation}
We will perform a little more quantitative computation of these complexified couplings later, but here we will first overview the global characteristics of these $\tau$'s with emphasis on the monodromies.

An important aspect of circle-compactified gauge theories that are often overlooked is how the periodic nature of the holonomy variable
\bea
h\sim \int_{\IS^1} A
\eea
affects the dimensional reduction. The latter means we replace the gauge field along the circle with an adjoint scalar field, which loses the memory of the periodic nature of the gauge field. In this process, one really expands around a particular gauge holonomy, so the dimensionally reduced theories are labeled by this holonomy value. For general gauge theory, quantum effect develops a bosonic potential $U(h)$ for the holonomy, and the vacuum expectation value of the latter is closely related to whether the gauge theory in question is confining or not. As such, detailed dynamics of the holonomy variable are very important and equally difficult.

With supersymmetries, we often control the problem to a manageable level. With four supercharges, one usually finds a discrete set of preferred holonomy values, around which one finds various $(d-1)$ dimensional supersymmetric theories from a single $d$ dimensional theory. Such places have been recently dubbed ``holonomy saddles," and explored much via various supersymmetric (twisted) partition functions on a circle \cite{Hwang:2017nop, Hwang:2018riu}. A closely related observation on how a single 4d/3d duality descends to 3d/2d dualities between multiple theories can be found in  Refs.~\cite{Aharony:2013dha, Aharony:2017adm}.

With eight supercharges, one finds the potential for the holonomy variable is entirely lifted, $\partial_h U(h)=0$. After all, 4d Seiberg-Witten theory comes with complex Coulombic moduli, imaginary half of which may be viewed as the remnant of $A_5$. One can say that one finds a continuous family of the holonomy saddles. However, at generic such saddle, the dimensionally reduced theory would be free Abelian theory with no matter content; charged degrees of freedom would acquire a mass $\langle A_5\rangle \sim 1/R_5$ which diverges at 4d limit and decouples. Only for some exceptional values of the holonomy, such as $R_5\langle A_5\rangle =0 $ or $R_5\langle A_5\rangle =\sigma_3/2 $ for rank 1 cases, one would find interacting Seiberg-Witten theory in the 4d limit.

This means that a close inspection of the 5d Seiberg-Witten theory on a small circle would generically yield multiple 4d counterparts, all embedded in the infrared end of the former. These multiple 4d theories would be separated by distance $\sim 1/R_5$ along the $A_5$ direction. One quantity that reflects this complicated structure is the asymptotic monodromies resulting from traversing one of the holonomy periods at the asymptotic infinity of the 5d theory on $\IS^1$. For generic theory, these monodromies will not agree with those anticipated from the asymptotic monodromies of $\tau_{\rm 4d}$ based on the single 4d theory found by the naive dimensional reduction at $R_5\langle A_5\rangle =0 $. Instead, multiple asymptotic $\tau_{\rm 4d}$ monodromies from the various interacting holonomy saddles would combine to generate the total monodromy at the cylindrical infinity.

In this section, we will illustrate this by drawing the global overview of a few rank 1 moduli spaces, say, F0, F1, and dP$_2$, in various parameter regimes. We will assume, for the sake of simplicity, that the 4d scales $\Lambda_{\rm QCD}$ of interacting holonomy saddles is much less than $1/R_5$ in the following. F0 and F1 are both pure Sp(1) gauge theories, with the difference being that F1 has the discrete 5d theta angle turned on. As such the latter is equipped with dyonic instantons with unit electric charges, equivalent to that of a quark. dP$_2$ is Sp(1) theory with a single quark hypermultiplet, so there is some similarity between F1 and dP$_2$.

For F0, the two interacting holonomy saddles, at $R_5\langle A_5\rangle =0 $ and $R_5\langle A_5\rangle =\sigma_3/2 $, are equivalent to each other since the only charged matter is the W boson with the electric charge 2. As such, the 5d Seiberg-Witten geometry with $\IS^1$ is a half-cylinder of the asymptotic periodicity $1/2R_5$, which closes at the infrared end where 4d Sp(1) Seiberg-Witten geometry sits. The asymptotic monodromy is $\tau_{\rm4d}\rightarrow \tau_{\rm4d}+4$, which is the same as the monodromy around the strong coupling region of 4d Sp(1) Seiberg-Witten theory in our convention. We refer readers to Section 6 for further details of the monodromy; there we investigate the large $R_5$ limit, but the discrete monodromies data are identical to the small $R_5$ limit.

\begin{figure}[pbth]
\centering
\includegraphics[scale=0.35]{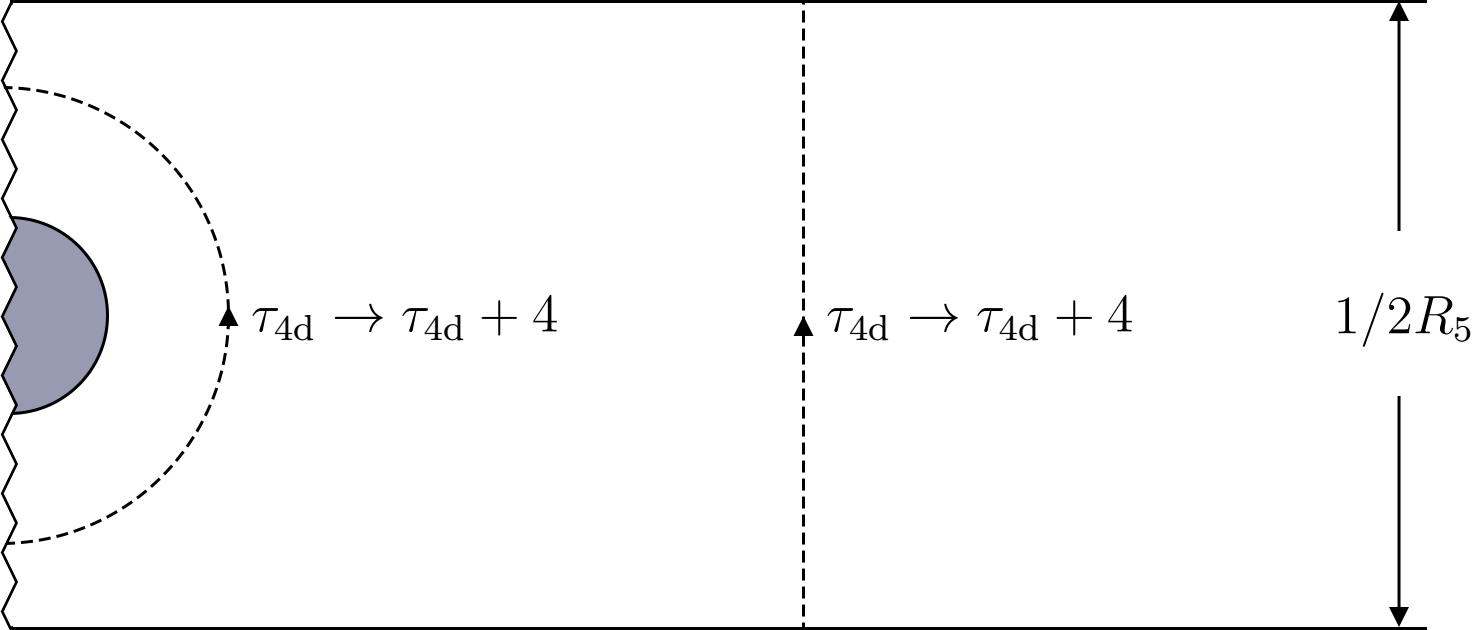}
\caption{Schematics of the F0 moduli space in terms of $\phi+iA_5$; the top and the bottom
boundaries should be identified while the left end should be folded in half. The shaded
region represent  4d Seiberg-Witten geometry of the pure Sp(1) theory. }
\label{Fig-F0-saddle}
\end{figure}

For dP$_2$, on the other hand, the two saddles $R_5\langle A_5\rangle =0 $ and $R_5\langle A_5\rangle =\sigma_3/2 $ are clearly distinct. With the 5d real mass $\mu_f$ for the quark, the holonomy complexifies this mass to
\bea
\sim \mu_f \pm i\frac{h_*}{R_5}\ ,
\eea
where $\pm h_*$ are the eigenvalues of $R_5\langle A_5\rangle$. At generic values of $h_*$, Sp(1) is strongly broken to U(1) by the holonomy itself and such generic holonomy saddles are noninteracting in the 4d limit.

At the trivial holonomy saddle, $h_*=0$, there is no such shift of the 4d mass relative to 5d mass, so the dimensional reduction gives the same content as 5d, i.e., Sp(1) theory with a quark hypermultiplet of bare mass $\mu_f$ which happen to be real. If one sits here and scales down $R_5\rightarrow 0$, the local geometry would appear to be that of 4d Sp(1) theory with a single quark hypermultiplet of the mass $\mu_f$.\footnote{The existence of these two distinct 4d limits was observed early on in Ref.~\cite{Ganor:1996pc}.}

\begin{figure}[pbth]
\centering
\includegraphics[scale=0.35]{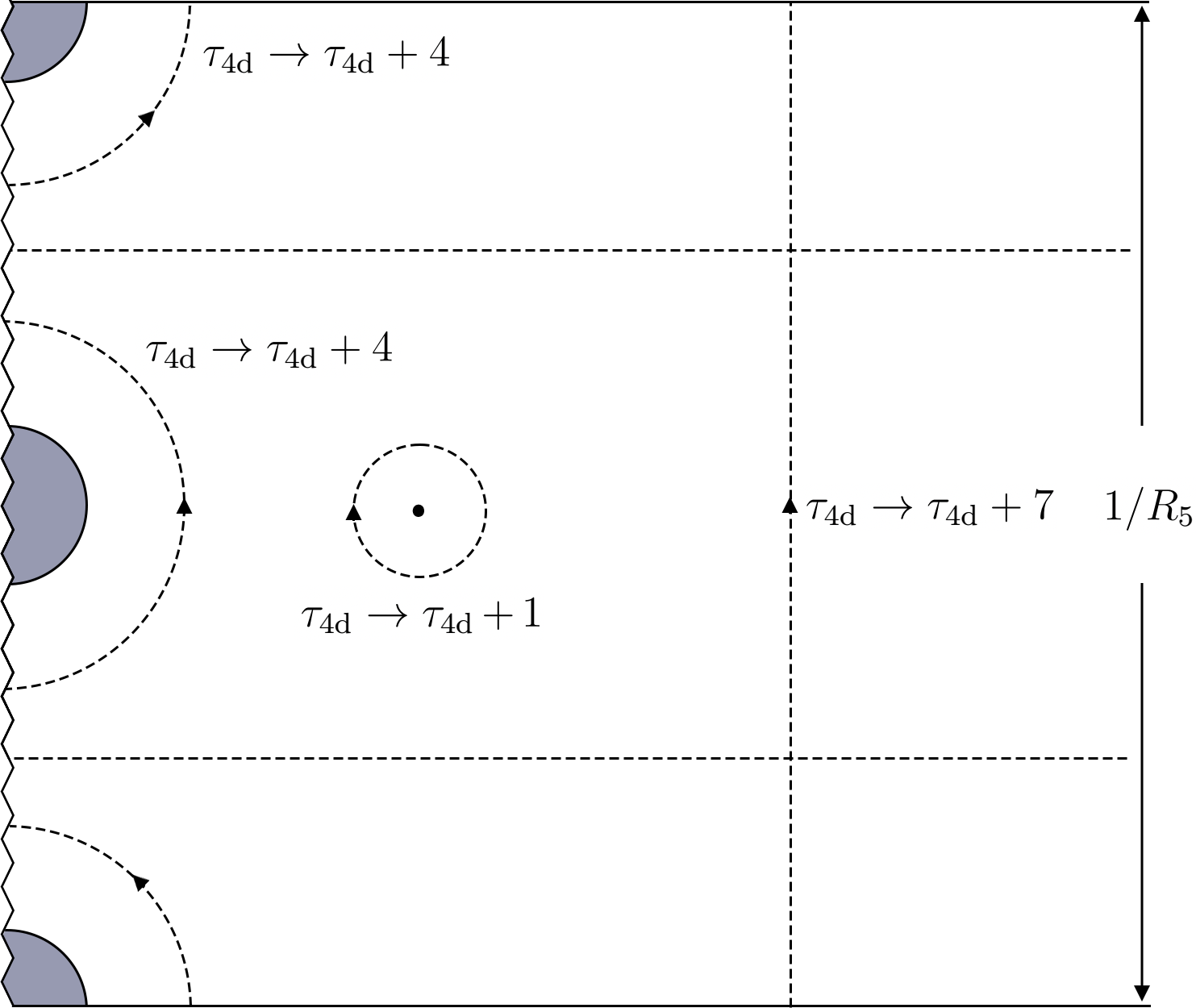}
\caption{Schematics of the dP$_2$ moduli space; again, the top and the bottom
boundaries should be identified while the left end should be folded in half.
The two holonomy saddles toward the left end are well approximated by a 4d Sp(1)
theory with a single flavor and a 4d pure Sp(1) theory, respectively. }
\label{Fig-dP2-saddle}
\end{figure}

In particular the monodromy around this local structure gives $\tau_{\rm4d}\rightarrow \tau_{\rm4d}+3$. At the other interacting saddle, $h_*=1/2$, one finds the bare 4d mass of the quark scaling as $1/R_5\rightarrow\infty$, so the reduced 4d theory is a pure Sp(1) gauge theory. The monodromy around this local region is the same old $\tau_{\rm 4d}\rightarrow \tau_{\rm 4d}+3$. The asymptotic monodromy for 5d dP$_2$ is $\tau_{\rm 4d}\rightarrow \tau_{\rm 4d}+7$, on the other hand. Clearly this decomposes in the 4d end as a pair of monodromies, $\tau_{\rm 4d}\rightarrow \tau_{\rm 4d}+3$ from the trivial holonomy saddle at $h_*=0$ and from the nontrivial holonomy saddle at $h_*=1/2$.

F1, despite its pure Sp(1) origin, has a more involved structure than F0. Because the 5d spectrum includes odd-integer charged dyonic instantons, the asymptotic monodromy is $\tau_{\rm 4d}\rightarrow \tau_{\rm 4d}+8$, twice that of a pure 4d Sp(1) theory. This is, of course, because pure 4d Sp(1) theory has only one charged particle, namely W boson of charge two, whereas the F1 theory carries the dyonic instanton with the electric charge one. As long as the latter dyonic instanton is very heavy everywhere, the schematic structure of F1 moduli space is roughly a double cover of the cigar-like F0 Seiberg-Witten moduli space. In the 4d end, one finds two copies of 4d pure Sp(1) Seiberg-Witten moduli space, sitting  at $h_*=0$ and $h_*=1/2$, respectively, each with the monodromy $\tau_{\rm 4d}\rightarrow \tau_{\rm 4d}+4$. Combined, these recover the asymptotic monodromy, $\tau_{\rm 4d}\rightarrow \tau_{\rm 4d}+8$.

\begin{figure}[pbth]
\centering
\includegraphics[scale=0.35]{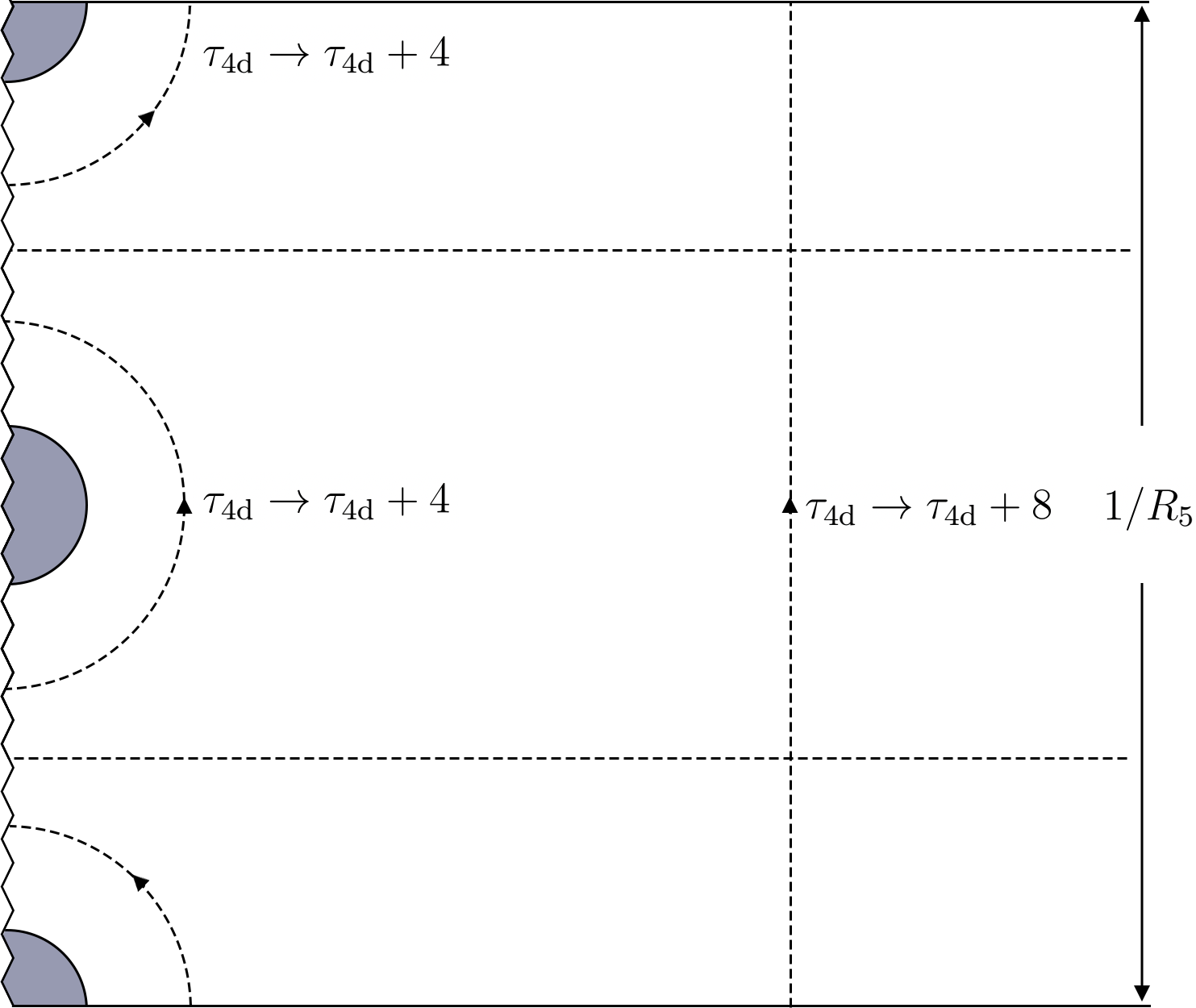}
\caption{Schematics of the F1 moduli space; again, the top and the bottom
boundaries should be identified while the left end should be folded in half.
The two shaded regions on the left end are both well approximated by
a 4d pure Sp(1) theory but react a little differently to $R_5\mu_0>0$. }
\label{Fig-F1-saddle}
\end{figure}

However, something else happens if 5d bare coupling-squared $\mu_0$ is negative. As we delineate in Section 6, the dyonic instanton with unit electric charge becomes massless somewhere in the moduli space. To the left of this dyonic instanton singularity, the monodromy increase by unit as $\tau_{\rm 4d}\rightarrow \tau_{\rm 4d}+9$, which is shared by three identical singularities in the infrared end, as will be discussed in Section 6.

\section{Wall-Crossing or Not}

When it comes to BPS spectra, a very distinctive feature of the 5d theory is the absence of wall-crossing for BPS particles. Let us see how this happens and what this absence means. From the low energy dynamics of BPS objects, the usual 4d wall-crossing occurs due to runaway Coulombic directions emerging at special values of FI constants \cite{Denef:2002ru, Denef:2007vg, Hori:2014tda}, which is in turn related to how the central charge phases are complex and phases of a pair can align at a co-dimension-one wall. With mutual intersection number, i.e., with nonzero Schwinger product of charges, this leads to a wall-crossing.

On the other hand,  the 5d BPS particles would be all represented by M2 wrapped on 2-cycles so that a pair of 2-cycles in a Calabi-Yau 3-fold cannot have a mutual intersection number. While D4 on a 4-cycle can have an intersection number against D2 on a 2-cycle, the former is an extended object in the form of the BPS string since it is really M5 wrapping a 4-cycle. In fact, most BPS objects that would have entered the wall-crossing quiver gauge quantum mechanics in the 4d limit are made up of magnetic strings.

Once compactified on a circle $\IS^1$ of sufficiently small radius $R_5$, however, the low energy dynamics are again well captured by BPS quiver quantum mechanics for D4-D2-D0 bound states, for which the wall-crossing are numerous. We will ask exactly how the wall-crossing turns itself off back in the limit $R_5\rightarrow \infty$. One quick answer to this is that since all magnetic objects become a string, as pointed out already, the question becomes moot if we ask for wall-crossing among 5d BPS particles. However, we can ask a little more by keeping track of BPS strings wrapped on $\IS^1$ considered as BPS particles. As we will see below, approaching the decompactification  limit from the compactified theory, the disappearance of wall-crossing occurs for multiple reasons, different for different BPS objects.

Let us recall how the 4d wall-crossing phenomena are tied to how asymptotic Coulombic flat direction opens up when certain combinations of FI constant approaches zero. With generic FI constants, the one-looped corrected D-term potential has the shape, schematically \cite{Denef:2002ru},
\bea
\left(\zeta_{ij}- \sum_a \frac{{\rm sgn}(Q_{a}^{ij})}{|\vec x_{ij}|} \right)^2
\eea
which creates a gap for large relative position $\vec x_{ij}=\vec x_i-\vec x_j$ along $\IR^3$. With $\zeta_{ij}=0$, however, the gap disappears, allowing wavefunctions to escape into the asymptotic infinity. This simply means, from the 4d field theory viewpoint, that the BPS bound state becomes unbound.

An important ingredient here is the sum, labeled by $a$, over the bifundamental chiral multiplets. If there were no such one-loop contributions from the chiral multiplets, there are no Coulombic $L^2$ supersymmetric bound states and no supersymmetric bound state to disappear into the Coulombic infinity at a special value of $\zeta_{ij}=0$. As such, the usual 4d wall-crossing, where the complex central charge $Z$ of the bound state aligns with  $Z_i$ and $Z_j$ and  decays as
\bea
Z\rightarrow Z_i+Z_j\ ,
\eea
is possible only if D-branes that constitute particles $i$ and $j$ share a net number of open strings attached to them. This is in turn counted by the intersection number between the two cycles wrapped by the D-branes.

In the $\IR^{3+1}$ gauge theory viewpoint, on the other hand, the intersection number translates to the Schwinger product of electromagnetic charges,
\bea
q_ig_j-q_jg_i\neq 0
\eea
so at least one of the two constituents must carry a magnetic charge. As such, one can imagine three logical possibilities for the 4d wall-crossing:
\begin{itemize}
\item
$Z$ is  non-magnetic, while both constituent $Z_i$ and $Z_j$ are magnetic,
\item
$Z$ and one of the two constituents, say $Z_i$, are magnetic,
\item
all three are magnetic.
\end{itemize}
Also, the marginal stability wall of such decay, found by equating phases of $Z_i$ to that of $Z_j$, would extend between $Z_i=0$ and $Z_j=0$ locus in the Seiberg-Witten moduli space.

Note that the claimed absence of 5d wall-crossings really refers to the first of these three where $Z$ represents a BPS {\it particle} in the 5d sense. It is clear that no particle-like state in 5d sense can possibly decay into a pair of BPS strings unless either $R_5$ is very small or the latter strings are nearly tensionless. If we take the absence of such wall-crossings as given, this in turn implies that no magnetically charged BPS strings should become tensionless, perhaps except at the boundary of the Coulombic phase.

In the $(p,q)$ web realization of IIB theory, which we briefly reviewed earlier, the tensionless limit of magnetic strings translates to a collapse of a face, so the above claim that the Coulomb phase ends where a magnetic BPS becomes tensionless is quite natural already. On the other hand, from the finite $R_5$ perspective, this place where magnetic strings become tensionless is a co-dimension-two singular hypersurface, and the Seiberg-Witten moduli space should be smoothly capped in its neighborhood. In the limit of $R_5\rightarrow \infty$, this capping of the moduli space turns into a boundary as the Coulomb moduli space reduces to real instead of complex. Later, we will explore these aspects of 5d theories microscopically by analyzing how the Seiberg-Witten theory on $\IS^1\times\IR^{3+1}$ approaches the decompactification limit. Depending on the value of 5d  bare ``coupling-squared," which could turn negative, one finds various different physics occurring at such ``endpoints" of the Coulomb phase.

For the second and the third, $Z$ would represent a magnetically charged BPS string wrapped on $\IS^1$.  The second type of marginal stability wall would emanate from a point in the moduli space where some components
of charged BPS particles become massless. For instance, imagine a quark hypermultiplet with mass $\mu_f$ in the defining representation of the gauge group. The central charges (densities) of any magnetically charged objects are by and large pure imaginary at finite $\phi$; Although we say that the 5d superalgebra admits real central charges only, this really applies to particle-like states. For strings, such as monopoles or dyons, the relevant superalgebra is that of 4d $\cN=2$.

At finite $R_5$, we obtain  particle-like states by wrapping these on $\IS^1$, so in the weak coupling limit, their central charge is almost pure imaginary
\bea
Z_i \; \sim\; Z_{\rm monopole} \; \approx\; \frac{16\pi^2 i R_5}{g_{5,\rm eff}^2} \phi\ .
\eea
Therefore, the wall emanating from the massless quark point following
\bea
{\rm Arg}(Z_i Z_j^*)=0
\eea
would initially extend along the circular Wilson-line direction of the small period  $1/R_5$. Taking the $R_5\rightarrow \infty$ limit, this means that there is a sense of discontinuity for magnetically charged strings across $\phi=\mu_f$.

Note that $\phi>\mu_f$ is precisely where an Sp(1) doublet fermion would contribute a Jackiw-Rebbi zero-mode \cite{Jackiw:1975fn} to the monopole. These zero-modes are responsible, for example, how monopoles and dyons in 4d Sp(1) gauge theory with $N_f$ flavors are in chiral isospinor representation under the flavor group $SO(2N_f)$. With $\phi< \mu_f$, this zero-mode is lifted, so the $d=1+1$ low energy dynamics of the monopole (dyon) strings change qualitatively across this point. This discontinuity translates to the conventional wall-crossing once the BPS string wraps a circle, $\IS^1$, which may be captured via the elliptic genus of such a magnetic string; later, we will review this with a concrete case of a magnetic string in the dP$_2$ theory, or Sp(1) theory with a single massive flavor. Also playing an important role here are the Kaluza-Klein modes along $\IS^1$, or D0 branes from the IIA viewpoint,  whose central charge is also pure imaginary.

Marginal stability walls of the third type where magnetic strings decay into a pair of magnetic strings deserve different considerations. In 4d, such decays involving three types of dyons are known, and these walls actually extend into the weak coupling region  \cite{Lee:1998nv}. However, for this type of wall-crossings, two independent adjoint scalar fields spanning the Coulombic moduli space are essential; The decaying states may be visualized by an analog of planar $(p,q)$ string web, which cannot be drawn when the adjoint scalar is real  \cite{Bergman:1998gs}. Such walls would not survive the decompactification limit since half of the Coulombic directions, corresponding to the Wilson line vev, collapses due to $R_5\rightarrow \infty$. Thus, such 3rd type of marginal stability would also turn off in the decompactification limit, leaving behind the simplest possible chamber.

\section{Sp(1) Theories on a Large $\IS^1$}

In the previous section, we have discussed three logical possibilities for marginal stability walls where a 4d BPS particle decay as $Z \rightarrow Z_i + Z_j$. For rank one field theory, 
we really have two types to explore:
\begin{itemize}
	\item $Z$ is non-magnetic, while both $Z_i$ and $Z_j$ are magnetic,
	\item $Z$ and one of the two constituents, say $Z_i$, are magnetic.
\end{itemize}
Here, we will analyze a pair of pure Sp(1) theories, namely F0 and F1 theories distinguished by the discrete theta angle, as well as the Sp(1) with a single flavor, say, dP$_2$ theory, and observe how the first kind of marginal stability walls collapses in the decompactification limit; the infrared boundary of 5d Coulomb phase occurs where a magnetic string becomes tensionless.\footnote{This fact has been noted and made use of in recent classification of 5d theories \cite{Jefferson:2017ahm, Jefferson:2018irk}.} We illustrate this in much detail for these rank one theories for both positive and negative bare coupling-squared.

While we give detailed expositions below, the crux of the matter is that, in the large radius limit, the 4d effective coupling is necessarily driven to be small thanks to the simple relation $g^2_{\textrm{4d}}\sim  g^2_{\textrm{5d}}/2\pi R_5$. As such, for most of the moduli space, the prepotential reduces to the familiar log behavior, except that in the argument of this log, we find a hyperbolic function reflecting the KK mode sum from the circle compactified nature of the theory \cite{Katz:1996fh, Lawrence:1997jr}. In particular, $1/R_5$ acts like a UV cut-off for 4d theory, at which scale the 4d coupling-squared vanishes linearly with $1/R_5$. This means that much of the strongly-coupled 4d Seiberg-Witten geometry is pushed well below $1/R_5$, so it must collapse entirely in the decompactification limit, which also implies that marginal stability walls for pure electric particles collapse and get pushed to the boundary.

One potential complication is the 5d bare inverse-coupling-squared $\mu_0$ that can become negative. For the F0 case, regardless of the sign of $\mu_0$, structures of the Coulomb phase remain essentially
the same, modulo a shift of the Coulombic variable below $1/R_5$ scale. However, this will not be the case for F1 with its discrete 5d theta-angle. In the middle subsection, we will consider the latter case in some depth and give the general overview of its Seiberg-Witten geometries, both at small and large $R_5$.

\subsection{F0 Theory}

\begin{figure}[pbth]
\centering
\includegraphics[scale=1]{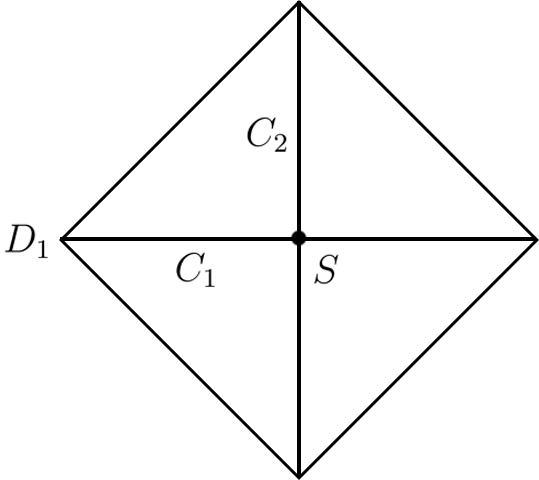}
\caption{The toric diagram for $F0$ theory.}
\label{Fig-F0-toric}
\end{figure}

We first consider the F0 theory, whose toric diagram is figure \ref{Fig-F0-toric}. There is only one compact divisor $S$, whose self-triple intersection number is 8, and one independent non-compact divisor $D_1$\footnote{In the toric diagram, the compact divisor corresponds to the internal vertex $S$ while the non-compact divisors correspond to the four vertexes at each corner in figure \ref{Fig-F0-toric}. However, these five divisors are not independent, and we can choose $S$ and $D_1$ as a basis, and the other three non-compact divisors can be written as a combination of these two; for example, the divisors correspond to the top, and the bottom vertices are $-S/2 - D_1$ while the divisor corresponds to the right vertex is $D_1$. The exact relations among these divisors in F0 and F1 models can be found in \cite{Closset:2018bjz}.}. There are also two independent compact 2-sphere $C_1$ and $C_2$, and their intersection number between $S$ are both -2. We summarize the geometric data in the following:
\begin{itemize}
	\item Divisors:
		\begin{equation}
	S \cdot S \cdot S = 8,
		\end{equation}
		\begin{equation}
	S \cdot S \cdot D_1 = -2,
		\end{equation}
		\begin{equation}
	S \cdot D_1 \cdot D_1 = 0.
		\end{equation}
	\item 2-cycles:
		\begin{equation}
			C_1 \cdot S = -2,\quad C_2 \cdot S = -2.
		\end{equation}
		\begin{equation}
			C_1 \cdot D_1 = 0,\quad C_2 \cdot D_1 = 1.
		\end{equation}
\end{itemize}
The K$\ddot{\textrm{a}}$hler form $[J]$ is then parameterized by:
\begin{equation}
	[J] = \mu_0 [D_1] - \phi [S],
\end{equation}
and the volume of two cycles $C_1$ and $C_2$ are given as:
\begin{equation}
	A_1 = M_W = J \cdot C_1 = 2\phi,\quad A_2 = M_I = J \cdot C_2 = \mu_0 + 2\phi,
\end{equation}
which are the same as the masses of the W-boson and dyonic instanton given in the brane web picture. The prepotential \eqref{5d-prepotential-toric} is then:
\begin{equation}
	\mathcal{F}_{\textrm{IMS}} (\phi,\mu_0) = -\frac{1}{6} J \cdot J \cdot J = \frac{4}{3} \phi^3 + \mu_0 \phi^2,
\end{equation}
where we have omitted the constant part $\sim D_1 \cdot D_1 \cdot D_1$. This prepotential from toric geometry is the same as the IMS prepotential given in \eqref{5d Sp(1) prepotential}.

After compactification, according to \eqref{Instanton prepotential}, the third derivative of the prepotential is given by:
\begin{equation}
	\frac{\partial^3 F_{\rm exact}(t,v)}{\partial t^3} = 8 + \sum_{m,n\geq 0} N_{m,n} \left( \frac{e^{-4\pi m i t}e^{2\pi n i (-2t + v) } }{1 - e^{-4\pi m i t}e^{2\pi n i (-2t + v)}} \right)  (-2m -2n)^3.
\end{equation}
When the volumes of the 2-cycles are large such that the instanton effect is small, the parameters $\{t,v\}$ can be read from the worldline instanton which is controlled by $\exp (-2\pi R_5 m M_W)$ and $\exp (-2\pi R_5 n M_I)$ :
\begin{align}
	e^{-4\pi m i t} = e^{-2\pi R_5 m M_W} \rightarrow \textrm{Im}(t) = - \phi R_5,\nonumber \\
	e^{2\pi n i (-2t + v)} = e^{-2\pi R_5 n M_I} \rightarrow \textrm{Im}(v) = \mu_0 R_5,
\end{align}
Since we are interested in the 5d theory where the mass is real, $v = i \mu_0 R_5$ remains purely imaginary
regardless of the compactification.

We need to rescale the $F_{\rm exact}(t,\mu)$ such that in the large radius limit $R_5\rightarrow \infty$, it reduce to $\mathcal{F}_{\textrm{IMS}}$ correctly. We redefine the flat coordinate $t$ as $a \equiv i t/R_5$ such that, in the asymptotic region of the Coulombic moduli space, we have $\textrm{Re}(a) \approx \phi$. Then we convert the dimensionless prepotential $F_{\rm exact}(t,\mu)$ by a scaling factor $( \frac{i}{R_5})^3$ to a dimensionful one, $\mathcal{F}_{\textrm{5d},\mathbb{S}^1}$, as
\begin{equation}\label{F0 prepotential}
	 \frac{\partial^3 \mathcal{F}_{\textrm{5d},\mathbb{S}^1}(a,\mu_0)}{\partial a^3} = 8 + \sum_{m,n\geq 0} N_{m,n} \left( \frac{e^{-4\pi m a R_5} e^{-2\pi n (2a + \mu_0)R_5 }}{1 - e^{-4\pi m a R_5} e^{-2\pi n (2a + \mu_0)R_5 }} \right)  (-2m -2n)^3,
\end{equation}
In the limit $R_5\rightarrow \infty$, all the instanton contributions vanish and $a \rightarrow \phi$ by definition, so
we recover the 5d result as
\begin{equation}\label{Relation-4d-5d-prepotential}
\mathcal{F}_{\textrm{IMS}}= \lim_{R_5\rightarrow \infty}  \mathcal{F}_{\textrm{5d},\mathbb{S}^1}(a,\mu_0)\ .
\end{equation}
We are mostly interested in the interface between 4d and 5d, so we take $R_5$ to be sufficiently large in the following discussion whereby $|g_5^2 / 2\pi R_5| \ll 1$.

\subsubsection*{Positive bare coupling-squared}
The brane web for positive bare coupling-squared is given in figure 2, and the 5d Coulomb moduli space is a half-line that ends at a point where the two D5-branes merge and the Sp(1) gauge symmetry is restored. When we compactify the theory on a large circle, one may expect the moduli space to be a thin cigar, which will shrink to the half-line in 5d limit $R_5\rightarrow \infty$. In this subsection, we will check this picture explicitly. There are three different regime depending on the magnitude of $a$ : $\textrm{Re}(a) \gg 1/2\pi R_5$, $\textrm{Re}(a) \sim 1/2\pi R_5$, $|a| \ll 1/2\pi R_5$.

Let's first consider the limit that $\textrm{Re} (a) \gg 1/2\pi R_5$, where $\textrm{Re}(a)$ can be identified with 5d Coulomb moduli $\phi$. In this regime, the compactification radius $R_5$ is sufficiently large compared to other length scales, the theory is effectively 5d and we can ignore the instanton contributions such that:
\begin{equation}
	\frac{\partial^3 \mathcal{F}_{\textrm{5d},\mathbb{S}^1}}{\partial a^3} \approx 8,
\end{equation}
which gives the monopole string tension and effective coupling via integration:
\begin{equation}\label{F0-tension-and-coupling-large-a}
	i T_{\textrm{mono}} = \frac{i}{2\pi} \frac{\partial \mathcal{F}_{\textrm{5d},\mathbb{S}^1}}{\partial a} \approx \frac{i a (2a + \mu_0)}{\pi},\quad  \tau_{\textrm{5d}} = \frac{i}{2\pi}\frac{\partial^2 \mathcal{F}_{\textrm{5d},\mathbb{S}^1}}{\partial a^2} \approx \frac{i (4 a + \mu_0)}{\pi},
\end{equation}
and they are the same as \eqref{5d Sp(1) coupling} and \eqref{5d Sp(1) monopole} replacing $a\rightarrow \phi$.
The moduli $a$ is complex with periodicity $a \rightarrow a+i/2R_5$ which can be read from the exponent of \eqref{F0 prepotential}, therefore the topology of the moduli space in this region is a cylinder, whose radius is $\sim 1/2R_5$, and will shrink to a line as $R_5\rightarrow \infty$. This translates to
\begin{equation}\label{F0-4d-coupling-large-a}
	\tau_{\textrm{4d}} = 2\pi R_5 \tau_{\textrm{5d}} \approx 2\mu_0 R_5 i + 8 a R_5 i,
\end{equation}
with monodromy $\tau_{\textrm{4d}} \rightarrow \tau_{\textrm{4d}} + 4$ around the circle.

Next, we consider $\textrm{Re}(a) \sim 1/2\pi R_5$, which means the energy scale is comparable with $1/2\pi R_5$ and the instanton contributions can not be omitted. Since we are working with large $R_5$ ($\mu_0 R_5 \gg 1$) , one can set $n=0$ in \eqref{F0 prepotential} to obtain:
\begin{equation}\label{F0 leading prepotential}
	\frac{\partial^3 \mathcal{F}_{\textrm{5d},\mathbb{S}^1}}{\partial a^3} \approx  8 +  \sum_{m\geq 0} N_{m,0} \left( \frac{e^{-4\pi m a R_5}}{1 - e^{-4\pi m a R_5}} \right)  (-2m)^3.
\end{equation}
The only non-zero $N_{m,0}$ is $N_{1,0} = -2$ for F0 \cite{Chiang:1999tz}, therefore we have:
\begin{equation}
	\frac{\partial^3 \mathcal{F}_{\textrm{5d},\mathbb{S}^1}}{\partial a^3} = 8 + 16 \left( \frac{e^{-4\pi a R_5}}{1 - e^{-4\pi a R_5}} \right),
\end{equation}
which gives the 5d coupling:
\begin{align}\label{F0-5d-coupling-middle-a}
	\tau_{\textrm{5d}} 
	&\approx \frac{i \mu_0}{\pi} + \frac{i}{\pi^2 R_5} \log 4 \sinh^2 (2\pi R_5 a),
\end{align}
or in terms of the 4d effective coupling $\tau_{\textrm{4d}} = 2\pi R_5 \tau_{\textrm{5d}}$:
\begin{equation}\label{F0 4d-coupling}
	 \tau_{\textrm{4d}} \approx 2\mu_0 R_5 i - \frac{4}{2\pi i} \log 4 \sinh^2 (2\pi R_5 a).
\end{equation}
It is clear from the above expressions that this approximate form of the effective coupling will blow up when $\textrm{Re}(a)$ is small enough.

As discussed before, the compactification introduces an energy scale $\sim 1/R_5$, which is the lowest KK mode mass, and if we are working below this scale, we cannot feel the existence of the compactified circle, and we may think the theory is effectively a 4d gauge theory. In this sense, it is natural to identify $1/2\pi R_5$ as the UV scale $\Lambda_{\textrm{UV}}$ in the 4d effective theory since the 4d description will break down here, which is also suggested in  \cite{Lawrence:1997jr}. Since we work with a large radius $R_5$, the first term in the second line, which is related to the 5d bare coupling, is much larger than the second term when $\textrm{Re}(a) \sim 1/2\pi R_5$, therefore we can  identify the 4d gauge coupling $1/g_{4,\textrm{UV}}^2$ at $\Lambda_{\textrm{UV}}$ as $\mu_0 R_5/4\pi $.

We define the $\Lambda_{\textrm{QCD}}$ in the 4d effective theory according to the 4d $\beta$-function as:
\begin{equation} \label{Lambda QCD}
	\frac{1}{g^2_{4,\textrm{UV}}} \sim \frac{1}{4\pi^2}\log \left(\frac{\Lambda_{\textrm{UV}}^2}{\Lambda_{\textrm{QCD}}^2}\right),
\end{equation}
and one has $\Lambda_{\textrm{QCD}} \ll \Lambda_{\textrm{UV}}$ since $\mu_0 R_5 \gg 1$ for large radius. Substitute it into \eqref{F0 4d-coupling} we have the 4d effective coupling in terms of the $\Lambda_{\textrm{QCD}}$ scale:
\begin{equation}\label{F0 4d-coupling-2}
	\tau_{\textrm{4d}} \approx i \frac{2}{\pi} \log \left(\frac{4 \sinh^2 (2\pi a R_5)}{(2\pi R_5 \Lambda_{\textrm{QCD}})^2} \right) .
\end{equation}
where we have identified $ \Lambda_{\textrm{UV}} \sim 1/2\pi R_5$. In particular, if $\textrm{Re}(a) \ll 1/2\pi R_5$ such that the 4d approximation can be trusted, \eqref{F0 4d-coupling-2} indicates:
\begin{equation}
	\frac{1}{g^2_{4,\textrm{eff}}} \approx  \frac{1}{4\pi^2} \log \left(\frac{(2a)^2}{\Lambda^2_{\textrm{QCD}}}\right),
\end{equation}
which is the correct 4d coupling implied by $\beta$-function, and the hyperbolic $\sinh$ in \eqref{F0 4d-coupling-2} can be seen as the corrections of KK-tower in 4d effective theory \cite{Katz:1996fh, Lawrence:1997jr}.

For completeness, one can check that the instanton sum in \eqref{F0 prepotential} is matched with the 4d Seiberg-Witten result following \cite{Katz:1996fh}. For $n=0$, only $N_{1,0}$ is non-zero and the summation is already given by \eqref{F0 4d-coupling-2}. For any fixed non-zero $n$, $N_{m,n}$ is generally not zero and the sum among infinite $m$ will be dominated by large $m$. A key observation in  \cite{Katz:1996fh} is when $m$ is large, the invariant $N_{m,n}$ for $F0$ can be split as:
\begin{equation}
	N_{m,n} \sim \gamma_n m^{4n-3},\quad (m\gg 1)
\end{equation}
where $\gamma_n$ is a negative factor independent of $m$. Using the approximation one can sum over the instanton contributions explicitly and 4d effective coupling $\tau_{\textrm{4d}} \equiv i R_5 \frac{\partial^2 \mathcal{F}_{\textrm{5d},\mathbb{S}^1}}{\partial a^2}$ is:
\begin{equation}\label{F0-positive-fullcoupling}
	\tau_{\textrm{4d}} \approx \frac{4 i}{\pi } \log \left(\frac{2 a }{\Lambda_{\textrm{QCD}}}\right) + \sum_{n\geq 1} i \gamma_n \frac{2 \Gamma(4n)}{\pi} \left(\frac{\Lambda_{\textrm{QCD}}}{2 a}\right)^{4n},
\end{equation}
where $\Lambda_{\textrm{QCD}}$ is the one defined in \eqref{Lambda QCD}. The result is matched with the $N=2$ 4d gauge coupling expansion derived via the Seiberg-Witten curve.

\begin{figure}[pbth]
\centering
\includegraphics[scale=0.5]{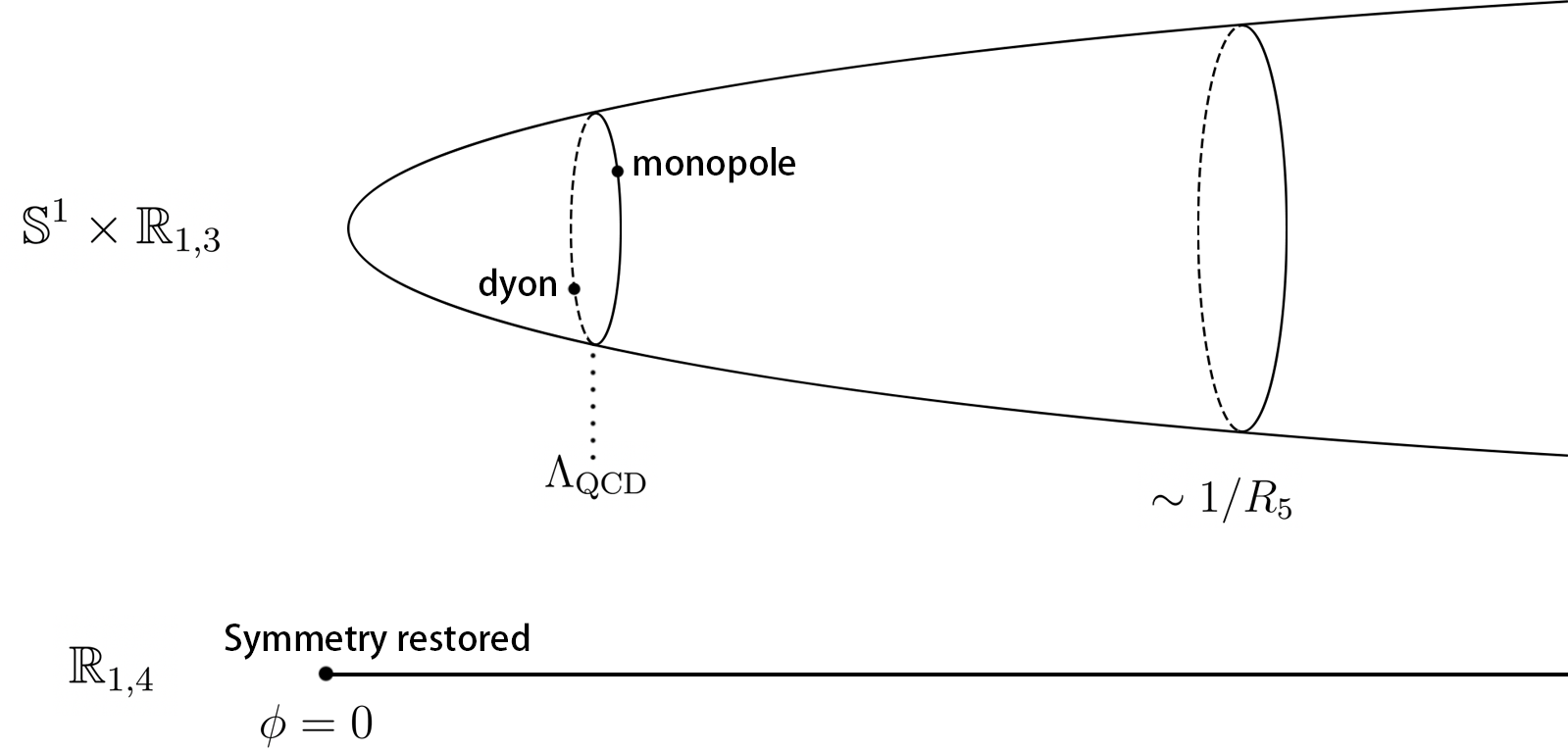}
\caption{Moduli space for F0-theory with positive coupling-squared. In the large $R_5$ limit, entire strong coupling region is pushed
to the far left of $1/R_5$, which itself is pushed toward $\phi=0$. No trace of the marginal stability wall is left behind. }
\label{Fig-F0-cigar}
\end{figure}

The moduli space for F0 theory with positive bare coupling-squared is depicted as figure \ref{Fig-F0-cigar}. There is a strongly coupled region $\phi \sim \Lambda_{\textrm{QCD}}$ near the tip of the cigar, and the theory is effectively the strongly coupled pure Sp(1) Seiberg-Witten theory. Around $\Lambda_{\textrm{QCD}}$ there are two singularities where the mass of $(0,1)$ monopole and $(2,-1)$ dyon will separately become zero, and there is a marginal stability wall connecting these two singularities as shown in the figure \ref{Fig-F0-cigar}.

Now let's see what happens in the decompactification limit $R_5\rightarrow \infty$. Since the instanton contribution vanishes in this limit, we can identify $\textrm{Re}(a) = \phi$. The period of $a$ is $a \sim a+i/2R_5$, therefore as $R_5\rightarrow \infty$ the imaginary part of $a$ is zero such that $a=\phi$ in this limit, and the cylinder will become the half-line in figure \ref{Fig-F0-cigar}. From \eqref{Lambda QCD} we can see as $R_5 \rightarrow \infty$, the $\Lambda_{\textrm{QCD}}$ becomes much smaller than $\Lambda_{\textrm{UV}}$, therefore in  $R_5\rightarrow \infty$ limit one has:
\begin{equation} \label{F0-RLimit}
	\Lambda_{\textrm{QCD}} \ll \Lambda_{\textrm{UV}}\sim \frac{1}{2\pi  R_5} \rightarrow 0,
\end{equation}
such that the two singularities and the marginal stability wall both shrink to the endpoint $\phi = 0$ in the 5d moduli space. At that point both W-bosons and monopoles become massless, therefore the Sp(1) gauge symmetry is restored.

We have discussed the first kind of marginal stability wall where the W-boson decays into a pair of $(0,1)$ monopole and $(2,-1)$ dyon and seen how the wall collapses to the endpoint of the 5d moduli space. This wall is indeed the only marginal stability wall for 4d pure Sp(1) Seiberg-Witten theory. However, that is not true for the compactified 5d version, where there are three {\rm U}(1) charges: electric charge $q_e$, instanton charge $q_i$ and KK-charge $k$ (or D0-charge in IIA picture). For any BPS particle with a central charge $Z(q_e,q_i,k)$, a marginal stability wall of the first kind may exist where it can decay into a pair of magnetic particles $Z_i$ and $Z_j$. Based on our previous analysis of the W-boson marginal stability wall, we will now argue that all these walls will collapse to the endpoint $\phi=0$ in the decompactification limit.

The central charges correspond to the three {\rm U}(1) charges are given by:
\begin{equation}
Z_{\textrm{elec}} = 2a,\quad Z_{\textrm{Inst}} = \mu_0, \quad Z_{\textrm{KK}} = \frac{i}{R_5},
\end{equation}
while the central charge for the monopole is:
\begin{equation}
	Z_{\textrm{mag}} = i R_5 \frac{\partial \mathcal{F}_{\textrm{5d},\mathbb{S}^1}}{\partial a},
\end{equation}
which is of order $\sim \Lambda_{\textrm{QCD}}$ in the core region or behaves as:
\begin{equation}\label{F0-magnetic-central-charge}
	Z_{\textrm{mag}}\sim i R_5 2a(2a+\mu_0),
\end{equation}
in the asymptotic region $\textrm{Re}(a) > 1/2\pi R_5$.

The first thing we learn  is that the marginal stability wall of the first kind cannot extend into the generic point of the Coulomb phase with nonzero $a(a+\mu_0)$, since there the magnetic mass would scale linearly with $R_5$. The wall, if any, must remain in the region where  $Z_{\textrm{elec}},Z_{\textrm{Inst}},Z_{\textrm{KK}}$ are comparable with $Z_{\textrm{mag}}$, end where the magnetic particles $Z_i$ or $Z_j$ become massless. $Z_{\textrm{elec}} \sim Z_{\textrm{mag}}$ requires  $a\sim \Lambda_{\textrm{QCD}}$ while $Z_{\textrm{Inst}} \sim Z_{\textrm{mag}}$ can happen only if $\textrm{Re}(a) \sim 1/2\pi R_5$. $Z_{\textrm{KK}}$ scales down as $1/R_5$ also. Therefore, no walls of the first kind can extend beyond $a\sim 1/2\pi R_5$, which in turn collapse to $\phi=0$ in  the decompactification limit.

\subsubsection*{Negative bare coupling-squared}

The brane web for negative bare coupling-squared ($\mu_0<0$) is given in figure 3; although the diagram suggests a 90-degree flip, hence an S-duality in IIB sense, the reality is simpler. The 5d gauge sector does not arise from D5 segments alone but a combination of D5 and NS5 segments, as can be seen from how the effective coupling is determined by the total length of the internal rectangle. In the 5d limit, the moduli space is still a half-line which ends at a point, but the endpoint is $\phi = |\mu_0|/2$ instead of $\phi = 0$, where a dyonic instanton becomes massless and the monopole string become tensionless. The dyonic instanton carries the same electric charge as a W-boson, and it restores the Sp(1) gauge symmetry by becoming massless at the boundary.

When we compactify the theory on a large circle,  the moduli space is again a cigar and we will analyse the three different regimes: $\textrm{Re}(a) - |\mu_0|/2 \gg 1/2\pi R_5$, $\textrm{Re}(a) - |\mu_0|/2 \sim 1/2\pi R_5$ and $|a - |\mu_0|/2| \ll 1/2\pi R_5$. Let us see how this explicitly by analyzing $F_{\rm exact}$. The regime $\textrm{Re}(a) - |\mu_0|/2 \gg 1/2\pi R_5$ is identical to the previous case, since the same old IMS prepotential governs the regime. The monopole string tension and the 5d effective couplings are, therefore,
\begin{equation}
	 i T_{\textrm{mono}} = \frac{i}{2\pi} \frac{\partial \mathcal{F}_{\textrm{5d},\mathbb{S}^1}}{\partial a} = \frac{i a (2a - |\mu_0|)}{\pi}, \quad	\tau_{\textrm{5d}} = \frac{i}{2\pi} \frac{\partial^2 \mathcal{F}_{\textrm{5d},\mathbb{S}^1}}{\partial a^2} = \frac{i (4 (a - |\mu_0|/2)+|\mu_0|)}{\pi} \ .
\end{equation}
although we wrote the latter differently; why we do so should become obvious below.

We begin to see a real difference at $\textrm{Re}(a) - |\mu_0|/2 \sim 1/2\pi R_5$. Here with negative $\mu_0$,
we  instead set $m=0$ in \eqref{F0 prepotential} and find
\begin{align}
	\frac{\partial^3 \mathcal{F}_{\textrm{5d},\mathbb{S}^1}}{\partial a^3} &\approx  8 + \sum_{n\geq 0} N_{0,n} \left( \frac{e^{-2\pi n (2a - |\mu_0|)R_5}}{1 - e^{-2\pi n (2a - |\mu_0| )R_5}} \right)  (-2n)^3 \nonumber \\
	&= 8 + 16 \left( \frac{e^{-2\pi n (2a - |\mu_0|)R_5}}{1 - e^{-2\pi n (2a - |\mu_0|)R_5}} \right),
\end{align}
where in the second line we use the fact that the only non-zero $N_{0,n}$ is again $N_{0,1} = -2$ \cite{Chiang:1999tz}.
The effective coupling is:
\begin{align}\label{F0 4d-coupling-negative}
	\tau_{\textrm{5d}} &= \frac{i}{2\pi} \frac{\partial^2 \mathcal{F}_{\textrm{5d},\mathbb{S}^1}}{\partial a^2} \approx -\frac{i|\mu_0|}{\pi} + \frac{4 i}{\pi}  \left(a  + \frac{1}{2\pi R_5 } \log \left(1-e^{-2\pi (2a - |\mu_0| )R_5} \right) \right) \nonumber \\
	&= \frac{i |\mu_0|}{\pi} + \frac{i}{\pi^2 R_5} \log 4 \sinh^2(\pi(2a-|\mu_0|) R_5).
\end{align}
Notice that the negative bare coupling-squared is corrected by the instanton contribution to positive. Note how the result is identical to \eqref{F0 4d-coupling}  but with $a$ replaced by $a - |\mu_0|/2$; the latter of course tells us that the dyonic instanton will play the role of the W-boson here.

The 4d effective coupling is, with $\Lambda_{\textrm{QCD}}$ defined parallel to \eqref{Lambda QCD},
\begin{equation}
	\tau_{\textrm{4d}} \approx i \frac{2}{\pi} \log \left(\frac{4 \sinh^2 (\pi(2a- |\mu_0|)R_5)}{(2\pi R_5 \Lambda_{\textrm{QCD}})^2} \right),
\end{equation}
which is the same as \eqref{F0 4d-coupling-2} by replacing $a$ to $a - |\mu_0|/2$. Therefore the analysis of the moduli space is parallel to the positive bare coupling-squared case, the cylindrical moduli space will begin to close at $\textrm{Re}(a) - |\mu_0|/2 \sim 1/2\pi R_5$ and eventually form a cigar.

Finally, as $|a- |\mu_0|/2| \ll 1/2\pi R_5$ the effective 4d theory is strongly coupled, and since the coupling \eqref{F0 4d-coupling-negative} is the same as the coupling \eqref{F0 4d-coupling} with $a$ replaced by $a - |\mu_0|/2$, it is natural to expect the theory is still an Sp(1) Seiberg-Witten theory. Indeed, the instanton sum in the asymptotic region gives the Seiberg-Witten prepotential following \cite{Katz:1996fh},  and the 4d effective coupling is:
\begin{equation}\label{F0-negative-fullcoupling}
	\tau_{\textrm{4d}} \approx \frac{4 i}{\pi } \log \left(\frac{2 (a- |\mu_0|/2) }{\Lambda_{\textrm{QCD}}}\right) + \sum_{m\geq 1} i \gamma_m \frac{2 \Gamma(4m)}{\pi} \left(\frac{\Lambda_{\textrm{QCD}}}{2 (a- |\mu_0|/2)}\right)^{4m},
\end{equation}
with $a$ replaced by $a - |\mu_0|/2$ compared to \eqref{F0-positive-fullcoupling}.

\begin{figure}[pbth]
\centering
\includegraphics[scale=0.5]{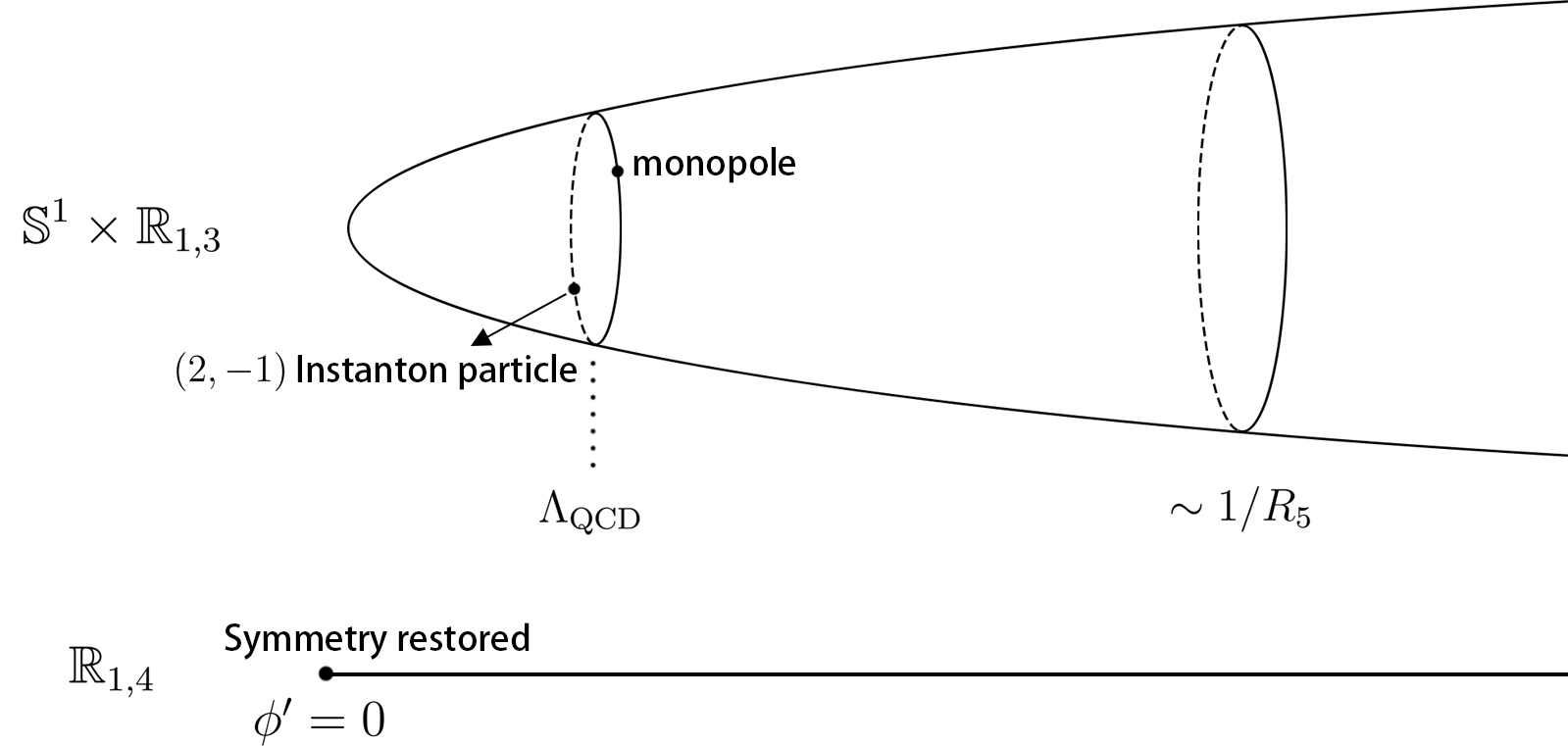}
\caption{Moduli space for F0-theory with negative coupling-squared.}
\label{Fig-F0-cigar-negative}
\end{figure}

For the ordinary Seiberg-Witten theory, the central charge of the massless state in the strongly coupled region is $a_D$ and $2 a - a_D$, which are (0,1) monopole and (2,-1) dyon. Now with $a$ replaced by $a - |\mu_0|/2$ in this case, the massless state should have central charge $a_D$ and $2(a - |\mu_0|/2) - a_D$. The first one is still the monopole, while the second one is a bound state with $+2$ electric charge, $-1$ magnetic charge and $+1$  (since $-|\mu_0| = + \mu_0$) instanton charge. The moduli space in this case is drawn as figure \ref{Fig-F0-cigar-negative}, where we use the parameters $\phi' \equiv \phi - |\mu_0|/2$ and $a' \equiv a - |\mu_0|/2$ instead of $\phi$ and $a$.

Figure \ref{Fig-F0-cigar} is equally applicable here, except for two differences. Firstly, the horizontal position is labeled not by $\phi$ but by $\phi'$ and, secondly, the dyon singularity is replaced by the singularity of the instanton particle carrying $+2$ electric charge and $-1$ magnetic charge. Again, in  $R_5\rightarrow \infty$ limit, the imaginary part of $a$ collapses, the cylinder becomes a half-line, and the two singularities and the marginal stability wall merge into the boundary in the 5d moduli space where an Sp(1) gauge symmetry would be restored. The dyonic instanton, rather than W-boson, becomes massless and restores the gauge symmetry. Although this involves a coherent state becoming massless, the phenomenon has little to do with the strong-weak duality of 4d. Other than this shift of the coordinate and the replacement of W-boson by the dyonic instanton, the physics of the Coulomb phase remains the same as the previous case.

\subsection{F1 Theory}

\begin{figure}[pbth]
\centering
\includegraphics[scale=0.25]{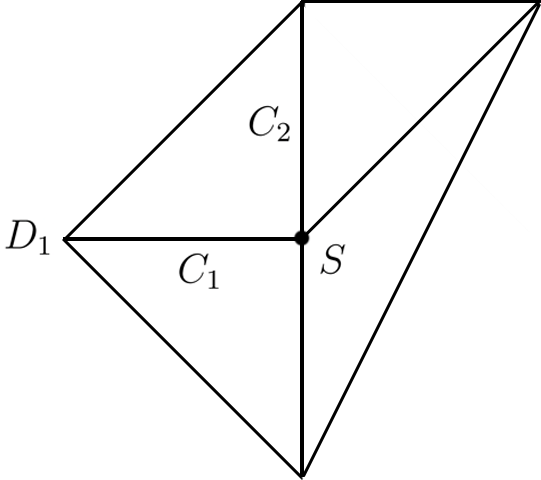}
\caption{The toric diagram for $F1$ theory.}
\label{Fig-F1-toric}
\end{figure}

Now we move to the F1 theory, whose toric diagram is given by figure \ref{Fig-F1-toric}. There is only one compact divisor $S$, whose self-triple intersection number is 8, and one independent non-compact divisor $D_1$. There are two compact 2-sphere $C_1$ and $C_2$, and their intersection number with $S$ are separately -2 and -1. We summarize the geometric data as \cite{Closset:2018bjz}:

\begin{itemize}
	\item Divisors:
		\begin{equation}
	S\cdot S \cdot S = 8,
		\end{equation}
		\begin{equation}
	S\cdot S\cdot D_1 = -2,
		\end{equation}
		\begin{equation}
	S\cdot D_1 \cdot D_1 = 0
		\end{equation}
	\item 2-cycles:
		\begin{equation}
			C_1 \cdot S = -2,\quad C_2 \cdot S = -1.
		\end{equation}
		\begin{equation}
			C_1 \cdot D_1 = 0,\quad C_2 \cdot D_1 = 1.
		\end{equation}

\end{itemize}
The K$\ddot{\textrm{a}}$hler form $[J]$ is then parameterized by:
\begin{equation}\label{F1-Kahler form}
	[J] =  \mu_0 [D_1] -\phi [S],
\end{equation}
and the volume of two cycles $C_1$ and $C_2$ are given as:
\begin{equation}
	A_1 =M_W = J \cdot C_1 = 2\phi ,\quad A_2 = M_I = J \cdot C_2 = \mu_0 + \phi,
\end{equation}
which are the masses of the W-boson and dyonic instanton. The prepotential \eqref{5d-prepotential-toric} is still:
\begin{equation}
	\mathcal{F}_{\textrm{IMS}} (\phi,\mu_0) = -\frac{1}{6} J \cdot J \cdot J = \frac{4}{3} \phi^3 + \mu_0 \phi^2,
\end{equation}
where we have omitted the constant part $\sim D_1 \cdot D_1 \cdot D_1$.

After compactification, the third derivative of the prepotential is given by \eqref{Instanton prepotential} as:
\begin{equation}
	\frac{\partial^3 F_{\rm exact}(t,v)}{\partial t^3} = 8 + \sum_{m,n\geq 0} N_{m,n} \left( \frac{e^{-4\pi m i t}e^{2\pi n i (-t + v) }}{1 - e^{-4\pi m i t}e^{2\pi n i (-t + v) }}  \right)  (-2m -n)^3.
\end{equation}
Proceeding as in F0 case, we redefine the flat coordinate $t$ as $a\equiv i t/R_5$ and rescale
$F_{\rm exact}(t,v)$ into
\begin{equation}\label{F1 prepotential}
		\frac{\partial^3 \mathcal{F}_{\textrm{5d},\mathbb{S}^1}(a,\mu_0)}{\partial a^3} = 8 + \sum_{m,n\geq 0} N_{m,n} \left( \frac{e^{-4\pi m a R_5}e^{-2\pi n (a + \mu_0)R_5 }}{1 - e^{-4\pi m a R_5}e^{-2\pi n (a + \mu_0 )R_5 }} \right)  (-2m - n)^3,
\end{equation}
In the large radius limit $R_5\rightarrow \infty$, this prepotential collapses to the usual IMS
prepotential, precisely for the same reason as in the F0 theory.

\subsubsection*{Positive bare coupling-squared}

For positive $\mu_0$, the discussions by and large parallel those of the F0 theory. The regime $\textrm{Re} (a) \gg 1/(2\pi R_5)$ is identical to the F0 case and the monopole string tension and the effective couplings are still given by \eqref{F0-tension-and-coupling-large-a} and \eqref{F0-4d-coupling-large-a}.
If $\textrm{Re}(a) \sim 1/2\pi R_5$ then the instanton contribution can not be omitted. Since $\mu_0 R_5 \gg 1$ for large $R_5$ it is sufficient to set $n=0$ in \eqref{F1 prepotential}, but we will also keep the next leading term with $n=1$ in order to distinguish the two different copies discussed in section 5.

Keeping the terms with $n=0,1$ in \eqref{F0-4d-coupling-large-a} gives:
\begin{align}
	\frac{\partial^3 \mathcal{F}_{\textrm{5d},\mathbb{S}^1}}{\partial a^3} \approx  8 &+ \sum_{m\geq 0} N_{m,0} \left( \frac{e^{-4\pi m a R_5}}{1 - e^{-4\pi m a R_5}} \right)  (-2m)^3 \nonumber \\ &+\sum_{m\geq 0} N_{m,1} \left(\frac{e^{-4\pi m a R_5} e^{-2\pi (a+\mu_0) R_5}}{1-e^{-4\pi m a R_5} e^{-2\pi (a+\mu_0) R_5}}\right),
\end{align}
where the only non-zero $N_{m,0}$ for F1 theory is still $N_{1,0} = -2$, and $N_{m,1} = 2m+1$ \cite{Chiang:1999tz}. The 5d effective coupling is:
\begin{align}
	\tau_{\textrm{5d}} \approx & \frac{i \mu_0}{\pi} + \frac{i}{\pi^2 R_5} \log 4 \sinh^2(2\pi R_5 a) - \sum_{m=0}^{\infty} \frac{i (2m+1)^3}{4\pi^2 R_5} \log \left(1-e^{-2\pi(2m+1) a R_5}e^{-2\pi \mu_0 R_5} \right), \nonumber \\
	=& \frac{i \mu_0}{\pi} + \frac{i}{\pi^2 R_5} \log 4 \sinh^2(2\pi R_5 a) \nonumber \\
	&+ \sum_{k=1}^{\infty} \frac{i}{8\pi^2 k R_5} e^{-2\pi k R_5 \mu_0} \left(\textrm{Li}_{-3}(e^{-2\pi k R_5 a}) - \textrm{Li}_{-3}(-e^{-2\pi k R_5 a}) \right),
\end{align}
where we have expanded the second logarithm in the first line and summed over $m$. The 4d effective coupling $\tau_{\textrm{4d}} = 2\pi R_5 \tau_{\textrm{5d}}$ is then:
\begin{align}
	\tau_{\textrm{4d}}  & = i \frac{2}{\pi} \log \left(\frac{4 \sinh^2 (2\pi a R_5)}{(2\pi R_5 \Lambda_{\textrm{QCD}})^2} \right) \nonumber \\ &+ \sum_{k=1}^{\infty} \frac{i}{4\pi k} (2\pi R_5 \Lambda_{\textrm{QCD}})^{4k} \left(\textrm{Li}_{-3}(e^{-2\pi k R_5 a}) - \textrm{Li}_{-3}(-e^{-2\pi k R_5 a}) \right) ,
\end{align}
where $\mu_0$ is replaced by $\Lambda_{\textrm{QCD}}$ via $e^{-2\pi R_5 \mu_0} = (2\pi R_5 \Lambda_{\textrm{QCD}})^4$, see the discussion in the previous F0 case.

If $\textrm{Re}(a) \gg 1/2\pi R_5$, the 	polylogarithm $\textrm{Li}_{-3}(\pm e^{-2\pi k R_5 a})$ are vanishing small. When $\textrm{Re}(a) \sim 1/2\pi R_5$, depending on which holonomy saddle we are located at, the instanton correction will be different. For example, let's consider the 4d region where the energy is much smaller that $1/2\pi R_5$ but still larger than $\Lambda_{\textrm{QCD}}$, at the trivial holonomy saddle $h_* = 0$, the first polylogrithm $\textrm{Li}_{-3}(e^{-2\pi k R_5 a})$ dominates and is approximated as $\frac{\Gamma(4)}{(2\pi R_5 k a)^{4}}$.

This gives the full 4d expression in the intermediate $a$ region, say $\Lambda_{\rm QCD}\ll |a| \ll 1/R_5$.
Following the discussion in \cite{Katz:1996fh}, we assume the growing of the $N_{m,n}$ for F1 is similar to such that $N_{m,n}$ can be split as:
\begin{equation}
	N_{m,n} \sim (-1)^n \gamma_n m^{4n-3},\quad (m \gg 1)
\end{equation}
where $\gamma_n$ is the same as that in F0 case and there is an additional sign factor depending on $n$. Following the same procedure, the 4d effective coupling with the full instanton sum is approximated as:
\begin{equation}\label{F1-positive-fullcoupling-1}
	h_*=0:\quad	\tau_{\textrm{4d}} \approx \frac{4 i}{\pi } \log \left(\frac{2 a }{\Lambda_{\textrm{QCD}}}\right) + \sum_{n\geq 1} i (-1)^n \gamma_n \frac{2 \Gamma(4n)}{\pi} \left(\frac{\Lambda_{\textrm{QCD}}}{2 a}\right)^{4n},
\end{equation}
for the trivial holonomy saddle, again reproducing the 4d instanton sum as in F0 theory, and,
\begin{equation}\label{F1-positive-fullcoupling-2}
	h_*=\frac{1}{2}:\quad	\tau_{\textrm{4d}} \approx \frac{4 i}{\pi } \log \left(\frac{2 \tilde{a} }{\Lambda_{\textrm{QCD}}}\right) + \sum_{n\geq 1} i  \gamma_n \frac{2 \Gamma(4n)}{\pi} \left(\frac{\Lambda_{\textrm{QCD}}}{2 \tilde{a}}\right)^{4n} + 4,
\end{equation}
for the holonomy saddle $h_* = 1/2$.

\begin{figure}[pbth]
\centering
\includegraphics[scale=0.3]{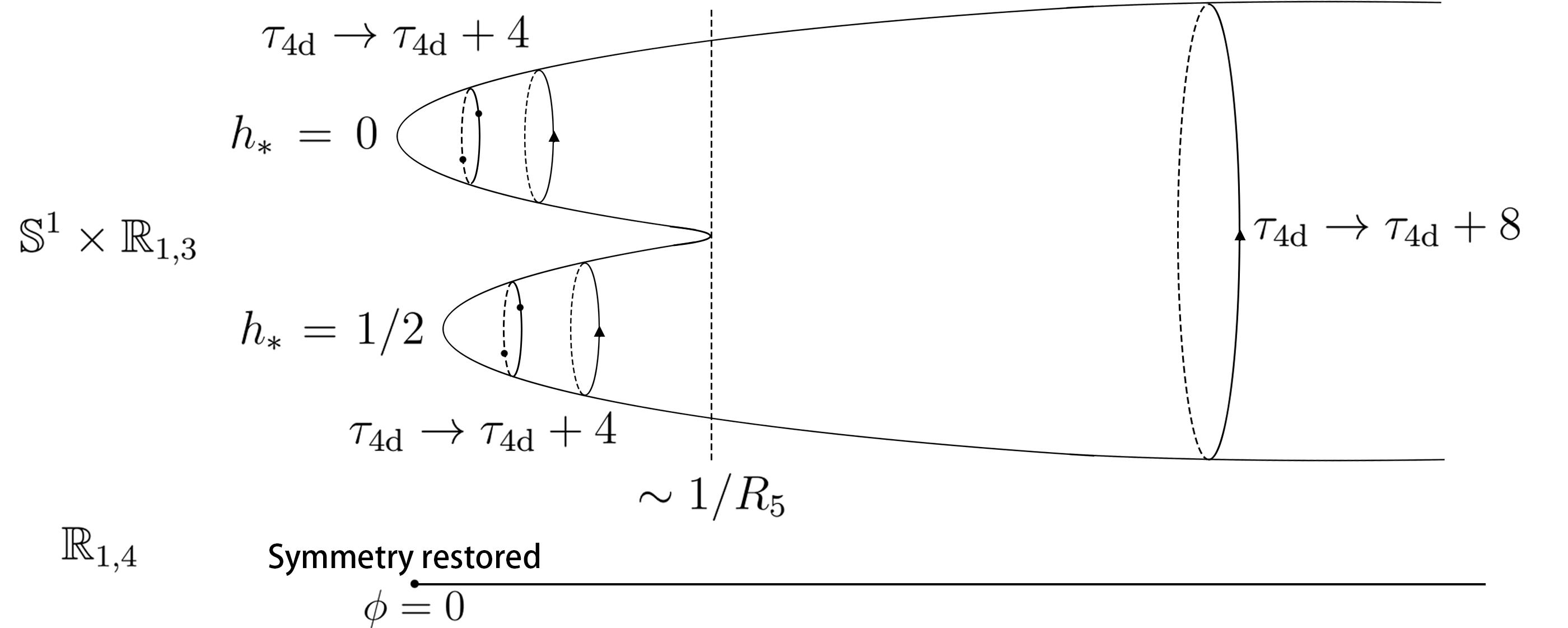}
\caption{Moduli space for F1-theory with positive coupling-squared. A pair of 4d Seiberg-Witten geometries,
or holonomy saddles, reside to the left of the scale $\sim 1/R_5$ and collapse to a point in the decompactification limit. }
\label{F1-cigar-positive}
\end{figure}

Note that the relative sign factor $(-1)^n$ between  $h_*=0$ and $h_*=1/2$ saddles can be compensate by shifting $\tilde a$ via $\tilde a\rightarrow \tilde a \exp(\pi i /4)$. Also the additive difference $+4$ is merely a $4\pi$ shift of the 4d theta angle and has no separated physical significance. This equivalence is as it should be since in the 4d limit, there is only one pure Sp(1) theory. On the other hand, the two saddles are
still very much distinct in the 5d theory context, since how we glue these 4d $a$ near $h_*=0$ or its cousin $\tilde a \exp(\pi i /4)$ near $h_*=1/2$ onto 5d $a$ with its periodic imaginary part are clearly very different. For this gluing the additive difference $+4$ is also significant.



This doubled period along the imaginary part of $a$ again confirms that the strong coupling end of the Coulombic moduli space is capped by two copies of pure Sp(1) 4d Seiberg-Witten geometry, as was delineated in Section 4. Although we are discussing here large $R_5$ limit primarily, the monodromy is of topological nature and interpolates intact between small and large $R_5$ limits.

\subsubsection*{Negative bare coupling-squared}

For negative $\mu_0$, the 5d moduli space is again a half-line, now divided into two parts $\phi > |\mu_0|$ and $\phi < |\mu_0|$. After compactification, one may still expect the moduli space to be a cigar, but the structure is different from the previous cases. First, the point $\phi = |\mu_0|$ will lift to a singularity in the moduli space where the mass of the $(1,0)$ dyonic instanton become zero; second, since the 5d endpoint corresponds to a $E_0$ theory rather than the symmetry restored Sp(1) gauge theory, after compactification the singularities and the monodromy on the Coulomb moduli space are also different. In fact, there are three singularities related by a $\mathbf{Z}_3$ symmetry \cite{Morrison:1996xf}, and they shrink to the endpoint of the 5d moduli space in the decompactification limit.

First let's consider $\textrm{Re}(a) > |\mu_0|$, and the prepotential \eqref{F1 prepotential} is:
\begin{equation}\label{F1 prepotential negative}
		\frac{\partial^3 \mathcal{F}_{\textrm{5d},\mathbb{S}^1}}{\partial a^3} = 8 + \sum_{m,n\geq 0} N_{m,n} \left( \frac{e^{-4\pi m a R_5}e^{-2\pi n (a - |\mu_0|)R_5 }}{1 - e^{-4\pi m a R_5}e^{-2\pi n (a - |\mu_0|)R_5 }}  \right)  (-2m - n)^3.
\end{equation}
The regime $\textrm{Re}(a) \gg |\mu_0|$ is still identical to the previous case and
the monopole string tension and 5d effective coupling are similarly given by:
\begin{equation}
	i T_{\textrm{mono}} = \frac{i}{2\pi} \frac{\partial \mathcal{F}_{\textrm{5d},\mathbb{S}^1}}{\partial a} = \frac{i a (2a - |\mu_0|)}{\pi},\quad	\tau_{\textrm{5d}} = \frac{i}{2\pi} \frac{\partial^2 \mathcal{F}_{\textrm{5d},\mathbb{S}^1}}{\partial a^2} = \frac{i (4 a - |\mu_0|)}{\pi}.
\end{equation}
The geometry of the moduli space is a cylinder and the period of $a$ is $a \sim a + i/R_5$ which can be read from \eqref{F1 prepotential negative}. The 4d effective coupling is:
\begin{equation}
	\tau_{\textrm{4d}} = 2\pi R_5 \tau_{\textrm{5d}} = -2 |\mu_0| R_5 i + 8 a R_5 i,
\end{equation}
which gives the monodromy $\tau_{\textrm{4d}} \rightarrow \tau_{\textrm{4d}} + 8$.

As $\textrm{Re}(a)$ approach $|\mu_0|$ the instanton contributions with non-zero $m$ in \eqref{F1 prepotential negative} are still vanishing small since we assume $R_5$ is large, but we can not omit those with $n\neq 0$ in \eqref{F1 prepotential negative}. Setting $m=0$ we have:
\begin{equation}\label{F1-prepotential-negative-original}
	\frac{\partial^3 \mathcal{F}_{\textrm{5d},\mathbb{S}^1}}{\partial a^3} \approx 8  + \sum_{n\geq 0} N_{0,n} \left( \frac{e^{-2\pi n (a - |\mu_0|)R_5 }}{1 - e^{-2\pi n (a - |\mu_0|)R_5 }} \right)  (-n)^3.
\end{equation}
The only non-zero $N_{0,n}$ for F1 theory is $N_{0,1} = 1$ \cite{Chiang:1999tz}, therefore we have:
\begin{equation}\label{F1-prepotential-negative-leading}
	\frac{\partial^3 \mathcal{F}_{\textrm{5d},\mathbb{S}^1}}{\partial a^3} = 8 -  \left( \frac{e^{-2\pi (a - |\mu_0|)R_5 }}{1 - e^{-2\pi (a - |\mu_0|)R_5 }} \right) ,
\end{equation}
and the 5d effective coupling is evaluated as:
\begin{equation}
\tau_{\textrm{5d}} = \frac{i}{2\pi} \frac{\partial^2 \mathcal{F}_{\textrm{5d},\mathbb{S}^1}}{\partial a^2} = \frac{17 }{4\pi} a i   - \frac{5 }{4\pi} |\mu_0| i - \frac{i}{8\pi^2 R_5} \log 4 \sinh^2 \pi (a-|\mu_0|)R_5,
\end{equation}
or in terms of the 4d effective coupling $\tau_{\textrm{4d}} = 2\pi R_5 \tau_{\textrm{5d}}$:
\begin{equation}\label{F1 4d-coupling-negative}
\tau_{\textrm{4d}} \approx \frac{17}{2} a R_5 i   - \frac{5}{2} |\mu_0| R_5 i + \frac{1}{4\pi i} \log 4 \sinh^2 \pi (a-|\mu_0|)R_5.
\end{equation}
The coupling suggests a singularity at $\textrm{Re}(a) = |\mu_0|$, where the mass of the dyonic instanton soliton in F1 theory vanishes.
Expand $a$ as $a - |\mu_0| \equiv \Delta a$ near the singularity, one has:
\begin{equation}
\tau_{\textrm{4d}} \approx \frac{17}{2}\Delta a R_5 i + 6 |\mu_0| R_5 i + \frac{1}{2\pi i} \log 2\pi \Delta a ,
\end{equation}
and as $\Delta a \rightarrow 0$, the 4d effective coupling will become vanishing small. Notice that the monodromy of the logarithm singularity is $\tau \rightarrow \tau + 1$ as $\Delta a \rightarrow e^{2\pi i} \Delta a$.

\begin{figure}[pbth]
\includegraphics[scale=1]{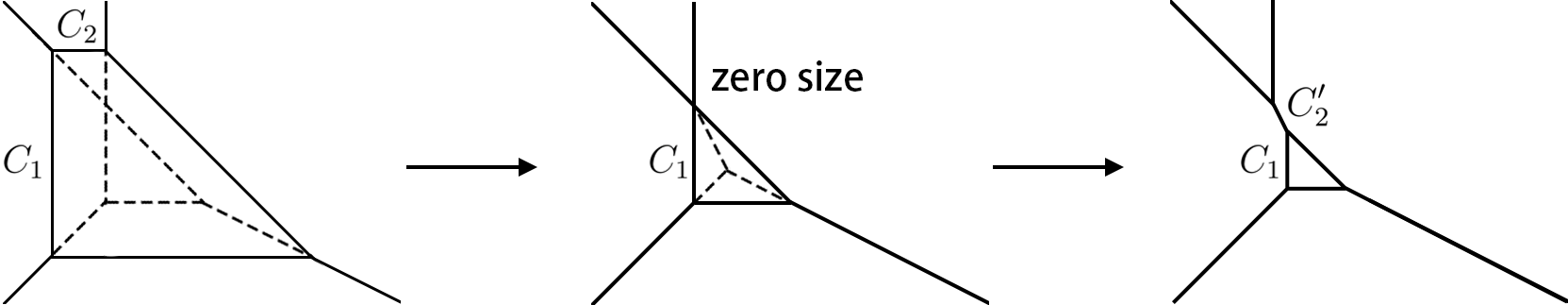}
\caption{The flop transition of F1 theory in terms of brane web.}
\label{Fig-F1-flop-transition-brane}
\end{figure}

Now let's consider what happens if we cross the circle $\textrm{Re}(a) = |\mu_0|$ and reach the region $\textrm{Re}(a) < |\mu_0|$. From the 5d brane web diagram (see figure \ref{Fig-F1-flop-transition-brane}), the top D5-brane $C_2$ will shrink to zero as $\phi$ approach $|\mu_0|$ and the 5d hypermultiplet given by the F1-string lying on $C_2$ will become massless. As $\phi$ becomes smaller than $|\mu_0|$, the zero size 'D5-brane' will blow up as another slanted 5-brane $C'_2$ and the hypermultiplet will be the instanton string lying on $C'_2$ \cite{Aharony:1997bh}. In terms of toric geometry, the 2-cycle $C_2$ shrinks to zero size as $\phi$ approach $|\mu_0|$ and the membrane wrapped on $C_2$ is massless. As $\phi$ becomes smaller than $|\mu_0|$, the $C_2$ will blow up in a geometrically distinct way, and we denote the new cycle as $C'_2$. After compactification, the singularity $\phi = |\mu_0|$ is lifted to $\textrm{Re}(a) = |\mu_0|$, and the membrane wrapped on the new cycle $C'_2$ will serve as the hypermultiplet which contributes to the singularity $\textrm{Re}(a) = |\mu_0|$ from the left side $\textrm{Re}(a) < |\mu_0|$ on the moduli space.

\begin{figure}[pbth]
\includegraphics[scale=1]{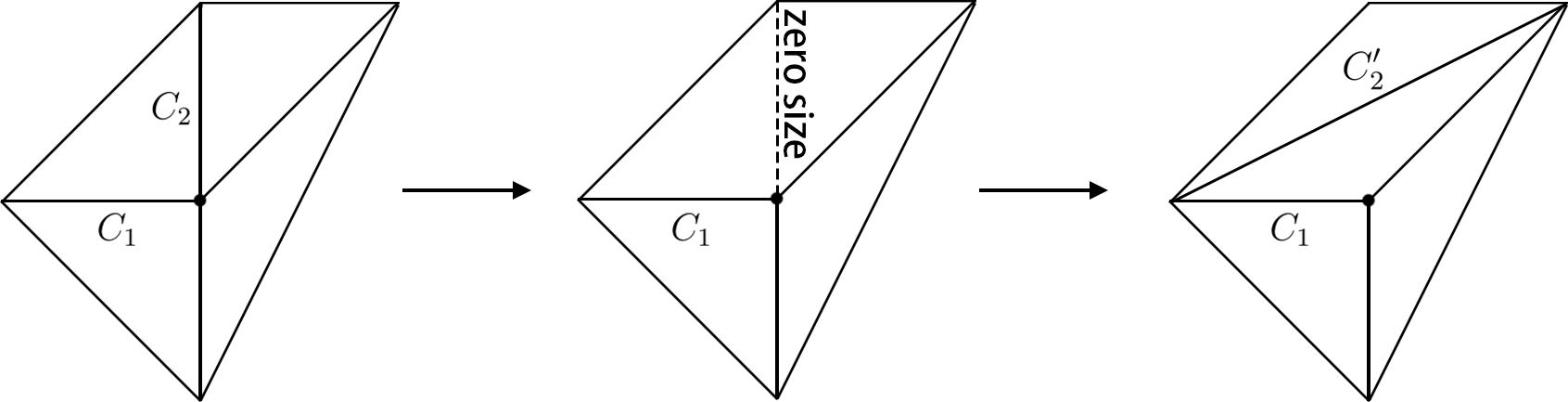}
\caption{The flop transition of F1 theory in terms of the toric diagram.}
\label{Fig-F1-flop-transition-toric}
\end{figure}

This is a flop transition in the Calabi-Yau 3-fold $\hat{X}$. The prepotential \eqref{F1-prepotential-negative-original} or \eqref{F1-prepotential-negative-leading} seem to be ill-defined when $\textrm{Re}(a) < |\mu_0|$, but it can be analytically continued to the region $\textrm{Re}(a) < |\mu_0|$ using the identity:
\begin{equation}
	\frac{e^{-2\pi n (a - |\mu_0|)R_5 }}{1 - e^{-2\pi n (a - |\mu_0|)R_5 }} = -1 - \frac{e^{2\pi n (a - |\mu_0|)R_5 }}{1 - e^{2\pi n (a - |\mu_0|)R_5 }},
\end{equation}
such that the prepotential \eqref{F1-prepotential-negative-original} is analytical continued as \cite{Katz:1996ht}:
\begin{align}\label{F1-prepotential-negative-flopped-original}
	\frac{\partial^3 \mathcal{F}_{\textrm{5d},\mathbb{S}^1}}{\partial a^3} &\approx \left( 8 - \sum_{n \geq 0} (-n)^3 N_{0,n} \right)  - \sum_{n\geq 0} N_{0,n} \left( \frac{e^{2\pi n (a - |\mu_0|)R_5 }}{1 - e^{2\pi n (a - |\mu_0|)R_5 }} \right)  (-n)^3 \nonumber \\
	&= 9 + \frac{e^{-2\pi n (|\mu_0| - a)R_5 }}{1 - e^{2\pi n (|\mu_0|-a) R_5 }},
\end{align}
where we have used $N_{0,1} = 1$ and others are zero. Notice that the monodromy becomes $\tau \rightarrow \tau + 9$.

\begin{figure}[pbth]
\centering
\includegraphics[scale=0.5]{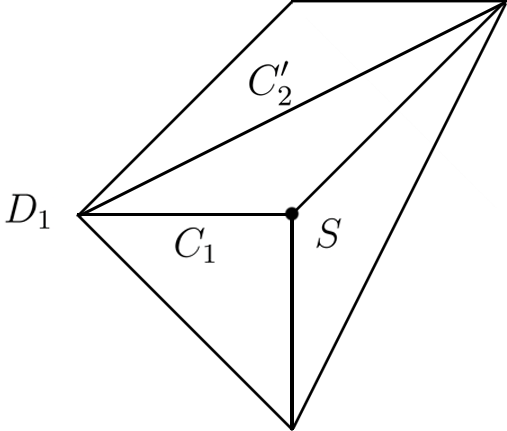}
\caption{The flopped toric diagram for F1 theory.}
\label{Fig-F1-flop-toric}
\end{figure}

\begin{figure}[pbth]
\centering
\includegraphics[scale=0.6]{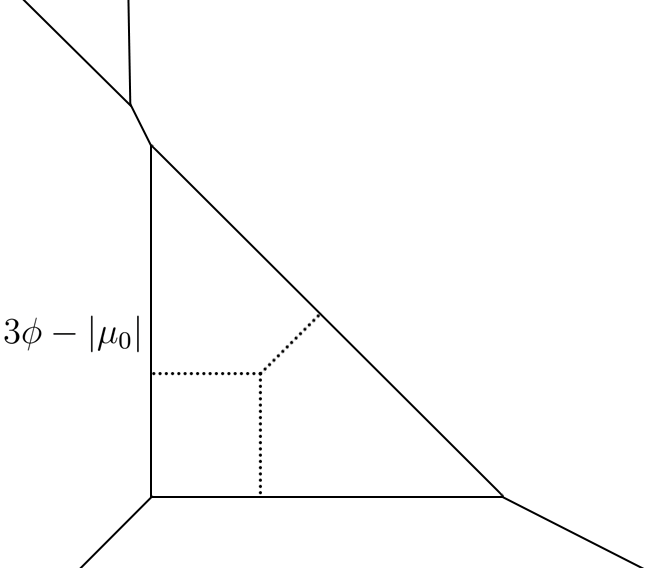}
\caption{The flopped F1 brane web and the junction string. Notice that the mass of the junction string also equals the length of the edge. }
\label{Fig-string-junction}
\end{figure}

Now let's analyse the region $\textrm{Re}(a) < |\mu_0|$ systemically from the flopped toric diagram, see figure \ref{Fig-F1-flop-toric}. The independent divisors are still compact divisor $S$ and non-compact divisor $D_1$, but their intersection data will change. We collect the useful geometric data as \cite{Closset:2018bjz}:
\begin{itemize}
	\item Divisors:
		\begin{equation}
	S\cdot S \cdot S = 9.
		\end{equation}
		\begin{equation}
	S\cdot S \cdot D_1 = -3.
		\end{equation}
		\begin{equation}
	S\cdot D_1 \cdot D_1 = 1.
		\end{equation}
	\item 2-cycles:
		\begin{equation}
			C_1 \cdot S = -3,\quad C'_2 \cdot S = 1.
		\end{equation}
		\begin{equation}
			C_1 \cdot D_1 =1,\quad  C'_2 \cdot D_1 = -1 .
		\end{equation}
\end{itemize}
The K$\ddot{\textrm{a}}$hler form is still given by \eqref{F1-Kahler form} which is:
\begin{equation}
	[J] = -|\mu_0| [D_1] - \phi [S],
\end{equation}
and the sizes of each cycle $C_1$ and $C'_2$ are:
\begin{equation}
	A_1 = J \cdot C_1 = 3\phi - |\mu_0|,\quad A'_2 = J \cdot C'_2 = |\mu_0| - \phi.
\end{equation}
The coefficient $'3'$ of $\phi$ in $A_1$ reflects that the M2-brane wrapped on the $C_1$ cycle gives rise to a BPS state with triple electric charges, which turns out to be the junction string state in the 5d brane web picture as shown in figure \ref{Fig-string-junction}.
The 5d prepotential for the flopped diagram is:
\begin{equation}
	\mathcal{F}_{\textrm{IMS}}(\phi,|\mu_0|) = - \frac{1}{6} J \cdot J \cdot J = \frac{3}{2} \phi^3 - \frac{3}{2} |\mu_0| \phi^2 + \frac{1}{2} |\mu_0|^2 \phi,
\end{equation}
where we have omitted the constant part given by $D_1 \cdot D_1 \cdot D_1$. The monopole string tension is:
\begin{equation}
	i T_{\textrm{mono}} = \frac{i}{2\pi} \frac{\partial \mathcal{F}_{\textrm{IMS}}(\phi,|\mu_0|)}{\partial \phi} = \frac{i}{4\pi} \left(3\phi - |\mu_0|\right)^2,
\end{equation}
which is proportional to the area of the triangle in figure \ref{Fig-string-junction} and the 5d gauge coupling is related to:
\begin{equation}
	\tau_{\textrm{5d}}= \frac{i}{2\pi} \frac{\partial^2 \mathcal{F}_{\textrm{IMS}}(\phi,|\mu_0|)}{\partial \phi^2} = 	\frac{i(9 \phi - 3 |\mu_0|)}{2\pi}.
\end{equation}
The endpoint of the moduli space is located at $\phi = |\mu_0|/3$ instead of $\phi=0$.

Let's briefly summarize the 5d result before we consider compactification. The prepotential is split into two parts depending on $\phi > |\mu_0|$ or $\phi < |\mu_0|$:
\begin{equation}
	\mathcal{F}_{\textrm{IMS}}(\phi,|\mu_0|) =
	\left\{ \begin{array}{l}
	\frac{4}{3}\phi^3 -  |\mu_0| \phi^2 \quad (\phi > |\mu_0|)\\
	\frac{3}{2}\phi^3 - \frac{3}{2}|\mu_0|\phi^2 + \frac{1}{2}|\mu_0|^2 \phi \quad (\frac{|\mu_0|}{3} < \phi < |\mu_0|)
	\end{array} \right.
\end{equation}
The monopole string tension is:
\begin{equation}
i T_{\textrm{mono}} = \frac{i}{2\pi} \frac{\partial \mathcal{F}_{\textrm{IMS}}(\phi,|\mu_0|)}{\partial \phi} = \left\{ \begin{array}{l}
i(2\phi^2 - |\mu_0| \phi)/\pi \quad (\phi > |\mu_0|) \\
i \left(3\phi - |\mu_0|\right)^2/4\pi \quad (\frac{|\mu_0|}{3} < \phi < |\mu_0|)
\end{array} \right.
\end{equation}
And the 5d coupling:
\begin{equation}
\tau_{\textrm{5d}} = \frac{i}{2\pi} \frac{\partial^2 \mathcal{F}_{\textrm{IMS}}(\phi,|\mu_0|)}{\partial \phi^2} = \left\{ \begin{array}{l}
i(4\phi - |\mu_0|)/\pi \quad (\phi > |\mu_0|) \\
i(9\phi - 3|\mu_0|)/2\pi \quad (\frac{|\mu_0|}{3} < \phi < |\mu_0|)
\end{array} \right.
\end{equation}
At the endpoint of the moduli space $\phi = |\mu_0|/3$, both the monopole string tension and the inverse gauge coupling-squared vanish, we will have a superconformal field theory which is usually denoted as $E_0$ theory.

Now let's consider the compactified theory on $\mathbb{S}^1$. From \eqref{Instanton prepotential} we can write down the prepotential directly as\footnote{The invariants $N_{m,n}$ here is no longer the same as the previous non-flopped F1.}
\begin{equation}
\frac{\partial^3 F_{\rm exact}(t,v)}{\partial t^3} = 9 + \sum_{m,n\geq 0} N_{m,n} \left(\frac{e^{2\pi m i (-3 t + v)}e^{2\pi n i (t - v)}}{1-e^{2\pi m i (-3 t + v)}e^{2\pi n i (t - v)}} \right) (-3m + n)^3.
\end{equation}
Setting $v = - i |\mu_0|R_5$ and redefine the parameter $a \equiv i t /R_5$ following the previous cases, one has
\begin{equation}\label{F1-prepotential-negative-flopped}
\frac{\partial^3 \mathcal{F}_{\textrm{5d},\mathbb{S}^1}(a,|\mu_0|)}{\partial a^3} = 9 + \sum_{m,n\geq 0} N_{m,n} \left(\frac{e^{-2\pi m (3 a - |\mu_0|)R_5}e^{-2\pi n (|\mu_0| - a)R_5 }}{1-e^{-2\pi m (3 a - |\mu_0|)R_5}e^{2\pi n (|\mu_0| - a) R_5}} \right) (-3m + n)^3.
\end{equation}

Start with \eqref{F1-prepotential-negative-flopped}, if we are closed to the singularity $\textrm{Re}(a) = |\mu_0|$ from the left side $\textrm{Re}(a) < |\mu_0|$ on the compactified 5d moduli space, then we can ignore the terms with $ m \neq 0$ in \eqref{F1-prepotential-negative-flopped} and obtain the former result \eqref{F1-prepotential-negative-flopped-original} since $N_{0,1}=1$ still holds for the flopped 2-cycle.

\setcounter{footnote}{0}

\begin{figure}[pbth]
\centering
\includegraphics[scale=0.5]{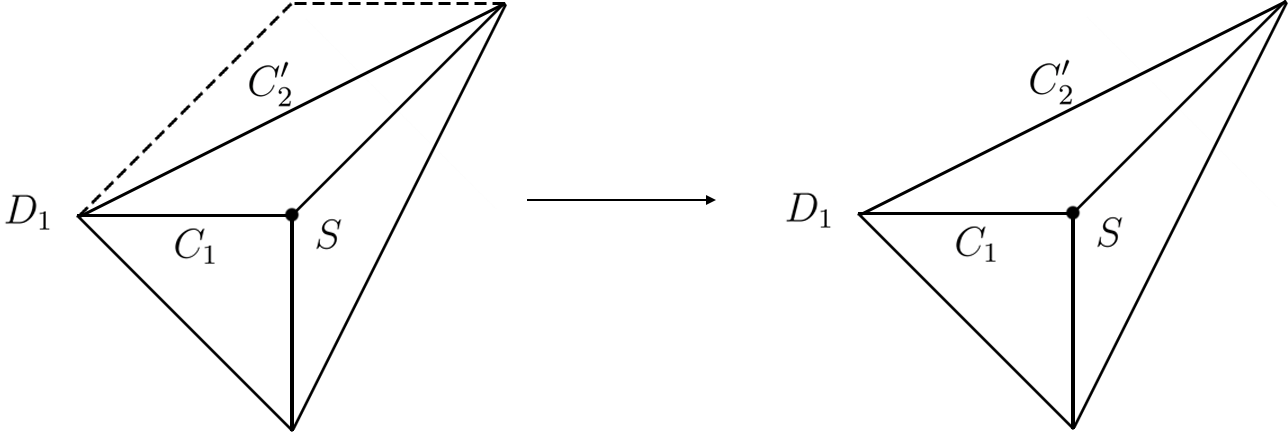}
\caption{The local $\mathbf{P}_2$ geometry as part of the flopped $F1$ geometry.}
\label{Fig-F1-P2}
\end{figure}

If we move to the core of the moduli space such that the volume of $C'_2$ becomes large, we can drop the terms with $n\neq 0$ in \eqref{F1-prepotential-negative-flopped} and the remaining is:
\begin{equation}
\frac{\partial^3 \mathcal{F}_{\textrm{5d},\mathbb{S}^1}(a,|\mu_0|)}{\partial a^3} \approx 9 + \sum_{m \geq 0} N_{m,0} \left(\frac{e^{-2\pi m (3 a - |\mu_0|)R_5}}{1-e^{-2\pi m (3 a - |\mu_0|)R_5}}\right)  (-3m)^3.
\end{equation}
If we redefine $a' \equiv (a - |\mu_0|/3)$, we will obtain the  prepotential for local $\mathbf{P}_2$ geometry:
 \begin{equation}\label{P2-prepotential}
\frac{\partial^3 \mathcal{F}_{\textrm{5d},\mathbb{S}^1}}{\partial a'^3} = 9 + \sum_{m\geq 0} N_{m} \left(\frac{e^{-2\pi m (3 a' R_5)}}{1-e^{-2\pi m (3 a' R_5)}}\right)  (-3m)^3,
\end{equation}
where the toric diagram is shown in figure \ref{Fig-F1-P2}: when $C'_2$ is large, we can decouple the dashed part and the remaining geometry is just a local  $\mathbf{P}_2$ . Here $N_m = N_{m,0}$ is the invariant of the $C_1$ cycle inside $\mathbf{P}_2$.

This theory is known to have three singularities related by a $\mathbf{Z}_3$ symmetry  \cite{Morrison:1996xf} in the strongly coupled region, and it is better to explore the structure of the moduli space from the IIB side. The IIB mirror curve, which can be found in Ref.~\cite{Aganagic:2006wq} for example,  is parameterized as:
\begin{equation}
	\sum_{i=1}^{3} x_i^3 - 3\psi \prod_{i=1}^{3} x_i = 0,
\end{equation}
$x_i (i=1,2,3)$ are complex coordinates and $\psi$ is the complex structure moduli in the IIB side which is dual to the K$\ddot{\textrm{a}}$hler moduli in the IIA side. There are three singularities for this curve and they are located at:
\begin{equation}
	\psi^3 = 1 \quad \rightarrow \quad \psi = 1, e^{\frac{2 \pi i}{3}}, e^{\frac{4 \pi i}{3}},
\end{equation}
and the point $\psi =0$ is the orbifold point and the local geometry in IIA side is $\mathbf{C}^3 / \mathbf{Z}_3$, which is also the endpoint of the moduli space. However, the moduli space is smooth since the mirror curve is well-behaved at $\psi = 0$.  The mirror map is given by:
\begin{equation}
	e^{-6\pi a' R_5} = z + \mathcal{O}(z^2),
\end{equation}
where  $z \equiv -1/(3\psi)^3$.  The three singularities at $z \sim \mathcal{O}(1)$ must be located around $\textrm{Re}(a') \sim 1/R_5$ on the moduli space. In  $R_5 \rightarrow \infty$ limit, the whole region will be pushed to the endpoint of the 5d moduli space at $\phi = |\mu_0|/3$, or at $\phi' = 0$ from the redefinition $a'\equiv a - |\mu_0|/3$.
\begin{figure}[pbth]
\includegraphics[scale=0.5]{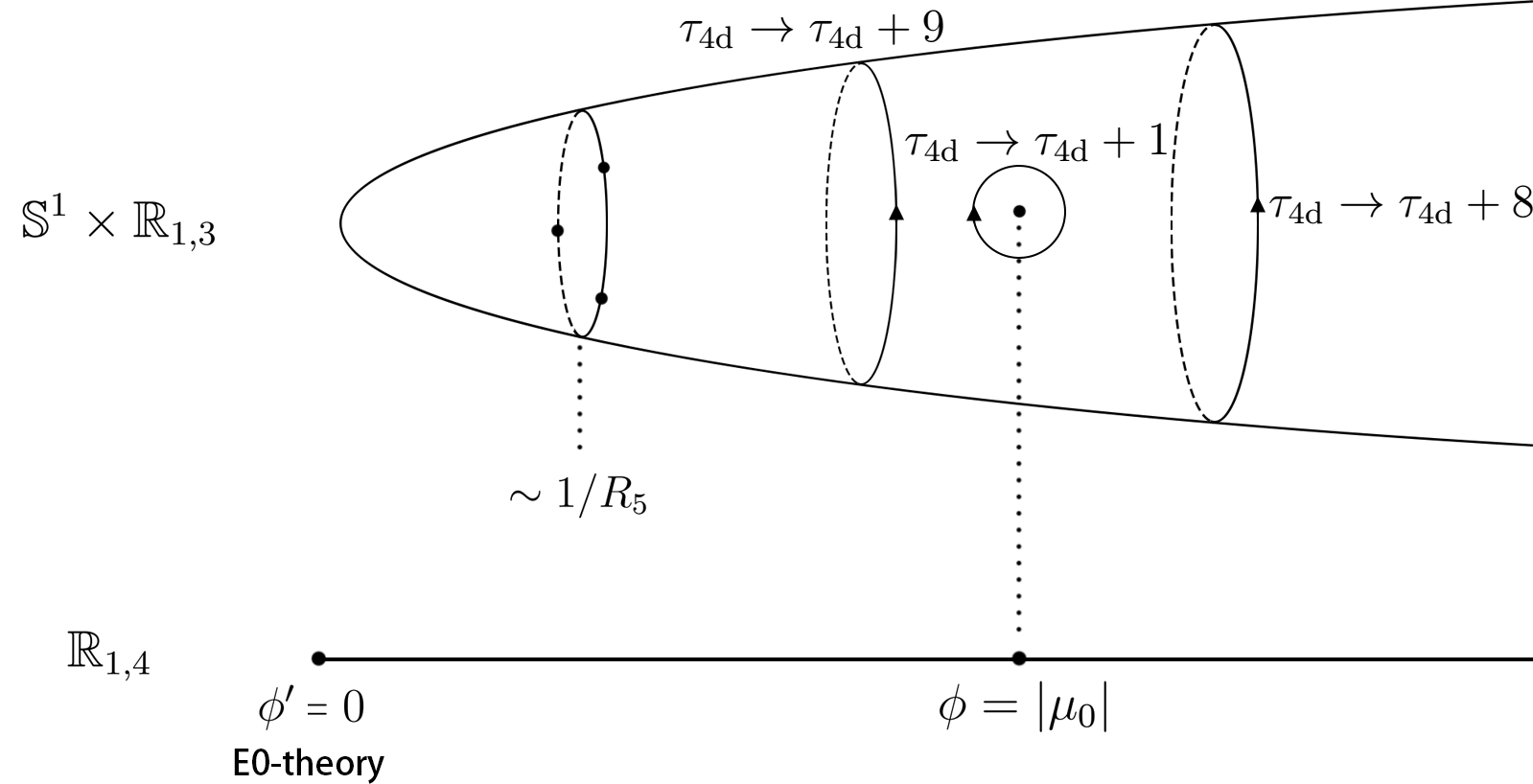}
\caption{Moduli space for F1-theory with negative coupling-squared.}
\label{Fig-F1-cigar-negative}
\end{figure}

The moduli space for this case is in figure \ref{Fig-F1-cigar-negative}. Near the end of the 5d moduli space, we are using coordinate $\phi' \equiv \phi - |\mu_0|/3$ such that the endpoint is at $\phi'=0$ as drawn in the figure. Here the three nodes around $\textrm{Re}(a') \sim 1/R_5$ are the three singularities in $\mathbf{P}_2$ theory, and the node in the middle is the logarithm singularity induced by the massless hypermultiplet of (1,0)-dyonic instanton. The monodromy  around the inner circle is $\tau_{\textrm{4d}} \rightarrow \tau_{\textrm{4d}} + 9$, while that around the logarithm singularity is $\tau_{\textrm{4d}} \rightarrow \tau_{\textrm{4d}} + 1$, they combine to the $\tau_{\textrm{4d}} \rightarrow \tau_{\textrm{4d}} + 8$ monodromy around the outer circle.

In $R_5\rightarrow \infty$ limit, the three $\mathbf{P}_2$ singularities shrink to the endpoint of the 5d moduli space $\phi'=0$ (or $\phi = |\mu_0|/3$), which corresponds to the $E_0$ theory. There are also marginal stability walls of the first kind connecting these three singularities, which will also shrink to the endpoint in the same way. Similarly, there could be other marginal stability walls of the first kind involving electric, flavour, and KK charges, and by a similar argument as the F0 case, all these walls will shrink to the $E_0$ point in the decompactification limit.

A remaining question is about the wall-crossing problem related to the singularity at $\phi = |\mu_0|$, which defines a marginal stability wall of the second kind. In the compactified theory, for a large $R_5$ 
the wall would extend along the circular Wilson-line direction of the small period $1/R_5$. Taking the $R_5 \rightarrow \infty$ limit, this naively suggests that magnetically charged strings would undergo a wall-crossing of some kind at $\phi = |\mu_0|$ here. We will return to this question in section 6.

\subsection{$\textrm{dP}_2$ Theory}

\begin{figure}[pbth]
\centering
\includegraphics[scale=0.4]{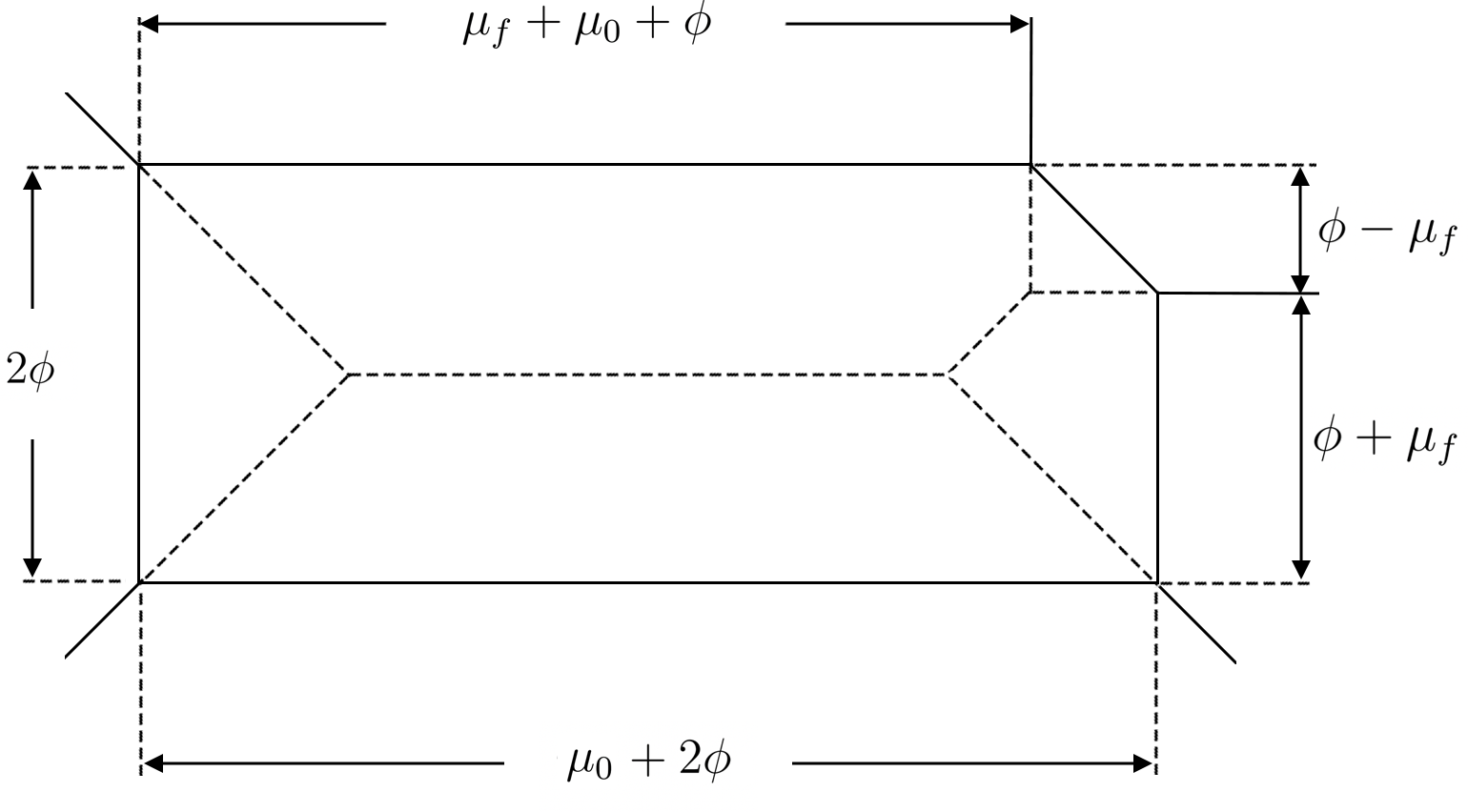}
\caption{The brane web diagram for dP$_2$ theory with positive $\mu_0$ and $\mu_f$.}
\label{Fig-Big-dP2}
\end{figure}

The brane web diagram for Sp(1) theory with a single flavor (or dP$_2$ theory) is given in figure \ref{Fig-Big-dP2}, which can be obtained by cutting a corner from the F0 brane web diagram. For simplicity, we work with a positive flavor mass $\mu_f$ and positive $\mu_0$. For the negative bare coupling squared, the discussion is parallel. We will give the toric diagram and 5d prepotential before and after the flop and briefly discuss the compactified theory's moduli space.

\begin{figure}[pbth]
\centering
\includegraphics[scale=0.3]{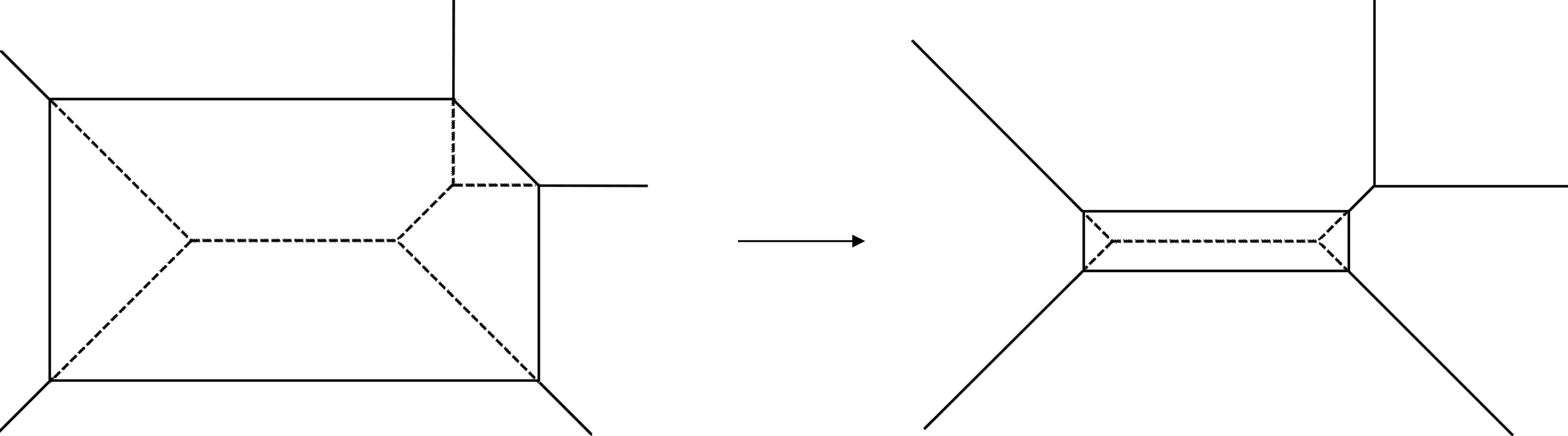}
\label{Fig-dP2-flop-brane}
\caption{The flop transition of dP$_2$ theory in terms of brane web.}
\end{figure}

The IMS prepotential for the brane web in figure \ref{Fig-Big-dP2} is:
\begin{align}\label{dP2-IMS-prepotential}
\mathcal{F}_{\textrm{IMS}}(\phi,\mu_0,\mu_f) &= \frac{4}{3} \phi^3 + (\mu_0 + \frac{1}{2} \mu_f)\phi^2 - \frac{1}{12} \left( (\phi + \mu_f)^3 + (\phi - \mu_f)^3 \right) \nonumber \\
&= \frac{7}{6} \phi^3 + (\mu_0 + \frac{1}{2}\mu_f)\phi^2 - \frac{1}{2}\mu_f^2 \phi,
\end{align}
which describes the Sp(1) theory with a single massive quark. The tension of the monopole string and the inverse 5d coupling-squared are:
\begin{align}
	i T_{\textrm{mono}} &= \frac{i}{2\pi} \frac{\partial \mathcal{F}_{\textrm{IMS}} }{\partial \phi} = \frac{i}{2\pi} \left[ 2\phi(2\phi + \mu_0) - \frac{1}{2}(\phi - \mu_f)^2\right],\nonumber \\
	  \tau_{\textrm{5d}} &= \frac{i}{2\pi} \frac{\partial^2 \mathcal{F}_{\textrm{IMS}}}{\partial \phi^2} = \frac{i}{2\pi} \left( 7\phi + 2 \mu_0 + \mu_f \right).
\end{align}
As the Coulomb moduli $\phi$ becomes smaller, the brane web will shrink along the dashed line in figure \ref{Fig-dP2-flop-brane}. At $\phi = \mu_f$, one component of the Sp(1) quark doublet will become massless, which is a singularity in the moduli space. Beyond this point, the brane web looks like the F0 brane web with an additional external vertex, which is a flop transition in terms of the Calabi-Yau geometry.
The prepotential for $\phi < \mu_f$ is similarly obtained from the IMS prepotential \eqref{5d IMS prepotential} as:
\begin{align}
\mathcal{F}_{\textrm{IMS}}(\phi,\mu_0,\mu_f) &= \frac{4}{3} \phi^3 + (\mu_0 + \frac{1}{2} \mu_f)\phi^2 - \frac{1}{12} \left( (\phi + \mu_f)^3 + (\mu_f - \phi)^3 \right) \nonumber \\
&= \frac{4}{3} \phi^3 + \mu_0 \phi^2 - \frac{1}{6} \mu_f^3
\end{align}
and it is the same as the pure Sp(1) IMS prepotential up to a constant $-\mu_f^3/6$, which does not enter the dynamics.

\begin{figure}[pbth]
\centering
\includegraphics[scale=1]{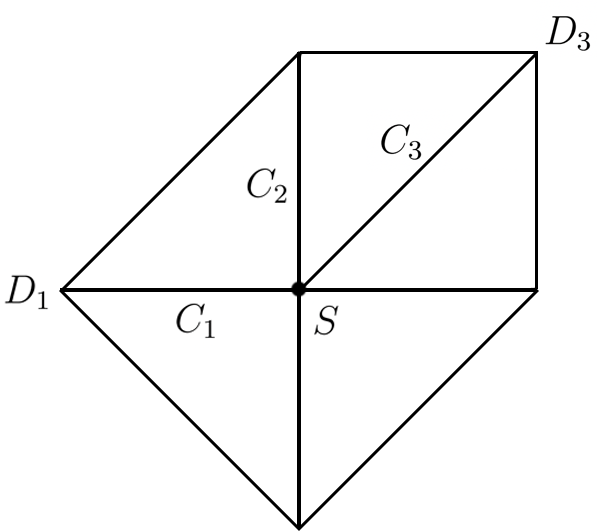}
\caption{The toric diagram for dP$_2$ theory.}
\label{Fig-dP2-toric}
\end{figure}

Now let's reinterpret the theory in terms of a toric diagram, which will also be useful in section 6. For $\phi > \mu_f$ the toric diagram is depicted in figure \ref{Fig-dP2-toric},
where we have three independent divisors $\{ S, D_1, D_3\}$ and among them $S$ is compact while $D_1,D_3$ are noncompact. There are also three compact 2-cycles dual to them: $\{C_1,C_2,C_3\}$. The intersection data among these divisors and 2-cycles are collected in the following \cite{Closset:2018bjz}:
\begin{itemize}
	\item Divisors:
		\begin{equation}
	S\cdot S \cdot S = 7.
		\end{equation}
		\begin{equation}
	S\cdot S \cdot D_1 = -2,\quad S\cdot S\cdot D_3 = -1.
		\end{equation}
		\begin{equation}
	S\cdot D_1 \cdot D_1 = 0,\quad S\cdot D_3 \cdot D_3 =  -1,\quad S\cdot D_1 \cdot D_3 = 0.
		\end{equation}
	\item 2-cycles:
		\begin{equation}
			C_1 \cdot S = -2,\quad C_2 \cdot S = -1,\quad C_3 \cdot S = -1.
		\end{equation}
		\begin{equation}
			C_1 \cdot D_1 =0,\quad  C_2 \cdot D_1 = 1,\quad C_3 \cdot D_1 = 0.
		\end{equation}
		\begin{equation}
			C_1 \cdot D_3 = 0,\quad C_2\cdot D_3 = 1,\quad C_3 \cdot D_3 = -1.
		\end{equation}
\end{itemize}
The K$\ddot{\textrm{a}}$hler form $[J]$ is expanded via:
\begin{equation}
	[J] = -\phi [S] + \mu_0 [D_1] + \mu_f [D_3],
\end{equation}
such that the volume of the 2-cycles $C_1,C_2,C_3$ are:
\begin{equation}
A_1 = J \cdot C_1 = 2\phi, \quad A_2 = J \cdot C_2 = \phi + \mu_0 + \mu_f,\quad A_3= J \cdot C_3 = \phi - \mu_f.
\end{equation}
The prepotential is similarly given via:
\begin{equation}
	\mathcal{F}_{\textrm{IMS}}(\phi,\mu_0,\mu_f) = -\frac{1}{6} J\cdot J \cdot J = \frac{7}{6} \phi^3 + (\mu_0 + \frac{1}{2}\mu_f)\phi^2 - \frac{1}{2}\mu_f^2 \phi,
\end{equation}
which is the same as \eqref{dP2-IMS-prepotential} up to some constant parts that we omitted.

\begin{figure}[pbth]
\centering
\includegraphics[scale=1]{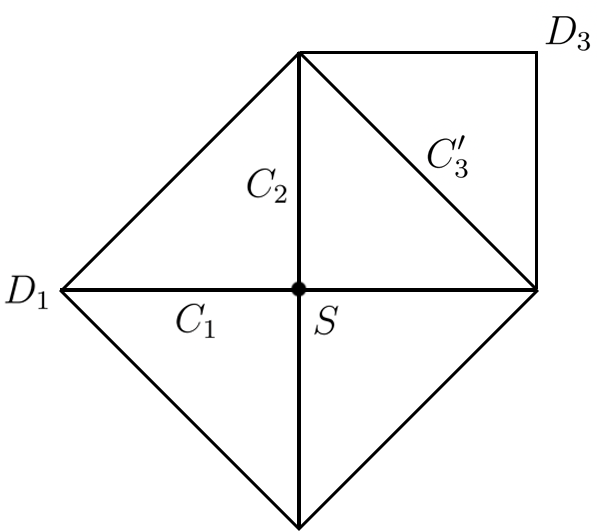}
\caption{The toric diagram for flopped $\widetilde{\textrm{dP}}_2$ theory.}
\label{Fig-dP2-toric-flop}
\end{figure}

When $\phi < \mu_f$, the geometry will become topologically distinct and the flopped toric diagram is depicted in figure \ref{Fig-dP2-toric-flop}\footnote{The flop transition is not unique for dP$_2$, one may also flop the curve $C_2$ to obtain F1 for example. Here we only consider the flop of $C_3$ curve which coincides the $(p,q)$ brane web. }, which will be called $\widetilde{\textrm{dP}}_2$ in the following discussions.
The independent divisors are still $\{S,D_1,D_3\}$ while the independent 2-cycles are $\{C_1,C_2,C'_3\}$ where $C'_3$ denotes the flopped cycle. The intersection data among them are \cite{Closset:2018bjz}:
\begin{itemize}
	\item Divisors:
		\begin{equation}
	S\cdot S \cdot S = 8.
		\end{equation}
		\begin{equation}
	S\cdot S \cdot D_1 = -2,\quad S\cdot S\cdot D_3 = 0.
		\end{equation}
		\begin{equation}
	S\cdot D_1 \cdot D_1 = 0,\quad S\cdot D_3 \cdot D_3 =  0,\quad S\cdot D_1 \cdot D_3 = 0.
		\end{equation}
	\item 2-cycles:
		\begin{equation}
			C_1 \cdot S = -2,\quad C_2 \cdot S = -2,\quad C'_3 \cdot S = 1.
		\end{equation}
		\begin{equation}
			C_1 \cdot D_1 =0,\quad  C_2 \cdot D_1 = 1,\quad C'_3 \cdot C_1 = 0.
		\end{equation}
		\begin{equation}
			C_1 \cdot D_3 = 0,\quad C_2\cdot D_3 = 0,\quad C'_3 \cdot D_3 = 1.
		\end{equation}
\end{itemize}
The K$\ddot{\textrm{a}}$hler form $[J]$ remains the same
and the volume of the 2-cycles $C_1,C_2,C'_3$ are:
\begin{equation}
A_1 = J \cdot C_1 = 2\phi, \quad A_2 = J \cdot C_2 = 2\phi + \mu_0,\quad A_3= J \cdot C_3 = \mu_f - \phi,
\end{equation}
in terms of new intersection data. Notice that $A_1$ and $A_2$ are the same as the F0 cases and they equal the mass of the W-boson and dyonic instanton. The 5d prepotential is then:
\begin{equation}
	\mathcal{F}_{\textrm{IMS}}(\phi,\mu_0,\mu_f) = -\frac{1}{6} J \cdot J \cdot J = \frac{4}{3} \phi^3 + \mu_0 \phi^2,
\end{equation}
which is again pure Sp(1) prepotential up to some constant parts that we omitted.

\begin{figure}[pbth]
\centering
\includegraphics[scale=0.3]{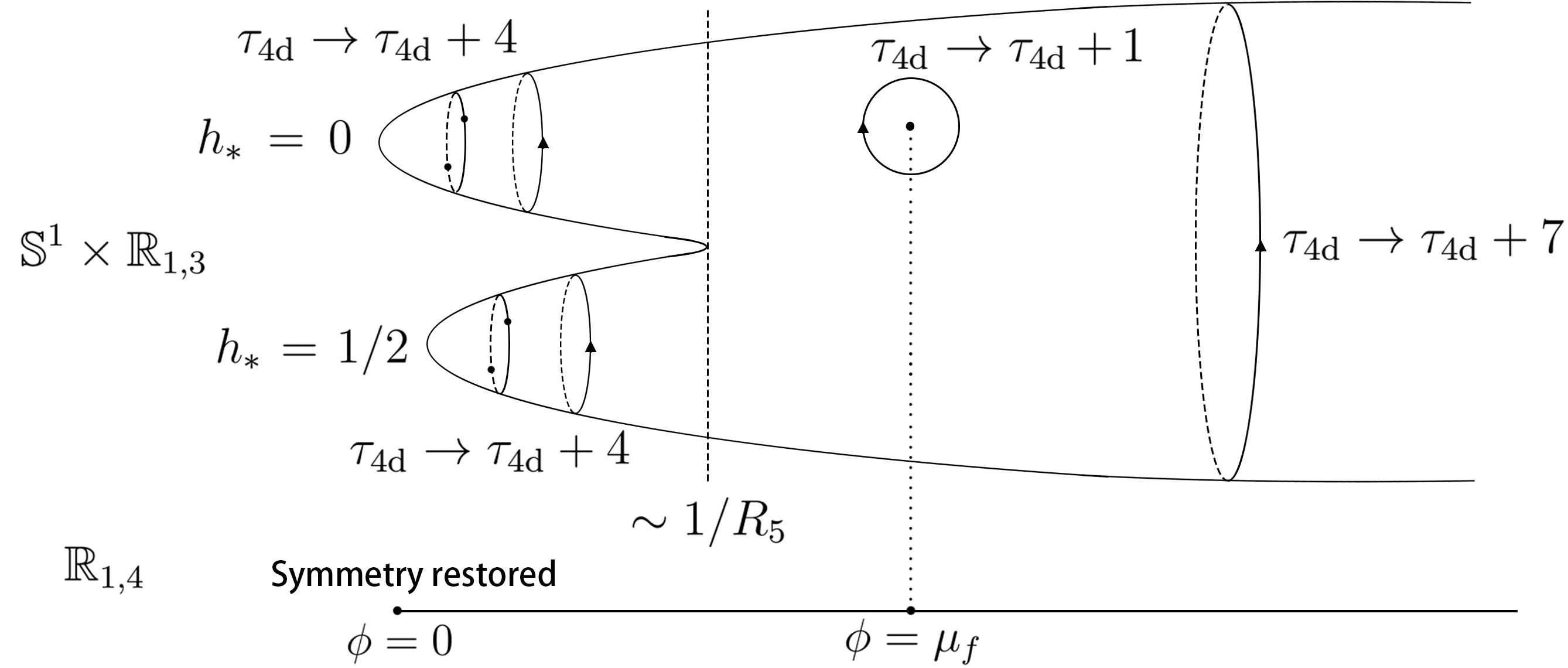}
\caption{Moduli space for dP$_2$-theory with $\mu_0,\mu_f > 0$.}
\label{Fig-dP2-cigar}
\end{figure}

The moduli space is depicted in figure \ref{Fig-dP2-cigar}, and there is an additional singularity compared to the pure F0 theory where one component of the quark becomes massless. Similar to the F1 case, a second kind marginal stability wall emanates from the massless quark point around the circle. After sending $R\rightarrow \infty$, the cigar will collapse to the ray of 5d moduli space, and all the first kind marginal stability walls will collapse to the endpoint.

\section{Wall-Crossing of a Magnetic BPS String}

In the previous section, we have seen that, for dP$_2$ theory and F1 theory with a negative bare coupling-squared,  a marginal stability wall of the second kind emanates from the point where a charged hypermultiplet becomes massless. These walls extend toward the imaginary and periodic direction, so, in the decompactification limit, will mark a dividing point in the Coulombic half-line. For these two examples, the dividing point is responsible for a flop transition from the Calabi-Yau viewpoint.

Here, we will discuss what happens to magnetic strings across such a flop transition by concentrating on dP$_2$ theory as a concrete example. The same idea can be applied straightforwardly to the flop transition that occurs for the F1 theory with negative bare coupling squared. The discontinuity we encounter here can be traced to how certain chiral fermion enters or does not enter the low energy dynamics of the magnetic string. The elliptic genus captures this from a geometric viewpoint, while an old-fashioned field theory argument also shows the same. Furthermore, this continuity can also be viewed as a by-now-familiar wall-crossing once we consider the BPS objects in question as D4-D2-D0 bound states. All three viewpoints yield the common
discontinuity, as we delineate in this last section.

\subsection{(0,4) Magnetic String and Modified Elliptic Genera}

Let's consider the dynamics of the monopole string given by the M5-brane wrapped on the compact divisor $S$ and study how it changes before/after the flop transition. The low-energy dynamics of the monopole string is a 2d (0,4) superconformal field theory that has been well-studied in various literature. We will briefly review them and refer the readers to the literature \cite{Maldacena:1997de, Minasian:1999qn} for further details.

The 2d (0,4) effective low energy dynamics arise from a reduction of the 6d (0,2) theory on the M5-brane wrapped on a compact divisor in the Calabi-Yau 3-fold $\hat{X}$. We will assume the divisor is rigid such that there are no deformation degrees of freedom, as appropriate for a unit monopole string. The five scalars in 6d produce three scalars in 2d, encoding $\IR^3$ positions of the monopole string. The others are chiral scalars obtained from the reduction of the chiral 2-form $B_2$ on the M5-brane along the harmonic 2-forms $H^2(S,\mathbb{R})$ on $S$, which is further decomposed into self-dual and anti-self-dual parts: $H^2(S,\mathbb{R}) = H^{2,+}(S,\mathbb{R}) \oplus H^{2,-}(S,\mathbb{R})$. Since $S$ is a K$\ddot{\textrm{a}}$hler manifold $H^{2,-}(S,\mathbb{R})$ is of Hodge type $(1,1)$, and $H^{2,+}(S,\mathbb{R})$ can be decomposed as:
\begin{equation}
	H^{2,+}(S,\mathbb{R}) = H^{2,0}(S,\mathbb{R}) \oplus H^{0,2}(S,\mathbb{R}) \oplus [J]_S,
\end{equation}
where $[J]_S$ is the K$\ddot{\textrm{a}}$hler form. The scalars associated to the self-dual 2-forms in $H^{2,+}(S,\mathbb{R})$ are right-moving while those associated to the anti-self-dual 2-forms in  $H^{2,-}(S,\mathbb{R})$ are left-moving.

The fermions living on the monopole string originate from 6d fermions in the $\mathbf{4}$ of USp(4),  which are also decomposed via $H^2(S,\mathbb{R})$: the odd forms lead to 2d left-moving fermions and the even forms lead to 2d right-moving fermions. There are no odd harmonic forms on the divisor $S$\footnote{If $S$ is a very ample divisor, that is true due to the Lefschetz hyperplane theorem, which says $b_1(S)$ is isomorphic to $b_1(\hat{X})$. In the present  $S$ are rigid divisors which are blow-ups of $\mathbf{P}^2$ or $\mathbf{P}^1 \times \mathbf{P}^1$, and $b_1(S)$ will also be zero. } and all the fermions are right-moving; therefore, only the right-moving sector forms supermultiplets.

In the following, we will consider the elliptic genus of the 2d (0,4) superconformal theory, which has been studied in various papers \cite{deBoer:2006vg,Gaiotto:2006wm,Gaiotto:2007cd,Denef:2007vg,Alim:2010cf}. In particular, we will mainly consider a single cover of M5-brane wrapped on a rigid divisor $S$ and we will assume $b_2^+(S)=1$, which means $H^2(S,\mathbb{R}) = H^{1,1}(S,\mathbb{R})$ and the only self-dual 2-from on $S$ is the K$\ddot{\textrm{a}}$hler form $[J]_S$.

After compactification on $R_5$, the M5-brane wrapped on the divisor $S$ can be treated at D4-brane. On the worldvolume of D4-brane, the {\rm U}(1) fluxes can be expanded as:
\begin{equation}
	F_2 = q^{A} [C_{A}]_S,
\end{equation}
which will generate D2-charges on the divisor $S$. Here $ [C_{A}]_S $ are harmonic 2-forms belong to the self-dual lattice $\Lambda_S \equiv H^2(S,\mathbb{Z})$ with $A = 1,\cdots,\dim \Lambda_S$, which are the Poincar$\acute{\textrm{e}}$ dual of the 2-cycles $C_{A}$ inside $S$. The flux over any 2-cycles $C_{B}$ is:
\begin{equation}
	\int_{C_{B}}  q^{A} [C_{A}]_S = q^{A} D_{A B} \equiv q_B,
\end{equation}
and should be quantized. We define the intersection matrix $D_{A B}$ on $\Lambda_S$ as $D_{AB} = C_A \cdot C_B |_S$, which will also serve as the metric on $\Lambda_S$. Therefore it seems the lattice $\ Lambda_S$ characterizes the D2-brane charges. But as discussed in \cite{Maldacena:1997de,Minasian:1999qn}, the physical D2-charges on $S$ are always labelled by the lattice $\Lambda \equiv i^* H^2(\hat{X},\mathbb{Z})$ which is generically a sublattice of $\Lambda_S$, where $i$ is the embedding $S\hookrightarrow \hat{X}$. If a 2-form $[C]$ in $S$ is exact in the ambient $\hat{X}$, then the corresponding 2-cycle $C$ must be a boundary of some surface $B$ such that $C = \partial B$, and it is possible to have a membrane instanton with worldvolume $B$ \cite{Minasian:1999qn}. That indicates a charge element in $\Lambda_S$ which is not in $\Lambda$ will decay to a state charged in $\Lambda$.

Therefore the physical flux should be expanded using the bulk 2-form $[D_a]$ and the charges are:
\begin{equation}
	\int_{C_{b}}  q^{a} [D_{a}]_S = q^{a} D_{ab}\equiv q_b,
\end{equation}
which belongs to the dual lattice $\Lambda^*$\footnote{There is an alternative way to see that. Here $D_{ab}=C_a \cdot D_b$ is also the Schwinger pairing between D2 and D4-branes. Since the D4-charges belong to $ H^2(\hat{X},\mathbb{Z}) = H_4(\hat{X},\mathbb{Z}) $, the D2-charges must lie in the dual lattice with respect to the Schwinger pairing, which is $\Lambda^*$.}. Here $ [D_{a}]_S = i^* [D_a] (a=1\cdots \dim \Lambda) $ form a basis of $\Lambda$ and $D_{ab} =  C_a \cdot D_b$ is the intersection matrix on $\Lambda$, which is also a block of $D_{AB}$. In the following, we will continue to use capital $A,B,C\cdots$ for indices in $\Lambda_S$ and small $a,b,c\cdots$ for indices in $\Lambda$ and $\Lambda^*$.

The modified elliptic genus that counts  the BPS states is given by \cite{deBoer:2006vg,Gaiotto:2006wm,Gaiotto:2007cd,Denef:2007vg,Alim:2010cf}:
\begin{equation}
	Z'(\tau,\bar{\tau},y) = \textrm{Tr}_R \frac{1}{2} F^2 (-1)^F e^{\pi i D_{ab}s^a q^b}  q^{L'_0 - \frac{c_L}{24}} \bar{q}^{\bar{L}'_0 - \frac{c_R}{24}} e^{2\pi i y_a q^a}.
\end{equation}
where the Virasoro generators $L'_0$ and $\bar{L}'_ 0$ can be obtained by reducing the energy-momentum tensor of M5-brane to the monopole string. In general, $S$ may not be a spin manifold, and in that case, there is the  Freed-Witten anomaly \cite{Freed:1999vc}, which requires one to turn on a half-integral flux $c_1(S)/2$ on $S$. Since $c_1(\hat{X})=0$, one has $c_1(S) = -[S]$ by adjunction formula and therefore the membrane charge $q^a$ is shifted by a vector $s^a/2$, where $s^a$ is the component of the 2-form $[S]$ pulled back to the divisor $S$ itself. $F$ is the fermion number counted by  $2 J_R^3$, where $J_R$ is the SO(3) R-symmetry which is also the spatial rotation group. The modified elliptic genus is weighted by an extra phase $\exp {\pi i D_{ab}s^a q^b}$, which is necessary for the modified elliptic genus to be modular invariant \cite{deBoer:2006vg}.

We can factorize the center of mass part by integrating out the momentum $\vec{p}_{\textrm{cm}}$. Rewriting the Virasoro generators $L'_0$ and $\bar{L}'_0$ as:
\begin{equation}
	L'_0 = \frac{1}{2} \vec{p}_{\textrm{cm}}^2 + L_0,\quad \bar{L}'_0 = \frac{1}{2} \vec{p}_{\textrm{cm}}^2 + \bar{L}_0,
\end{equation}
the modified elliptic genus is then:
\begin{align}
	Z' (\tau,\bar{\tau},y) &= \int d^3 p_{\textrm{cm}} e^{\pi i (\tau - \bar{\tau}) \vec{p}_{\textrm{cm}}^2} Z(\tau,\bar{\tau},y),\\
	&\sim (\textrm{Im}(\tau))^{-3/2} Z(\tau,\bar{\tau},y),
\end{align}
and we will mainly focus on the $Z(\tau,\bar{\tau},y)$ and simply call that elliptic genus in the following.

The elliptic genus $Z(\tau,\bar{\tau},y)$ is subject to a $\theta$-function decomposition which has been studied in various papers \cite{deBoer:2006vg,Gaiotto:2006wm,Gaiotto:2007cd,Denef:2007vg,Alim:2010cf,Kraus:2006nb,Dabholkar:2005dt}. We leave the details in the appendix and summarize the results in the following. The elliptic genus can be split as:
\begin{equation}\label{dP2-ellitpic-genus}
Z(\tau,y) = \sum_{\gamma \in \Lambda^* / \Lambda} f_{\gamma}(\tau) \theta_{\gamma}(\tau,y).
\end{equation}
Here $\gamma \in \Lambda^* / \Lambda$ is called shift vector, $f_{\gamma}(\tau)$ is a holomorphic function:
\begin{equation}
	f_{\gamma}(\tau) = \sum_{\hat{q}_0}  d_{\gamma}(\hat{q}_0) e^{2\pi i \tau (\hat{q}_0 - \chi(S)/24)},
\end{equation}
where $\chi(S)$ is the Euler number of the divisor $S$, $d_{\gamma}(\hat{q}_0)$ is related to the BPS index $\Omega(q_0,q_a)$ as:
\begin{equation}
	d_{\gamma}(\hat{q}_0) = \Omega\left(\hat{q}_0 - \frac{1}{2}\left( \gamma + \frac{s}{2}\right)^2 - \frac{\chi(S)}{24} , \frac{s^a}{2}+\gamma^a \right),
\end{equation}
and $\hat{q}_0$ is defined in the appendix as \eqref{dP2-D0-charges-spliting}. The $\theta_{\gamma}(\tau,y)$ is:
\begin{equation}
	\theta_{\gamma}(\tau,y) = \sum_{k \in \Lambda + [S]/2} e^{\pi i D_{ab}s^a (k + \gamma)^b} e^{-\pi i \tau (k+\gamma)^2} e^{2\pi i y_a (k+\gamma)^a},
\end{equation}
and we have set $\tau = \bar{\tau}$ in the above expression.

\subsection{Discontinuity } 

\begin{figure}[pbth]
\centering
\includegraphics[scale=0.5]{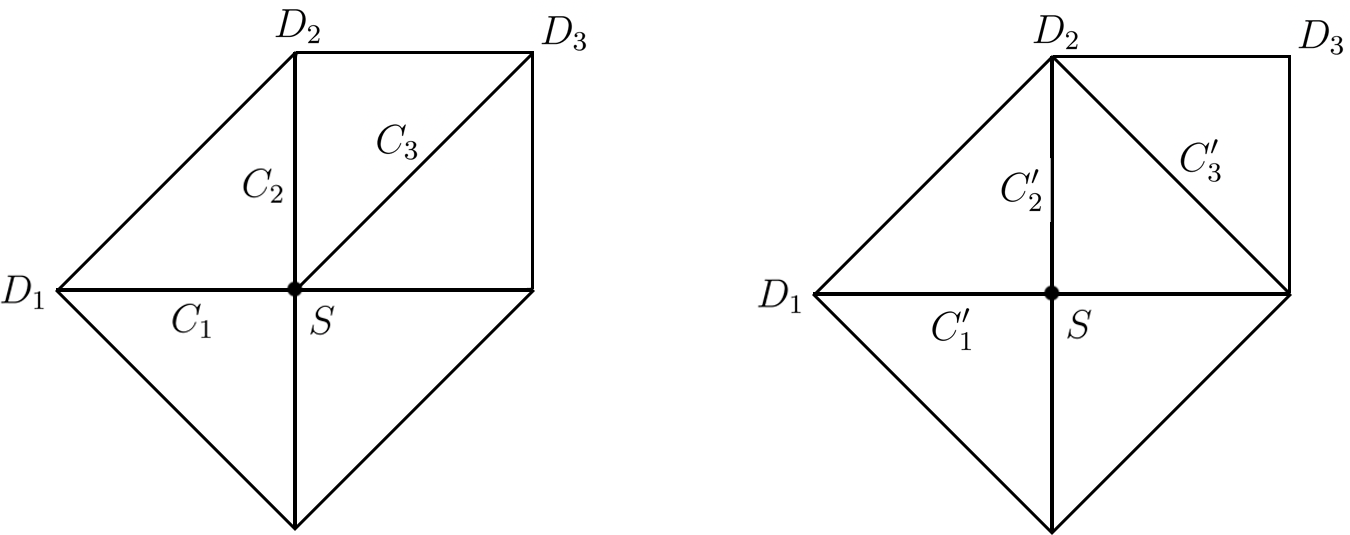}
\caption{The toric diagram for dP$_2$ and $\widetilde{\textrm{dP}}_2$. }
\label{Fig-dP2-toric-all}
\end{figure}

In this subsection we will use the formula \eqref{dP2-ellitpic-genus} to calculate the elliptic genus for dP$_2$ and $\widetilde{\textrm{dP}}_2$ geometry, where the toric diagrams are figure \ref{Fig-dP2-toric-all} and we have labelled all 2-cycles in the flop geometry $\widetilde{\textrm{dP}}_2$ as $C'$ in order to distinguish those from dP$_2$. The intersection data are given in the previous section. The compact divisor $S$ is rigidly embedded in the ambient space, the (co)homology group on $S$ can be read from the toric diagram directly, where we have $h^{0,0} = h^{2,2} =1$ and $h^{1,1}$ equals the number of independent 2-cycles inside $S$, the rest Hodge numbers are zero\footnote{These surfaces are obtained from $\mathbf{P}_2$ or $\mathbf{P}_1 \times \mathbf{P}_1$ by blowing up some holomorphic 2-spheres, and therefore cannot have any non-trivial odd cycles or $H_{2,0}$ and $H_{0,2}$ cycles.}.

We collect the geometry and CFT data in the following table for $\textrm{dP}_2$ and $\widetilde{\textrm{dP}}_2$.
\begin{table}[pbth]
\centering
\begin{tabularx}{1\textwidth}{>{\centering\arraybackslash}X|>{\centering\arraybackslash}X|>{\centering\arraybackslash}X|>{\centering\arraybackslash}X|>{\centering\arraybackslash}X|>{\centering\arraybackslash}X}
&  $b_2^-(S)$  & $b_2^+(S)$ & $\chi(S)$ & $c_L$ & $c_R$  \\
\hline
dP$_2$ & 2 & 1 &5& 5 & 6 \\
\hline
$\widetilde{\textrm{dP}}_2$   & 1 & 1 & 4&4 & 6 \\
\end{tabularx}
\label{SCFT-data}
\caption{Geometry and CFT data for $\textrm{dP}_2$ and $\widetilde{\textrm{dP}}_2$.}
\end{table}
For dP$_2$ there are three independent 2-cycles $C_1,C_2,C_3$ inside the divisor $S$ such that $b_2(S)$=3, and during the flop transition $C_3$ shrinks to zero and then blows up transversely into $C'_3$, therefore for $\widetilde{\textrm{dP}}_2$ there are only two independent 2-cycles $C'_1,C'_2$ inside $S$ and $b_2(S)=2$. The 2-forms are Poincar$\acute{\textrm{e}}$ dual of those 2-cycles in $S$ and among them one particular combination is the self-dual the K$\ddot{\textrm{a}}$hler form, the others are anti-self-dual. The right-moving sector is a (0,4) supermultiplet which contains three bosonic fields arising from $\IR^3$ positions  of the monopole string, one bosonic field due to the reduction of the chiral $B_2$-field on the self-dual K$\ddot{\textrm{a}}$hler form and four fermionic fields from the decomposition of the $\mathbf{4}$ of USp(4) on the 0-from, while the left-moving sector is non-supersymmetric and contains the left-moving part of  $\IR^3$ position scalars and $b^-_2(S)$ left-moving scalar fields. The Euler number $\chi(S)$ is given by $\chi(S) = b_2(S) +2$, which is the same as the left moving central charge, where the latter is contributed by the left-moving scalars which is $3+b_2^+(S) = 2 + b_2(S)$.

We begin with the dP$_2$ theory, in which one has $H^2(X,\mathbb{Z}) = i^*H^2(X,\mathbb{Z})=H^2(S,\mathbb{Z})$, therefore the two lattices $\Lambda_S$ and $\Lambda$ are the same. The intersection matrix $C_a \cdot C_b |_S$ can be read from the bulk triple intersection numbers $D_a \cdot D_b \cdot S$ which is:
\begin{equation}
	D_{ab} \equiv C_a \cdot C_b |_S = \left( \begin{array}{ccc}
	0&1&0\\
	1&-1&1\\
	0&1&-1
	\end{array}\right).
\end{equation}
The determinant of the intersection matrix is unit, which means the charge lattice is self-dual (as it should be since $\Lambda_S$ is always self-dual), therefore we need not to consider about the gluing vector $\gamma$. For a single D4-brane the summation over $\hat{q}_0$ gives \cite{Gottsche:1990,Alim:2010cf}:
\begin{equation}
	\sum_{\hat{q}_0} d(\hat{q}_0) e^{2\pi i \tau (\hat{q}_0 - \chi(S)/24)} = \frac{1}{\eta(\tau)^{\chi(S)}},
\end{equation}
where $\eta(\tau) = q^{-1/24} \prod_{n=1}^{\infty} (1-q^n) $ is the Dedekind eta function and $q \equiv \exp(2\pi i \tau)$. This part is contributed by the left-moving oscillating modes of the scalar fields on the monopole string. In the dP$_2$ case that is $\eta(\tau)^{-5}$ and the elliptic genus is:
\begin{equation}
	Z_{\textrm{dP}_2}(\tau,y) = \frac{\theta_{\textrm{dP}_2}(\tau,y)}{\eta(\tau)^5} ,
\end{equation}
where $\theta_{\textrm{dP}_2}(\tau,y)$ is given by:
\begin{equation}
	\theta_{\textrm{dP}_2}(\tau,y) = \sum_{k \in \Lambda + [S]/2} e^{\pi i D_{ab}s^a k^b} e^{-\pi i \tau k^2} e^{2\pi i y_a k^a}.
\end{equation}
We need to determine the 2-form $[S]$ in the dP$_2$ theory. The definition of the intersection matrix tell us:
\begin{equation}
	\int_S [C_A]_S \wedge [C_B]_S = \int_{\hat{X}} [D_A] \wedge [D_B] \wedge [S],
\end{equation}
which means the 2-forms correspond to each divisor $[D_A]$ will pull back to the 2-forms $[C_A]_S$ on the divisor $S$. $S$ is related to the divisor $D_1,D_2,D_3$ by $S = -2D_1 - 2 D_2 - D_3$ which can be read from the toric vectors, therefore it will pull back to $[S]_S = -2 [C_1]_S - 2[C_2]_S - [C_3]_S$ on the divisor $S$ itself. One can then read the charge vector $s^a$ to be $s^a = (-2,-2,-1)$. Taking that into account, we find that $k^1,k^2 \in \mathbb{Z}$ and $k^3$ is shifted by half such that $k^3 \in \mathbb{Z}+1/2$, and the $\theta$ part is then:
\begin{equation}\label{dP2-theta-term}
	\theta_{\textrm{dP}_2}(\tau,y) = \sum_{k^1,k^2 \in \mathbb{Z}, k^3 \in \mathbb{Z}+1/2}  e^{-\pi i (k^2 + k^3) } e^{-\pi i (2k^1 k^2 + 2k^2 k^3 - (k^2)^2 - (k^3)^2)}  e^{2\pi i (y_1  k^1 + y_2 k^2 + y_3 k^3)}.
\end{equation}

For the $\widetilde{\textrm{dP}}_2$ theory the discussion is similar. $H^2(\hat{X},\mathbb{Z})$ is larger than $H^2(S,\mathbb{Z})$ since the basis for $H^2(\hat{X},\mathbb{Z})$ is  $\{[D_1], [D_2], [D_3]\}$ while the basis for $H^2(S,\mathbb{Z})$ is $\{[C'_1]_S, [C'_2]_S\}$, where $C'_3$ is outside the divisor $S$. However, the intersection between $D_3$ and $S$ is zero in the $\widetilde{\textrm{dP}}_2$ geometry, which indicates the pull back of $[D_3]$ to $S$ is trivial and one still has $i^* H^2(\hat{X},\mathbb{Z}) = H^2(S,\mathbb{Z})$ and thus $\Lambda = \Lambda_S$. In fact, the dynamics of the monopole string only depends on the geometry of $S$ itself, we can simply ignore the $D_3$ node in the toric diagram and treat the geometry as F0.

One still has:
\begin{equation}
	f_{\widetilde{\textrm{dP}}_2}(\tau) = \frac{1}{\eta^4(\tau)},
\end{equation}
where the Euler number of $S$ has changed to $\chi(S) = 4$ and the partition function $Z_{\widetilde{\textrm{dP}}_2}(\tau,y)$ is then:
\begin{equation}
	Z_{\widetilde{\textrm{dP}}_2}(\tau,y) = \frac{\theta_{\widetilde{\textrm{dP}}_2}(\tau,y)}{\eta(\tau)^4}.
\end{equation}
The intersection matrix $D'_{a b}$ is:
\begin{equation}\label{dP2-flop-intersection-matrix}
D'_{a b} \equiv C'_a \cdot C'_b |_S =  \left( \begin{array}{cc}
0&1\\1&0
\end{array}\right).
\end{equation}
The divisor $S$ is still equivalent to $S=-2 D_1 - 2D_2 - D_3$ and it will pull back to $[S']_S = -2 [C'_1]_S - 2[C'_2]_S$ on the divisor $S$ itself, where we have used the fact $S\cdot D_3 =0$. Therefore the shift vector is $s'^a = (-2,-2)$ and the $\theta$-part is:
\begin{equation}\label{dP2-flop-theta-term}
	\theta_{\widetilde{\textrm{dP}}_2}(\tau, y') = \sum_{k'^1,k'^2 \in \mathbb{Z}} e^{-2\pi i \tau  k'^1 k'^2} e^{2\pi i (y'_1 k'^1 + y'_2 k'^2)}.
\end{equation}

We still need to relate these two results together. During the flop transition, we can treat the new 2-cycles $\{ C'_a \}$ as a linear transformation of the original basis $\{ C_a \}$, such that the intersection numbers of $\{ C'_a \}$ with all divisors in $\hat{X}$ are correctly determined by those of $\{ C_a \}$. From this point of view, the linear transformation can be written as:
\begin{equation}\label{dP2-2cycles-transformation}
\left( \begin{array}{c}
C'_1\\C'_2\\C'_3
\end{array}\right) = \left(\begin{array}{ccc}
1&0&0\\0&1&1\\0&0&-1
\end{array}\right) \left( \begin{array}{c}
C_1\\C_2\\C_3
\end{array}\right).
\end{equation}
We start with the partition function $Z_{\textrm{dP}_2}(\tau,y)$ and change the parameters $\{k^a,y_a\}$ according to the new basis $\{C'_a\}$ as:
\begin{equation}
	(k'^1,k'^2,k'^3)  = (k^1,k^2,k^3) \left(\begin{array}{ccc}
1&0&0\\0&1&1\\0&0&-1
\end{array}\right)^{-1} = (k^1, k^2, k^2 - k^3),
\end{equation}
here one can see that $k'^1,k'^2 \in \mathbb{Z}$ and $k'^3 \in \mathbb{Z}+1/2$. The chemical potentials $\{y_a\}$ changes as:
\begin{equation}
\left( \begin{array}{c}
y'_1\\ y'_2\\ y'_3
\end{array}\right) = \left(\begin{array}{ccc}
1&0&0\\0&1&1\\0&0&-1
\end{array}\right) \left( \begin{array}{c}
y_1\\y_2\\y_3
\end{array}\right) = \left( \begin{array}{c}
y_1\\y_2+y_3\\-y_3
\end{array}\right).
\end{equation}
Moreover, the intersection matrix in terms of the $\{C'\}$ basis is:
\begin{equation}
	D'_{a b} = \left( \begin{array}{cc|c}
	0&1&0\\
	1&0&0\\
\hline
	0&0&-1
	\end{array}\right).
\end{equation}
Here we need to stress that what we have done is merely a change of basis in dP$_2$ geometry, and the intersection matrix above in terms of the new basis should not be mixed up with the intersection matrix for $\widetilde{\textrm{dP}}_2$ in \eqref{dP2-flop-intersection-matrix}, although the $2\times 2$ block correctly reproduce \eqref{dP2-flop-intersection-matrix}.

Replacing the parameters $\{k^a,y_a\}$ in terms of $\{k'^a,y'_a\}$, one can rewrite the $\theta$-term \eqref{dP2-theta-term} in elliptic genus $Z_{\textrm{dP}_2}$ as:
\begin{equation}
\theta_{\textrm{dP}_2} (\tau, y') =  \sum_{k'^1,k'^2 \in \mathbb{Z}}e^{-2\pi i k'^1 k'^2}e^{2\pi i (y'_1 k'^1 + y'_2 k'^2)} \sum_{k'^3 \in \mathbb{Z}+1/2}  e^{\pi i k'^3 } e^{\pi i \tau (k'^3)^2}  e^{2\pi i y'_3 k'^3},
\end{equation}
which can be factorized into two combinations. We recognize the first one as the $\theta$-term \eqref{dP2-flop-theta-term} for $Z_{\widetilde{\textrm{dP}}_2}$, and recall the definition of $\theta_{11}$-function:
\begin{equation}
	\theta_{11}(\tau,y) = i \sum_{n} (-1)^n \left( e^{\pi i \tau} \right)^{(n+1/2)^2} \left(e^{2\pi i y}\right)^{n+1/2},
\end{equation}
one immediately gets:
\begin{equation}
\theta_{\textrm{dP}_2} (\tau, y') = \theta_{\widetilde{\textrm{dP}}_2} (\tau, y') \theta_{11}(\tau,y'_3).
\end{equation}
Taking into account of the difference between $f_{\textrm{dP}_2}(\tau)$ and $f_{\widetilde{\textrm{dP}}_2}(\tau)$, one obtains the relations between the elliptic genus $Z_{\textrm{dP}_2}$ and $Z_{\widetilde{\textrm{dP}}_2}$:
\begin{equation}\label{dP2-relation-of-elliptic-genus}
	Z_{\textrm{dP}_2} = - Z_{\widetilde{\textrm{dP}}_2} \times \frac{ \theta_{11}(\tau,y_3)}{\eta(\tau)},
\end{equation}
where we have used the relation $y'_3 = -y_3$ and $\theta_{11}(\tau,-y) = -\theta_{11}(\tau,y)$.

We use the dP$_2$ and $\widetilde{\textrm{dP}}_2$ as a illustration of the change of elliptic genus under flop transition. Actually, one can repeat the procedure in this subsection for other geometries, like F1 geometry in the last section, and it turns out the above relation is quite universal. In general, during the flop transition, the dimension of the $H^2(S,\mathbb{Z})$ of the compact divisor $S$ will decrease by one since there is one particular 2-cycle flopping out of the divisor $S$, which indicates we are losing one left-moving bosonic degree of freedom on the monopole string whose partition function will give the difference between two elliptic genera.

Although we performed the above computation in the geometric context of the elliptic genera, there is a simpler old-fashioned field theory physics behind \eqref{dP2-relation-of-elliptic-genus}. The venerable Jackiw-Rebbi zero-mode story  \cite{Jackiw:1975fn} is applicable for the hypermultiplet fermions, and a well-known result from the Callias index theorem \cite{Callias:1977kg} is that an Sp(1) doublet fermion produces a zero-mode on the unit monopole for $\phi > \mu_f$ while it is lifted on the other side, $\phi < \mu_f$. In 4d, the zero mode induces a double degeneracy of the monopole, so the wall of marginal stability emanating from $a=\mu_f$ is closely related to this Jackiw-Rebbi phenomenon.

For a 5d monopole string, the same zero-mode exists along the length of the monopole string such that it really produces a two-dimensional chiral fermion on the string. As such, the effective two-dimensional theory of the monopole string would differ by this chiral fermion on the two sides of the flop transition. The latter fermion, which does not come with supersymmetry partners, would contribute multiplicatively to the partition function, so one must expect discontinuity of the elliptic genus.

Once the monopole string wraps the circle $\IS^1$, the chiral fermion will split into a tower of fields labelled by the KK-level $\psi_{n},\bar{\psi}_n$, and they satisfy the hermitian conjugate relations:
\begin{equation}
	\psi_n^{\dagger} = \bar{\psi}_{-n},\quad \bar{\psi}_n^{\dagger} = \psi_{-n}.
\end{equation}
The vacuum state has a double degeneracy due to the Clifford algebra of $\{ \psi_0,\bar{\psi}_0\}$, with flavour charges $-1/2$ and $+1/2$, in particular.
The partition function for such a single left-moving fermion reads
\bea
Z_{\psi}(\tau,y_3)& = &\textrm{Tr} \ q^{L_0+a_{\psi}} e^{\pi i s^a D_{a3} q^3} e^{2\pi i y_3 q^3} \cr\cr
&=& q^{\frac{1}{12}} (ie^{-\pi i y_3} -ie^{\pi i y_3}) \prod_{n=1}^{\infty} (1 - e^{2\pi i y_3} q - e^{-2\pi i y_3} q + q^2)\cr\cr
&=& -\frac{\theta_{11}(\tau,y_3)}{\eta(\tau)},
\eea
where the phase factor $\exp(\pi i s^a D_{a3} q^3)$ is inserted to fit the modified elliptic genus and $a_{\psi}$ is the normal ordering constant. This handsomely explains \eqref{dP2-relation-of-elliptic-genus}.

Of course, this field theory understanding is an alternate view on the same surplus of chiral degrees of freedom on one side of the flop. In the geometrical language, we have characterized these as chiral scalars, but of course, in 2d the usual bosonization makes the distinction moot.  This merely reminds us that such discontinuity is inherently a piece of the field theory and does not require geometric engineering.

\subsection{D4-D2-D0 Wall-Crossing Revisited}

With the monopole string wrapped on $\IS^1$, the elliptic genus must capture the degeneracies of a single D4 bound with D2's and D0's. In the large $R_5$ limit, the monopole central charge and D0 central charge are both almost pure imaginary while that D2 central charge goes as $\phi-\mu_f$ plus an imaginary part. As such, the marginal stability walls for D4-D2-D0 would emanate from singularities in the vicinity of $a=\mu_f$. In terms of the central charge plane, these singularities would be spaced at an even interval $\sim 1/R_5$, along the imaginary direction. In the limit of very large $R_5$, on the other hand, the monopole central charge dominates over the D0 central charges, so these singularities, labeled by the D0 charge, will eventually collapse toward $\phi=\mu_f$ in the Coulomb phase.

Given how these elliptic genera compute the degeneracies of certain D4-D2-D0 systems, the same discontinuity we saw above should also follow from the Kontsevich-Soibelman wall-crossing algebra \cite{Kontsevich:2008fj, Kontsevich:2009xt}. One must accumulate an infinite number of wall-crossings, once for each singularity and the wall thereof, and sum up degeneracies of D2's and D0's into a generating function to reach the discontinuity of the elliptic genus across a flop at $\phi-\mu_f$. This task has been performed elsewhere \cite{Nishinaka:2010qk}, but here, we briefly review the computation for the sake of completeness. Since we wish to connect the result to the elliptic genera above, we must give a little more care to how charges are matched across the transition and how the generating function is related to the elliptic genera of the previous subsection.

With a single D4-brane wrapped on $S$ to which various D2/D0-branes are attached,
the charge vector is
\begin{equation}
	\Gamma = [S] + q^a [C_a] - q_0 [d V],
\end{equation}
where $[S] \in H^2(X)$, $[C_a] \in H^4(X)$ and $[d V] \in H^6(X)$ is the volume element. Assuming $R_5$ is finite but sufficiently large, the central charge will be dominated by the D4-brane $[S]$ which is purely imaginary. Let's consider the decaying process:
\begin{equation}
	\Gamma \rightarrow \Gamma_1 + \Gamma_2,
\end{equation}
where $\Gamma_1$ is chosen to be:
\begin{equation}
	\Gamma_1 = \beta^a [C_a] - \delta [dV],
\end{equation}
which is a generic D2-D0 bound state, and $\Gamma_2 = \Gamma - \Gamma_1$ is the remaining part which consists of the D4-brane.

The central charge of D0-brane is also imaginary, but they are proportional to the inverse radius $R_5^{-1}$ and can be omitted in the $R_5 \rightarrow \infty$ limit. Therefore the central charge of $\Gamma_1$ totally depends on those of D2-branes, and near the massless quark point (where $C_3$ shrinks to zero) in the moduli space, we have a marginal stability wall emanating from that point and extend along the circular Wilson-line direction of the small period $1/R_5$\footnote{More accurately, there is a wall emanating from the massless quark points $\beta^a [C_a] - \delta [dV]$ for any number of D0-charge $\delta$, and they overlap when $R_5$ is sufficiently large. One can introduce a phase to the mass parameter $\mu_f$ and $\mu_0$ to separate these walls apart as the authors did in  \cite{Nishinaka:2010qk}.}.

An important simplification in the present case is that, for a membrane wrapped on the 2-cycle $C_3$, the only non-trivial contribution to the index $\Omega$ are:
\begin{equation}
\Omega(\pm [C_3] - \delta [dV]) = 1,\quad \Omega(\delta [dV]) = -2,
\end{equation}
which means we can focus on the special cases that $\beta^3 = \pm 1,0$. Moreover, if $\beta^3 = 0$, then $\Gamma_1$ consists only of D0-branes and the Dirac pairing between $\Gamma_1$ with $\Gamma_2$ will be zero since there is no D6-brane involved in the problem, and one cannot have a wall-crossing like that. Therefore it is suffice to consider $\beta^3 = \pm 1$.

\begin{figure}[pbth]
\centering
\includegraphics[scale=0.5]{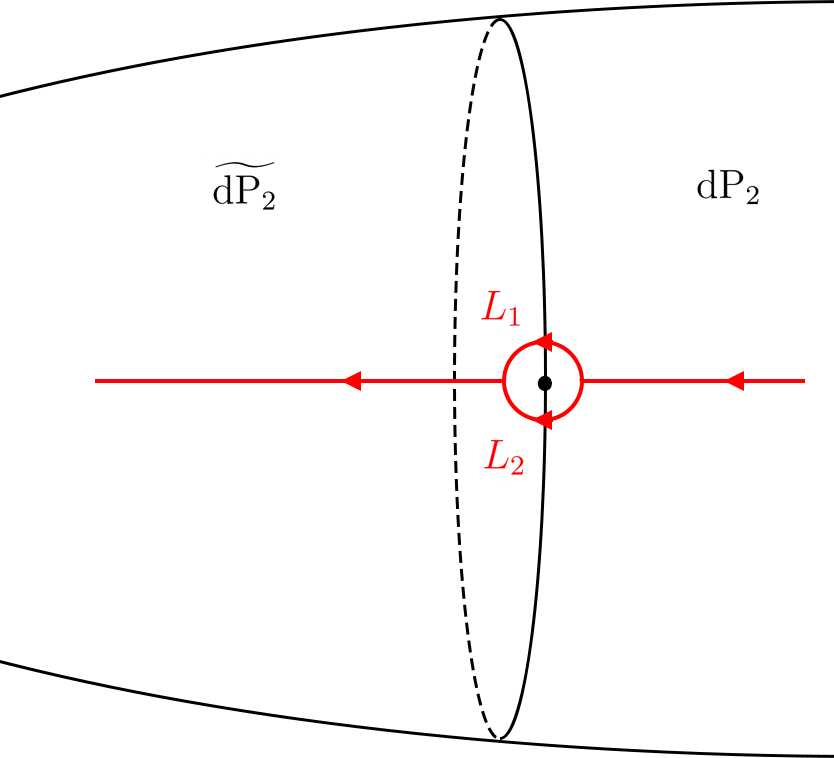}
\caption{Two paths crossing the marginal stability wall.}
\label{Fig-dP2-wall-crossing}
\end{figure}

Now we can solve the wall-crossing problem using the Kontsevich-Soibelman algebra. Assuming the geometry of the region above the wall is dP$_2$ and the region below the wall is $\widetilde{\textrm{dP}}_2$ as shown in figure \ref{Fig-dP2-wall-crossing}, one defines the generating function $\mathcal{G}$ in the dP$_2$ region as:
\begin{equation}
	\mathcal{G}(\tau,y^a) = \sum_{q_0,q^a} \Omega^{\pm} \left(q^a,q^0\right) e^{i\pi D_{ab}q^a s^b} e^{2\pi i \tau q^0} e^{2\pi i q^a y_a},
\end{equation}
which will be eventually related to the elliptic genera $Z$ of the previous two subsections.
The phase factor $\exp (i\pi D_{ab}q^a s^b)$ is inserted to match the elliptic genus, i.e.,
\begin{equation}
\mathcal{G}_{\textrm{dP}_2}(\tau,y)=Z_{\textrm{dP}_2}(\tau,y)
\end{equation}
on the dP$_2$ side of the flop.

We will consider two paths,  denoted as $L_1$ and $L_2$ in figure \ref{Fig-dP2-wall-crossing}, they differ by how one moves around the quark singularity.
The details can be found in  \cite{Nishinaka:2010qk}, the generating functions in the $\widetilde{\textrm{dP}}_2$ region $\mathcal{G}_{\widetilde{\textrm{dP}}_2}$ is related to $\mathcal{G}_{\textrm{dP}_2}$ as:
\begin{equation}
\mathcal{G}_{\textrm{dP}_2}(\tau,y) = \mathcal{G}_{\widetilde{\textrm{dP}}_2}(\tau,y)  \times (1-e^{\pm 2\pi i y_3}) \prod_{n=1}^{\infty} \left(1 - e^{2\pi i n \tau} e^{-2\pi i y_3} \right)\left(1 - e^{2\pi i n \tau} e^{2\pi i y_3} \right),
\end{equation}
where the minus sign is for the path $L_1$ and plus sign is for $L_2$. Recall the alternative definition of the $\theta_{11}$-function:
\begin{equation}
	\theta_{11}(\tau,q) = -2 q^{\frac{1}{8}} \sin(\pi y) \prod_{n=1}^{\infty} (1-q^n) (1-2\cos(2\pi y)q^n + q^{2n}),
\end{equation}
where $q = \exp 2\pi i \tau$. One can rewrite the difference of the partition function as:
\begin{equation}\label{dP2-relation-of-elliptic-genus-wallcrossing}
\mathcal{G}_{\textrm{dP}_2}(\tau,y) = -\mathcal{G}_{\widetilde{\textrm{dP}}_2}(\tau,y) \times \left(\mp i e^{\pm i \pi y_3} q^{-\frac{1}{12}} \frac{\theta_{11}(\tau,y_3)}{\eta(\tau)} \right),
\end{equation}
where the minus sign is for $L_1$ and plus sign is for $L_2$.

The jump we see here from the wall-crossing picture \cite{Nishinaka:2010qk} appears slightly different from that between the elliptic genera \eqref{dP2-relation-of-elliptic-genus} by an extra factor $\mp i e^{\pm i \pi y_3} q^{-\frac{1}{12}}$.
Recall that in the dP$_2$ geometry the flux on the D4-brane is shifted by $\frac{1}{2}[S] = -[C_1]_S - [C_2]_S - \frac{1}{2}[C_3]_S$. Using the basis transformation described in \eqref{dP2-2cycles-transformation}, in terms of new basis $\{ C'_a \}$ the flux is $\frac{1}{2}[S] = -[C'_1]_S - [C'_2]_S -\frac{1}{2} [C'_3]_S$. Also, we have seen that in the $\widetilde{\textrm{dP}}_2$ geometry the additional flux is $\frac{1}{2} [S'] = - [C'_1]_S - [C'_2]_S$, therefore during the flop transition the D2 charge on the D4-brane will shift implicitly by $-\frac{1}{2} [C'_3]_S$. At the same time, the geometry of $S$ is changing during the flop transition, and Euler number $\chi(S)$ will also jump by $-1$, which will shift the D0 charge by $1/24$ through the dependence of $\eta$-function.

The problem is merely that the generating function $\mathcal{G}$ did not take into account these overall fractional
shifts of D2 and D0 charges across the flip in a manner the elliptic genus does. Once we take these shifts into account, the generating function on the other side is related to $Z_{\widetilde{\textrm{dP}}_2}$ is
\begin{equation}
\mathcal{G}_{\widetilde{\textrm{dP}}_2} = Z_{\widetilde{\textrm{dP}}_2} \times \left( q^{-\frac{1}{24}} \right)\times \left(e^{i\pi D'_{33}(-\frac{1}{2})(-1)} e^{- i\pi \tau D'_{33}(-\frac{1}{2})^2} e^{2\pi i y'_3 (-\frac{1}{2})} \right),
\end{equation}
where the $q^{-1/24}$ in the first parenthesis is because $\mathcal{G}_{\widetilde{\textrm{dP}}_2}$ still believes the D0 charge is shifted by $-5/24$ instead of $-4/24$, therefore this factor shifts the Euler number dependence from $q^{-4/24}$ in $Z_{\widetilde{\textrm{dP}}_2}$ to $q^{-5/24}$. The second parenthesis is due to that $\mathcal{G}_{\widetilde{\textrm{dP}}_2}$ would think there is still a D2-charge $-\frac{1}{2}[C'_3]_S$ remaining after the wall-crossing, which comes from the difference between $[S]$ and $[S']$.
Collecting them together and further recalling that $D'_{33} = -1$ and $y'_3 = -y_3$, one finds
\begin{equation}
\mathcal{G}_{\widetilde{\textrm{dP}}_2} = Z_{\widetilde{\textrm{dP}}_2} \times  (-i)q^{\frac{1}{12}}e^{\pi i y_3} .
\end{equation}
Substituting it into \eqref{dP2-relation-of-elliptic-genus-wallcrossing} one gets:
\begin{equation}
Z_{\textrm{dP}_2}(\tau,y) = -Z_{\widetilde{\textrm{dP}}_2}(\tau,y) \times \frac{\theta_{11}(\tau,y_3)}{\eta(\tau)},
\end{equation}
for the path $L_2$, as desired. The other path $L_1$ gives
\begin{equation}
Z_{\textrm{dP}_2}(\tau,y) = -Z_{\widetilde{\textrm{dP}}_2}(\tau,y) \times \left(-e^{2\pi i y_3}\right) \frac{\theta_{11}(\tau,y_3)}{\eta(\tau)},
\end{equation}
which, relative to $L_2$, merely reflects the monodromy $q^3 \rightarrow q^3 + 1$ around the singularity.

\section*{Acknowledgments}

We would like to thank Xin Wang, Zhihao Duan for discussions. QJ and PY were supported by KIAS Individual Grants (PG080801 and PG005704) at Korea Institute for Advanced Study.

\appendix
\section{The $\theta$-Decomposition of Elliptic Genus}

In this appendix, we give the details of the $\theta$-decomposition of the elliptic genus.  Due to the insertion of factor $(-1)^F F^2$ in the definition of the elliptic genus, only the short BPS multiplets will contribute to the elliptic genus. That has a strong restriction to the right-moving (supersymmetric) sector, and the BPS condition is:
\begin{equation}
	\bar{L}_0 - \frac{1}{2} q_R^2 -\frac{c_R}{24} = 0,
\end{equation}
here $\vec{q}_R$ is the right-moving momentum of the scalar fields along the monopole string which satisfies:
\begin{equation}\label{dP2-right-moving-momentum}
	q_R^2 = \frac{(D_{AB}q^A \lambda^B)^2}{D_{AB}\lambda^A \lambda^B},
\end{equation}
where $\lambda^A$ are the components of the K$\ddot{\textrm{a}}$hler form $[J]_S$ and $q^A$ are D2-charges, which are also the winding numbers of scalar fields on the monopole string.  On the other hand, the left-moving sector is non-supersymmetric and is not constrained by the BPS condition. The D0-brane charge $q_0$ as the momentum along the string is $(L_0 - \bar{L}_0)-(\frac{c_L - c_R}{24})$ by definition, and is also given by:
\begin{equation}\label{dP2-D0-charges}
	q_0 = - \frac{1}{2} \int_S F_2 \wedge F_2 - \frac{\chi(S)}{24} + N,
\end{equation}
where the first two terms come from the Chern-Simons action of D4-brane and $N$ is the number of isolated D0-branes which are not carried by $F_2$ fluxes\footnote{With a suitable non-zero B-field turned on, they can be realized as smooth non-commutative {\rm U}(1) instantons. }. The Euler number $\chi(S) = b_2^+(S) + b_2^-(S) + 2$ is equal to the left-moving central charge $c_L = b_2^-(S) + 3$ for $b_2^+=1$, which also holds for general cases with $b_2^+ \neq 1$, see  \cite{Maldacena:1997de, Minasian:1999qn}.

The elliptic genus $Z(\tau,\bar{\tau},y)$ has a good property that allows it to be decomposed into theta functions. Generically, the lattices $\Lambda$ and $\Lambda^*$ are sublattices of $\Lambda_S$; we will denote $\Lambda^{\bot}$ as the lattice in $\Lambda_S$ which is orthogonal to $\Lambda$ with respect to the intersection matrix $D_{A B}$. Then $\Lambda \oplus \Lambda^{\bot}$ is a sublattice of $\Lambda_S$ and one can decompose the $F_2$ fluxes in $S$ as \cite{Denef:2007vg}:
\begin{equation}
	F_2 = \frac{[S]}{2} + f^{\|} + f^{\bot} + \gamma,
\end{equation}
where $f^{\|} \in \Lambda$, $f^{\bot}\in \Lambda^{\bot}$ and $\gamma \in \Lambda_S / (\Lambda \oplus \Lambda^{\bot})$ is called the gluing vector which can be further decomposed into:
\begin{equation}
	\gamma = \gamma^{\|} + \gamma^{\bot}.
\end{equation}
Since $\Lambda_S$ is self-dual, by the Nikulin primitive theorem  \cite{Nikulin:1980} one has the isomorphism:
\begin{equation}
	\Lambda_S / (\Lambda \oplus \Lambda^{\bot}) \approx \Lambda^* / \Lambda \approx (\Lambda^{\bot})^* / \Lambda^{\bot},
\end{equation}
such that $\gamma^{\bot}$ and $\gamma$ will be uniquely determined after $\gamma^{\|}$ is given, and we can identify the gluing group  $\Lambda_S / (\Lambda \oplus \Lambda^{\bot})$ with the determinant group $\Lambda^* / \Lambda$. The D0-brane charge induced by the $F_2$ fluxes is then:
\begin{equation}
	-\frac{1}{2}\int_S F_2 \wedge F_2 = -\frac{1}{2}(\frac{[S]}{2} + f^{\|} + \gamma^{\|})^2 - \frac{1}{2}(f^{\bot} + \gamma^{\bot})^2,
\end{equation}
and the total D0-brane charge is:
\begin{align}\label{dP2-D0-charges-spliting}
q_0 &= -\frac{1}{2}(\frac{[S]}{2} + f^{\|} + \gamma^{\|})^2 - \frac{1}{2}(f^{\bot} + \gamma^{\bot})^2 - \frac{\chi(S)}{24} + N \nonumber \\
	&= -\frac{1}{2}(\frac{[S]}{2} + f^{\|} + \gamma^{\|})^2 - \frac{\chi(S)}{24} + \hat{q}_0,
\end{align}
with $\hat{q}_0 \equiv N - (f^{\bot} + \gamma^{\bot})^2/2$.

The partition function can be written as:
\begin{align}
Z(\tau,\bar{\tau},y) = &\sum_{\gamma \in \Lambda^* / \Lambda } \sum_{k \in \Lambda + [S]/2} \  \sum_{\hat{q}_0} \Omega(q_0,k^a + \gamma^a)  \nonumber \\
&\times e^{\pi i D_{ab}s^a \cdot (k + \gamma)^b} e^{2\pi i \tau q_0} e^{\pi i (\tau - \bar{\tau}) q_R^2} e^{2\pi i y_a (k+\mu)^a},
\end{align}
where one has to sum over the directions along $(\Lambda^{\bot})^*$ to obtain the index $\Omega(q_0,k^a + \gamma^a)$, which only depends on the physical D2-brane charges in $\Lambda^*$. The K$\ddot{\textrm{a}}$hler moduli enters the elliptic genus through the right-moving momentum $\vec{q}_R$ via \eqref{dP2-right-moving-momentum}. In the following, we will simply set $\tau = \bar{\tau}$ such that the elliptic genus will be locally independent of the K$\ddot{\textrm{a}}$hler moduli and counts the degeneracies of short BPS states. Of course, the elliptic genus will still change as we vary the moduli to cross the marginal stability wall.

Now the key is that there is a shifting symmetry of D0 and D2 charges $q_0$ and $q_a$:
\begin{align}
	q^a &\rightarrow q^a + t^a \\
	q_0 &\rightarrow q_0 - D_{ab} q^a t^b - \frac{D_{ab}t^a t^b}{2}
\end{align}
such that the index is invariant:
\begin{equation}
	\Omega (q_0 - D_{ab} q^a t^b - \frac{D_{ab}t^a t^b}{2} , q^a + t^a) = \Omega(q_0,q^a),
\end{equation}
where $t^a \in \Lambda$. This can be understood from the monodromies in the moduli space of the Calabi-Yau $\hat{X}$ in the large volume limit. Using the shift symmetry, we can shift the vector $k^a \in \Lambda + [S]/2$ in the index $\Omega(q_0,k^a + \gamma^a)$ to $s^a/2$ and obtain:
\begin{equation}
	\Omega(q_0,k^a + \gamma^a) = \Omega\left(\hat{q}_0 - \frac{1}{2}\left( \gamma + \frac{s}{2}\right)^2 - \frac{\chi(S)}{24} , \frac{s^a}{2}+\gamma^a \right) \equiv d_{\mu}(\hat{q}_0),
\end{equation}
which is independent of $k^a$ and we will denote it as $d_{\mu}(\hat{q}_0)$. Substitute that into the elliptic genus one has:
\begin{equation}
Z(\tau,y) = \sum_{\gamma \in \Lambda^* / \Lambda } \sum_{k \in \Lambda + [S]/2} \  \sum_{\hat{q}_0} d_{\gamma}(\hat{q}_0)  e^{\pi i D_{ab}s^a \cdot (k + \gamma)^b} e^{2\pi i \tau (\hat{q}_0 - \chi(S)/24)} e^{-\pi i \tau (k+\gamma)^2} e^{2\pi i y_a (k+\gamma)^a},
\end{equation}
which can be factorized as:
\begin{equation}
Z(\tau,y) = \sum_{\gamma \in \Lambda^* / \Lambda} f_{\gamma}(\tau) \theta_{\gamma}(\tau,y),
\end{equation}
where $f_{\gamma}(\tau)$ is holomorphic:
\begin{equation}
	f_{\gamma}(\tau) = \sum_{\hat{q}_0}  d_{\gamma}(\hat{q}_0) e^{2\pi i \tau (\hat{q}_0 - \chi(S)/24)},
\end{equation}
and $\theta_{\gamma}(\tau,y)$ is:
\begin{equation}
	\theta_{\gamma}(\tau,y) = \sum_{k \in \Lambda + [S]/2} e^{\pi i D_{ab}s^a (k + \gamma)^b} e^{-\pi i \tau (k+\gamma)^2} e^{2\pi i y_a (k+\gamma)^a}.
\end{equation}

\end{document}